\documentclass[extra]{gji}
\usepackage{amssymb,color,graphicx,longtable,mathrsfs,times,url}
\usepackage[fleqn]{amsmath}
\usepackage{listings}
\usepackage{lineno}
\usepackage{xspace}
\usepackage{setspace}
\usepackage{verbatim}

\newcommand{\unboxed}{}
\newcommand{\V}{$\vert$}
\newcommand{\argu}{\frac{2\nu^\half}{\pi\rho}r}

\newcommand{\Sbar}{\bar\mcS}

\newcommand{\also}{\qquad\mbox{and}\qquad}

\newcommand{\for}{\qquad\mbox{for}\qquad}

\newcommand{\sQ}{\mathsf{Q}}

\newcommand{\bbg}{{\mbox{\boldmath$\bar{\gamma}$}}}

\newcommand{\bmcF}{\mbox{\boldmath$\mathcal{F}$}}

\newcommand{\bnabla}{\mbox{\boldmath$\nabla$}}

\newcommand{\btheta}{\boldsymbol{\theta}}
\newcommand{\bthetap}{\boldsymbol{\theta^{\prime}}}

\newcommand{\hbtheta}{\mbox{\boldmath$\hat\theta$}}
\newcommand{\btruth}{\btheta_0}

\newcommand{\bzero}{\mbf{0}}

\newcommand{\cov}{\mbox{cov}}

\newcommand{\Dx}{\Delta x}
\newcommand{\Dy}{\Delta y}

\newcommand{\bgth}{\bar{\gamma}_{\theta}}
\newcommand{\bgthp}{\bar{\gamma}_{\theta'}}

\newcommand{\half}{\frac{1}{2}}
\newcommand{\hbt}{{\mbox{\boldmath$\hat{\theta}$}}}
\newcommand{\hsomm}{\hspace{-0.1em}}
\newcommand{\hsom}{\hspace{0.1em}}

\newcommand{\hspst}{\hspace{0.25em}}
\newcommand{\hsps}{\hspace{0.75em}}

\newcommand{\inlaw}{\stackrel{\text{law}}{\longrightarrow}}

\newcommand{\km}{\mathrm{km}}
\newcommand{\kbp}{\mbf{k'}}
\newcommand{\kb}{\mbf{k}}

\newcommand{\m}{\mathrm{m}}
\newcommand{\mbf}{\mathbf}
\newcommand{\mcC}{{\mathcal{C}}} 
\newcommand{\mcF}{{\mathcal{F}}}
\newcommand{\mcH}{{\mathcal{H}}} 

\newcommand{\mcL}{{\mathcal{L}}}
\newcommand{\mcN}{{\mathcal{N}}}

\newcommand{\mcS}{{\mathcal{S}}}

\newcommand{\bmth}{\bar{m}_{\theta}}
\newcommand{\bmthp}{\bar{m}_{\theta'}}

\newcommand{\nff}{\left(\frac{\Dy\Dx}{\Ny\Nx}\right)^\half}
\newcommand{\nffs}{\left(\frac{\Dy\Dx}{\Ny\Nx}\right)}

\newcommand{\norml}{\frac{1}{K}\sum_{\kb}}
\newcommand{\normltu}{\frac{1}{K^2}\sum_{\kb}\sum_{\kbp}}

\newcommand{\Nx}{\mathrm{N_x}}
\newcommand{\Ny}{\mathrm{N_y}}

\newcommand{\ofkp}{(\kbp)}
\newcommand{\ofk}{(\kb)}

\newcommand{\ofsk}{(k)}

\newcommand{\ofsy}{(y)}

\newcommand{\ofth}{(\hbt)}
\newcommand{\oftr}{(\btruth)}

\newcommand{\oft}{(\btheta)}

\newcommand{\ofst}{_{\btheta}}
\newcommand{\ofstp}{_{\bthetap}}
\newcommand{\ofx }{(\xb)}
\newcommand{\ofy }{(\yb)}
\newcommand{\ofxp}{(\xb')}

\newcommand{\p}{\phantom{0}}
\newcommand{\hp}{\hspace{0.25em}}

\newcommand{\pl}{\partial}

\newcommand{\stx}{s^2_X}
\newcommand{\st}{{\sigma^2}}

\newcommand{\var}{\mbox{var}}

\newcommand{\xb}{\mbf{x}}
\newcommand{\yb}{\mbf{y}}

\onecolumn
\title[Statistical analysis of irregularly and incompletely sampled fields]
      {Irregularly and incompletely sampled random fields in the Earth sciences:
        Analysis and synthesis of parameterized covariance models}
\author[O.~L.~Walbert et al.]{
  \begin{minipage}[]{\textwidth}{Olivia L.~Walbert$^{1}$, Frederik J.~Simons$^{1,2}$, Arthur P.~Guillaumin$^3$, and Sofia C.~Olhede$^4$\\[-0.5em]}\end{minipage}\\
    $^1$ Department of Geosciences, Princeton University, Princeton, NJ 08544, USA. E-mail: olwalbert@princeton.edu\\
    $^2$ Program in Applied \& Computational Mathematics, Princeton University, Princeton, NJ 08544, USA\\
    $^3$ School of Mathematical Sciences, Queen Mary University of London, E1 4NS, London, UK\\
    $^4$ Institute of Mathematics, Ecole Polytechnique F\'ed\'erale de Lausanne, Switzerland}

\begin{document}
\maketitle
\begin{summary} 
Geophysical data observed in the Earth, planetary, and lunar sciences are
spatially limited to sampling plans designed to span a feature of interest whose
resolution and regularity in discretization are bound by site, observational,
and instrumental constraints. Common wisdom tells us that the more observations
we make, the less variable our estimates of statistical properties of measured
physical quantities will be. However, given the limited spatial extent and
possibly irregular sampling of any dataset, how confident can we be in our
estimates of the parameters of statistical processes underlying such fields?  We
study how sampling geometry contributes to uncertainty in modeling spatial
geophysical observations as sampled random fields characterized by stationary,
isotropic, parametric covariance functions. We incorporate the signature of
discrete spatial sampling patterns into an asymptotically unbiased spectral
maximum-likelihood estimation method along with analytical uncertainty
calculation. We illustrate the broad applicability of our modeling through
synthetic and real data examples with sampling patterns that include irregularly
bounded contiguous regions of interest, structured sweeps of instrumental
measurements, and missing observations dispersed across the domain of a field,
which spur behaviors in the estimator.  We find through asymptotic studies
that allocating samples following a growing-domain strategy rather than a
densifying, infill scheme best reduces estimator bias and (co)variance, whether
the field has been sampled regularly or not. As our modeling assumptions, too,
shape how (well) an observed random field can be characterized, we study the
effect of covariance parameters assumed \textit{a priori}. We demonstrate the
desirable behavior of the general Mat\'ern class and show how to interrogate
goodness-of-fit criteria to detect departures from the null hypothesis of
Gaussianity, stationarity, and isotropy.
\end{summary}


\section{I~N~T~R~O~D~U~C~T~I~O~N}

How does sampling (discretization over limited domains) affect what can be
learned about the statistical properties of a spatial dataset?  Geophysical data
are collected following spatial patterns designed to best survey a region of
interest within practical experimental constraints, such as sampling cost, site
accessibility, and anticipated measurement quality. Datasets in practice are
always finite and discrete, often departing from regular (rectangular or
circular), evenly spaced, full grids. Properly accounting for the finiteness and
discrete nature of spatial data when attempting to characterize the
second-order, i.e., correlation, structure, of the underlying random field, is
a must.

Writing as \cite{Simons+2026}, we argued for the merits of modeling in the
spectral domain, showing how to recover the parameters of the spectral density
of locally stationary, isotropic, processes. Working entirely in the space
domain, \cite{Goff+88} developed a procedure for the parameterized estimation of
the spatial covariance of geophysical fields. Both the spatial covariance and
the spectral density are ``covariance functions'', Fourier pairs that carry the
same information and are governed by the same parameters, but estimation
procedures anchored in the spatial domain contend with different issues than
approaches rooted in the spectral domain. For the former, based on spatial
likelihood maximization or least-squares fitting of sample spatial covariances,
the size of matrix inversion invariably becomes a computational bottleneck,
scaling as $\mathcal{O}(K^3)$ with the size of the data~$K$. With the latter,
often based on maximizing a spectral pseudo-likelihood, unbiased parameter
recovery and incorporating irregular sampling patterns were typically seen as
challenging, especially in higher dimensions. The attractive $\mathcal{O}(K\log
K)$ scaling of Fourier-based algorithms, however, stimulated our work on
addressing both of these problems. Adopting the debiased spatial Whittle
maximum-likelihood estimation strategy of \cite{Guillaumin+2022}, we first
showed, for geospatial data with structure across multiple physical scales, that
aliasing, finite-field, and edge effects can be treated, resulting in the
unbiased estimation of isotropic Mat\'ern covariance function parameters.  The
Mat\'ern covariance is a class of covariance functions that possess finite
variance, non-negative correlation, and monotonic decay.

Initially, we limited our treatment and analysis to completely and regularly
sampled rectangular domains, see \cite{Simons+2026}. Our present work broadens
the applicability of (univariate) spectral estimation to the more challenging
settings familiar to the geophysical community, namely: realistic data sampling
patterns defined by irregular boundaries, structured tracks, and general
`missingness' that may arise as voids, sparsity, or random deletions. We provide
geometric and asymptotic insight into the effects of sampling on the spatial
statistics inferred through the spectral estimation process, and shed light on a
number of other issues, such as how and when (not) to fix parameters in the
inversion, how to recognize the effect of departures from isotropy and
stationarity, and how to design geophysical sampling experiments.

\begin{figure}
  \centering
  \includegraphics[width=0.8\linewidth]{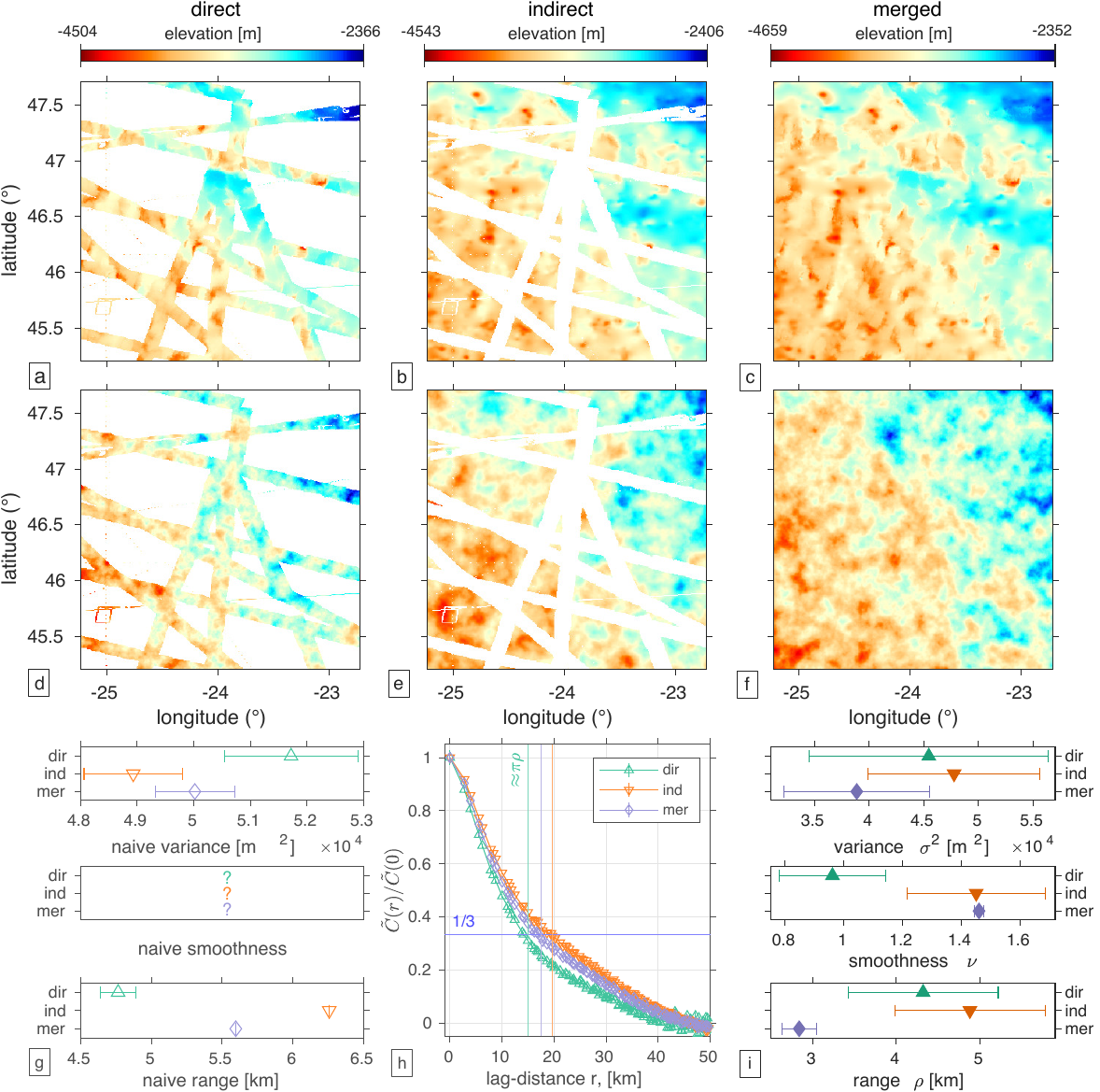}
  \caption{\label{fig:bath2synth}
    Analyzing geophysical random fields on domains with disjoint support. The
    debiased Whittle maximum-likelihood analysis developed in this paper is able
    to estimate statistical parameters of random field covariance structure,
    suitable for simulation, out-of-sample extension, evaluation,
    interpretation, and uncertainty appraisal. Bathymetric data from the
    northeast Atlantic \textit{(a--c)}. Synthetic realizations \textit{(d--f)}
    that share the first- and second-order statistics of the superjacent data
    within their window of observation. In triptych~\textit{(i)}, debiased
    Whittle likelihood Mat\'ern model parameter estimates (for variance
    $\sigma^2$, smoothness $\nu$, and range $\rho$) and their one standard
    deviation uncertainties for the direct (\textsf{dir}, green upward
    triangles), indirect (\textsf{ind}, orange downward triangles), and merged
    (\textsf{mer}, purple diamonds) data. The parameter estimates for the
    merged, nonstationary, dataset are not simple averages of those of the
    underlying component processes, and their uncertainties are irreconcilable
    due to model misspecification. In triptych \textit{(g)}, naive estimators
    (unfilled symbols) and their uncertainties: sample variance of the 6$\times$
    decimated dataset, and correlation length obtained from the sample
    autocovariance~\textit{(h)}.}
\end{figure}
Here, in modeling geophysical data as spatial random fields, we continue to work
under the hypothesis that the data observed are a realization of a locally
stationary, isotropic, Gaussian process whose statistical structure is estimable
through three parameters characterizing features related to the size, scale, and
shape of the field through its covariance structure. In the case of the Mat\'ern
density, those are variance $\st$, smoothness $\nu$, and range $\rho$. We
emphasize the connection between the general Mat\'ern covariance and its special
cases \cite[]{Guttorp+2006} which are frequently assumed within the geophysical
community as descriptions of the process that underlies an observed field,
either from rules-of-thumb or through domain-specific analysis and precedent
\cite[e.g.,][]{Przybilla+2009,Valentine+2020,Beyaztas+2021,TunnicliffeWilson+2022,Schwaiger+2023,Zhang+2023}. We
illustrate the relationship between the general and special cases to bring
awareness to the implications of selecting, e.g., a specific smoothness, in
terms of the bias it imparts when chosen incorrectly.

We show how parameter uncertainty can be analytically predicted as a function of
the sampling pattern, allowing us to study the asymptotic behavior of the
estimator and its covariance. We explore these attributes of our estimator in
the context of fixed and growing domains with regular and irregular sampling,
with the intent of informing the experimental design of sampling patterns. We
demonstrate how structure within the spectral-domain residuals can be
informative of when, how, and at what wavelength data defy the assumptions of
the null hypothesis, and we quantify model goodness-of-fit through a number of
carefully crafted diagnostics. We demonstrate the broad applicability of our
methodology through several synthetic simulation studies with realistic sampling
strategies common to the geosciences, and through several actual case studies
that use datasets from multiple subdisciplines. We provide all developed
computational tools openly as a complete software suite for simulation,
estimation, and diagnostics of spatial datasets across scientific domains.

To lead with a worked example while relegating all the details of the theory,
implementation, and interpretational framework to the remainder of this paper,
we show in Fig.~\ref{fig:bath2synth} how modeling the statistical properties of
composite datasets presents the geoscientist with the need for acknowledging
how the (supposedly ``noiseless'', correctly ``positioned'') ``data'',
themselves are (geophysical) ``models''. In the case of sea level
\cite[e.g.,][]{Slobbe+2012}, or seafloor bathymetry
\cite[e.g.,][]{Smith+97,Sandwell+2022}, for example, spatially distinct domains
may incorporate observational strategies with very different bandwidths and
resolutions. When these are spatially merged and presented as a \textit{data
  product} \cite[e.g.,][]{GEBCO2024}, acknowledgment must be made of their
disjoint spatial footprints prior to analysis. In the case of parameterized
covariance estimation, by working directly in the spatial domain, \cite{Goff+88}
naturally achieve spatial data selectivity in the inversion, but it is not
\textit{a priori} clear how modeling carried out in the spectral domain can be
made to be similarly aware of data provenance. And yet this is among the
merits of the debiased spatial Whittle likelihood \cite[]{Guillaumin+2022}: that
it can leverage the scaling and speed of the Fast Fourier Transform in the
spectral domain while running on incompletely observed spatial
grids. Fig.~\ref{fig:bath2synth} shows a portion of northeast Atlantic
bathymetry, identifying the scope of shipboard
(Fig.~\ref{fig:bath2synth}\textit{a}), satellite-augmented
(Fig.~\ref{fig:bath2synth}\textit{b}), and merged values
(Fig.~\ref{fig:bath2synth}\textit{c}).  Along the structured sampling tracks of
Fig.~\ref{fig:bath2synth}(\textit{a}) and within their complement
(Fig.~\ref{fig:bath2synth}\textit{b}), examples of the bounded contiguous and
disperse sampling patterns that we study throughout this work meld. Our modeling
is able to recover the underlying Mat\'ern parameters from those (dis)joint
subdomains, and simulate synthetic realizations on the corresponding subdomain
(Fig.~\ref{fig:bath2synth}\textit{d--f}), or anywhere else. The simulated
realizations preserve the (mean and deterministic trends) first-order (obtained
via least-squares first-order polynomial fitting in the space-domain) and
(two-point correlation) second-order statistics (estimated parametrically from
the data using the spectral Whittle likelihood in a manner that is unbiased by
the sampling process). Fig.~\ref{fig:bath2synth}(\textit{i}) puts forward that
distinct domains can be compared, but conclusions based on their unseparated,
merged mixture are meaningless.  Parameter inference and uncertainty estimation
lead to models that are ``right'' (different, testable) when they are
individually informed from the correct spatial support (``\textsf{dir}'' and
``\textsf{ind}''), while mixing and mingling of data sources (``\textsf{mer}'')
leads to models, incorrectly posited as stationary, that are (demonstrably,
verifiably) ``wrong''. For comparison, Fig.~\ref{fig:bath2synth}(\textit{g})
shows space-based naive estimators for the variance and range of the three data
domains, with their uncertainties. The sampling variance was calculated on a
6$\times$ decimated observed grid, and the range lengths estimated empirically
from the drop-off to $1/3$ in the autocorrelation
(Fig.~\ref{fig:bath2synth}\textit{h}), ignoring the effect of smoothness in that
measure. Naive estimators for smoothness are not readily available.  We will
return to the directly measured dataset (Fig.~\ref{fig:bath2synth}\textit{a}) as
one of our real-world data examples later in this paper.

\section{T~H~E{\hsps}C~O~M~P~L~E~T~E{\hsps}M~E~T~H~O~D~O~L~O~G~Y{\hspst}:{\hsps}M~A~T~\'E~R~N{\hsps}M~E~E~T~S{\hsps}W~H~I~T~T~L~E}

Our Whittle maximum-likelihood method for the spectral-domain covariance estimation
of data drawn from a stationary, isotropic Mat\'ern class
\cite[]{Whittle54,Matern60} grew out of developments in a (geo)physical context
made by \cite{Simons+2013}, and was put on firm statistical footing by
\cite{Sykulski+2019} for time series and \cite{Guillaumin+2022} for spatial
data in any number of dimensions~$d$.

\cite{Simons+2026} fully established the debiased spatial Whittle
maximum-likelihood method for univariate analysis as used here, but we did not
then study its behavior under realistic sampling scenarios, which is the subject
of this paper. To make the present contribution self-contained, we review the
essential elements of the methodology in this section. We identify special cases
of the Mat\'ern covariance, introduce a new measure for understanding spatial
correlation, discuss the simulation of spatial realizations and the unbiased
recovery of parameters via the Whittle likelihood, and we review its quality
assessment and uncertainty quantification.

We conclude this section with practical take-aways for data preparation and
preprocessing and a reader's guide to the rest of the paper.

\subsection{Stationary, isotropic parametric covariance models of spatial random fields}
\label{sec:parametriccovariancemodels}

The Mat\'ern class of covariance functions offers flexibility and generality for
stationary processes \cite[]{Stein99}. Adopting the parameterization of
\cite{Handcock+94}, the space-domain representation of the isotropic Mat\'ern
covariance for spatial separation lag distances~$r$ is
\begin{equation}
\label{eq:spatmatern}
    \mcC\ofst(r)= \sigma^2 \frac{2^{1-\nu}}{\Gamma\left(\nu\right)}
    \left(\argu\right)^\nu K_\nu \left( \argu \right) \for r \geq 0\,,\,\st>0\,,\,\nu>0\,,\,\mathrm{and }\,\,\rho>0
    .
\end{equation}
The three Mat\'ern parameters $\btheta=[\st,\,\nu,\,\rho]^{T}$ flexibly
characterize the covariance function, encoding the amplitude of the field in the
variance, $\sigma^2$, its mean-squared differentiability
\cite[]{Adler1981,Christakos92} through the smoothness, $\nu$, and its lateral
correlation scale through the range, $\rho$. Two special functions appear in
eq.~(\ref{eq:spatmatern}): the Gamma function $\Gamma(z)$ defined in eq.~(6.1.1)
of \cite{Abramowitz+65}, and the modified Bessel function of the second kind of
order $\nu$, $K_{\nu}(z)$, see eq.~(9.6.1) of \cite{Abramowitz+65}.

In the stationary, isotropic case, the Mat\'ern spatial covariance possesses a
Wiener-Khintchine Fourier-pair relationship with the Mat\'ern spectral density,
itself the covariance of orthogonal spectral increments \cite[]{Cramer42},
specified in~$d$ dimensions for wavenumbers~$k$ as
\begin{equation}\label{eq:specmaterndd}
  \mcS_{\btheta}^d\ofsk = \sigma^2 \frac{\Gamma\left( \nu + d/2
    \right)}{\Gamma\left( \nu \right)} \frac{1}{\pi^{d/2}} \left(
  \frac{4\nu}{\pi^2 \rho^2} \right)^{\nu} \left( \frac{4\nu}{\pi^2 \rho^2} + k^2
  \right)^{-\nu - d/2} \for k\in\mathbb{R}^+.
\end{equation}
All that follows will be in terms of two-dimensional random
fields with generality to higher dimensions straightforward. Evaluating
eq.~(\ref{eq:specmaterndd}) for the two-dimensional case, we drop the
dimensionality superscript and express the Mat\'ern spectral density as
\begin{equation}\label{eq:specmatern2d}
  \mcS\ofst\ofsk=
  \st\frac{\pi\rho^2}{4}\left(\frac{4\nu}{\pi^2\rho^2}\right)^{\nu+1}
  \left(\frac{4\nu}{\pi^2\rho^2}+k^2\right)^{-\nu-1} \also k\in\mathbb{R}^+
  .
\end{equation}

\begin{figure}\centering
  \includegraphics[width=0.91\textwidth,trim=2.75cm 2.5cm 2.25cm 2.75cm,clip]{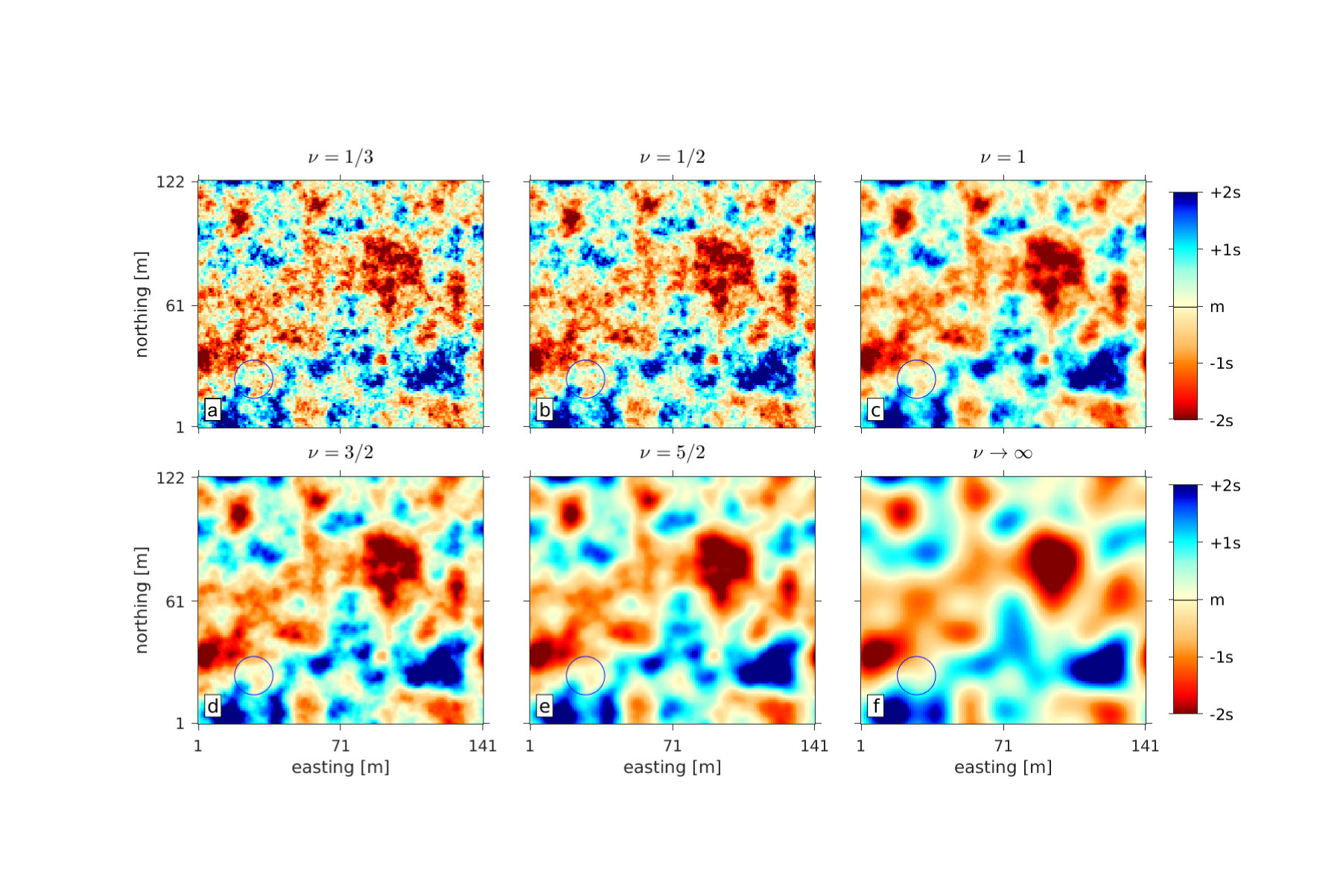}
  \caption{\label{fig:specialcasessim}
    Spatial random fields simulated from special cases of the Mat\'ern class,
    for a common variance ($\sigma^2=1\,\m^2$) and range ($\rho=5\,\m$) over
    a rectangular, regular, grid, $122\,\m \times 141\,\m$, with even $1\,\m$
    spacing, for different smoothness parameters $\nu$. \textit{(a--c)}
    Smoothness parameter $\nu=1/3$, $1/2$, and $1$; \textit{(d--f)}
    $\nu=3/2$, $5/2$, and $\infty$ (see Table~\ref{tab:MaternSpecial} for
    their analytic forms). Blue circles mark the approximate $1/{}3$
    decorrelation length of the field by their radii $\pi\rho$, all of which are
    equal here due to the common range $\rho$ of the six models. Colormaps range
    between $\pm2$ standard deviations~$\mathsf{s}$ of the fields with
    means~$\mathsf{m}$.}
\end{figure} 

The three-parameter form of the Mat\'ern class lends to its generality, with
several, well-known two-parameter special cases being simplifications of
eqs~(\ref{eq:spatmatern}) and (\ref{eq:specmaterndd}) where the value of the
smoothness parameter, $\nu$, is fixed, and the two parameters that remain free
characterize the shape and scale of the covariance. Often, the special value
of~$\nu$ reduces the $K_{\nu}(z)$ term to a more wieldy function, such as the
product of an exponential and a polynomial term when $\nu$ is half integer.
While many of these two-parameter covariance functions are widely adopted
throughout the geosciences, the Mat\'ern class that they generalize from is less
commonly recognized as such. We present the analytical expressions for several
special models in Tables~\ref{tab:MaternSpecial} and~\ref{tab:MaternSpecialAll},
as direct simplifications of the parameterization of the two- and
$d$-dimensional Mat\'ern forms in
eqs~(\ref{eq:spatmatern})--(\ref{eq:specmatern2d}).

In Fig.~\ref{fig:specialcasessim}, we display random fields generated for
several special cases with increasing smoothness values~$\nu$ and a common
variance~$\sigma^2$ and range~$\rho$, on a fully observed, evenly spaced,
rectangular sampling grid. The distinction between these covariance models is
evident: fields characterized by a larger~$\nu$ display more gentle, undulous
transitions in amplitude between spatially sampled values. To make this figure,
as the only example in this paper where we wish to preserve the dominant spatial
patterns, we scaled a single set of complex normal Gaussian variates with
spectral densities of varying smoothness. The omission of wavenumber correlation
that that entails causes wrap-around periodicity and loss of
stationarity. Hence, for simulations everywhere else, we will use circulant
embedding, see Section~\ref{sec:sim}.

\begin{figure}\centering
    \includegraphics[width=0.91\textwidth,trim=0cm 6.5cm 0cm 0cm,clip]{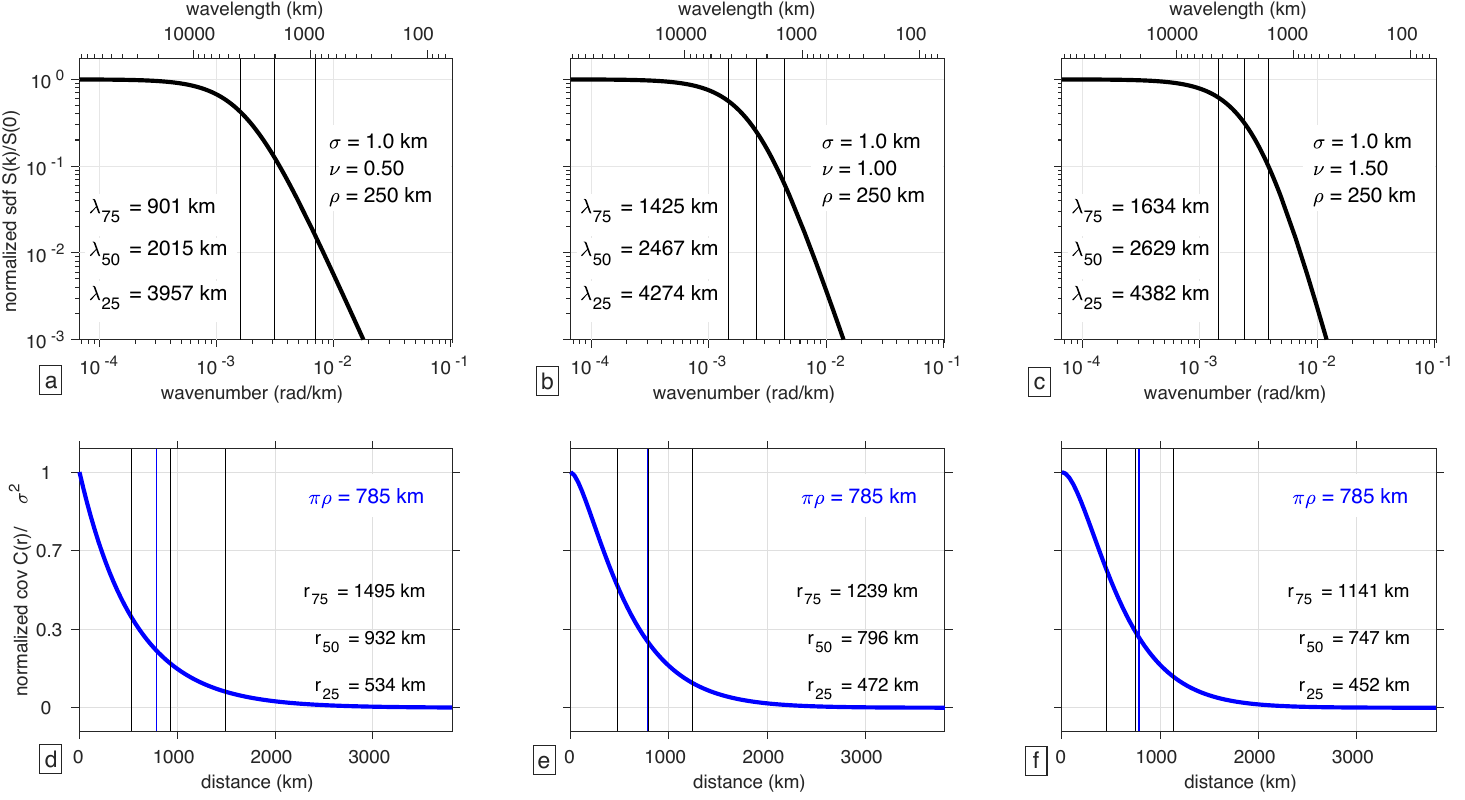}
    \caption{\label{fig:specialsdfcov}
      Normalized, radial spectral densities $\mcS\ofst\ofsk/\mcS\ofst(0)$ (\emph{a--c},
      black) as a function of wavenumber~$k$, and normalized spatial covariances
      $\mcC\ofst(r)/\st$ (\emph{d--f}, blue) as a function of lag
      distance~$r$. We show three special cases of smoothness of the
      Mat\'ern covariance class (see Tables~\ref{tab:MaternSpecial}--\ref{tab:MaternSpecialAll}), including
      $\nu=1/{}2$ (\emph{a, d}), $\nu=1$ (\emph{b, e}), and $\nu=3/{}2$
      (\emph{c, f}), for a common variance $\st=1\,\km^2$ and range
      $\rho=250\,\km$ over a regular rectangular 
      $3840\,\km\times3840\,\km$ grid with even $30\,\km$ spacing. (\emph{a--c})
      Vertical black lines at $\lambda_\alpha$ correspond to 
      percentages of cumulative spectral variance, $\lambda_{25},\,\lambda_{50},\,\lambda_{75}$, appearing left to right on the top wavelength axis. (\emph{d--f}) Vertical
      black lines at $r_\alpha$ correspond to percentages of
      cumulative spatial variance (eq.~\ref{eq:rootcov}), $r_{25},\,r_{50},\,r_{75}$, appearing left to right on the bottom distance axis, and vertical blue
      lines indicate a distance of $\pi\rho=785\,\km$ where the correlation
      reaches about one-third, depending on~$\nu$.}
\end{figure} 

As Fig.~\ref{fig:specialsdfcov} reveals, in their spectral representations,
higher-smoothness processes show reduced high-frequency content, and their
correlation functions take off from the origin at gentler angles.
\cite{Simons+2026} quantified the first notion by quoting the cumulative power
of the isotropic Mat\'ern form integrated from 0 to positive $k$. They defined
$\lambda_\alpha$, their eq.~(18), as the wavenumber (practically, limited by the
Nyquist frequency) at which the power reaches $100\times\alpha$ per cent of the
total, which we plotted in the top row of Fig.~\ref{fig:specialsdfcov}. Here, as
to the second notion, analogously, in the space domain, we find that the
cumulative spatial covariance as a function of lag distance~$r'$ (in practice, limited
by the diagonal length of the domain) is given by
\begin{equation}\label{eq:cumulativecov}
      \int_0^{r'} r\hsom\mcC_{\btheta}(r) \,dr
       = 
       \frac{1}{2}\st(\pi\rho)^2
       \left[1 - \frac{2^{-\nu}}{\Gamma(\nu+1)}
                    \left(\frac{2\nu^{\frac{1}{2}}}{\pi\rho} r'\right)^{\nu+1}
             K_{\nu+1}\left(\frac{2\nu^{\frac{1}{2}}}{\pi\rho} r'\right)
        \right]
       .
\end{equation}
We obtained the above result via integration by parts \cite[see also,
  eq.~7.14.1.3, of][]{ErdelyiII1953}. The total covariance accumulated over all
lag distances can be expressed \cite[using eq.~6.561.16 of][]{Gradshteyn+2000}
as
\begin{equation}\label{eq:auccov}
    \int^\infty_0r\hsom\mcC_{\btheta}(r) \,dr
    = 
    \frac{1}{2}\st(\pi\rho)^2
    =\frac{\mcS_{\btheta}(0)}{2\pi}
    .
\end{equation} 
To determine the lag distance~$r_\alpha$ that is associated with $\alpha$\% of
the total cumulative covariance, we implement a bisection root-finding strategy
to evaluate eqs~(\ref{eq:cumulativecov})--(\ref{eq:auccov}) numerically, as we
cannot isolate $r_\alpha$ from the argument of the Bessel function analytically,
per
\begin{equation}\label{eq:rootcov}
  \int^{r_{\alpha}}_0r\hsom\mcC_{\btheta}(r) \,dr=
  \alpha\int^\infty_0r\hsom\mcC_{\btheta}(r) \,dr
  .
\end{equation}
In Fig.~\ref{fig:specialsdfcov}, the bottom row is annotated at lag
distances~$r_\alpha$ that correspond to 25\%, 50\%, and 75\% of the total
spatial covariance by thin, black vertical lines. Likewise, using a thin, blue
vertical line, we mark the lag distance~$\pi\rho$, commonly quoted as the point
at which the correlation decays to $1/3$. We find that the quality of that
approximation, as to whether $\pi\rho\approx r_{33}$, is strongly dependent on
the smoothness~$\nu$.

\subsection{Simulation of spatial random fields from parametric covariance models}\label{sec:sim}

We generate synthetic spatial random fields $\mcH\ofx$ as a stationary Gaussian
process through the circulant embedding \cite[]{Chan+99,Dietrich+97,Kroese+2015}
of the spatial covariance model (eq.~\ref{eq:spatmatern},
Table~\ref{tab:MaternSpecial}) on the sampling grid (for an embedding that is in
practice $2^d\times$ as large). For a two-dimensional, $\Ny\times\Nx$,
observation window discretized by a spacing of $\Delta y\times \Delta x$, we
define $w\ofx$ as the data taper that indicates the presence or absence of data
observed on the discretized grid, i.e., $w\ofx\in\{0,1\}^{\Ny\times\Nx}$, or as
a smoothing function $w\ofx\in[0,1]^{\Ny\times \Nx}$ used as nonbinary
weights. Hence, we consider irregularities and incompleteness in the sampling of
$\mcH\ofx$ as modifications to an otherwise unitary $w\ofx$ that spans an
encompassing regular grid. We define $K=\sum{w\ofx}\le \Ny\Nx$.

The circulant embedding method is advantageous in that it produces, on average,
a zero-mean random field without the wrap-around effects nor omission of
wavenumber correlations that can burden spectral-domain simulation techniques
\cite[e.g., via Cholesky decomposition of spectral densities,][]{Dietrich+97}.
Successful embedding requires that the circulant matrix of the spatial
covariance is positive semi-definite, a condition met in practice by grids that
are sufficiently large relative to the correlation length. Independent
realizations generated by the circulant embedding method make for sampled,
spatial random fields that possess the requisite second-order statistics.

\subsection{Maximum-likelihood estimation of parametric covariance models from spatial random fields}\label{sec:mle}

We estimate the parametric covariance of an observed field, $\mcH\ofx$, as the 
parameters $\hbtheta$ that maximize the blurred, debiased Whittle likelihood
\begin{equation}\label{eq:dwl}
   \bar{\mcL}\oft=-\norml 
        \left[ \ln\Sbar\ofst\ofk +\frac{|H\ofk|^2}{\Sbar\ofst\ofk}\right]
     \quad\mbox{where}\quad 
   H\ofk \equiv \frac{1}{2\pi}\nff\sum_\xb w\ofx\mcH\ofx e^{-i\kb\cdot\xb},
\end{equation}
following eq.~(5) of \cite{Guillaumin+2022} and eqs~(48)--(49) of
\cite{Simons+2026}. All summations are limited to the spatial grid $\mathbf{x}$
and the Nyquist grid $\kb$.  By `modifying' the data $\mcH\ofx$ by the spatial
observation window~$w\ofx$, sampling bias for regular and irregular data can be
corrected for in the periodogram as a non-parametric estimator
\cite[e.g.,][]{Dahlhaus+1987,Simons+2000,Deb+2017}, but in the likelihood, that
term, $|H(\kb)|^2$, must not be directly compared to the theoretical
parameterized spectral density~$\mcS\ofst\ofk$ unless it is smoothed, e.g., via
convolution with the spectral window $|w\ofk|^2$, turning it into the `blurred'
spectral density denoted with the overbar as $\Sbar\ofst\ofk$. The blurring
operation earns eq.~(\ref{eq:dwl}) the designation \textit{debiased}. Compared
to space-based estimation methods which require the construction and
\textit{inversion} of the empirical spatial covariance matrix \cite[]{Goff+88},
after Fast Fourier Transformation to the wave vector domain, the simple spectral
\textit{division} in eq.~(\ref{eq:dwl}) is efficiently evaluated.

Calculating the debiased Whittle likelihood function requires the
window-modified empirical periodogram $|H(\kb)|^2$, formed from the
Fourier-transformed data vector $H(\kb)$. In eq.~(\ref{eq:dwl}), the periodogram
is equivalent \textit{in expectation} to the blurred spectral density
$\Sbar\ofst\ofk$ by which it is divided, as shown by Lemma~1 of
\cite{Guillaumin+2022} and eqs~(28)--(30) of \cite{Simons+2026}. It is
calculated exactly as
\begin{equation}
\label{eq:blurredspec}
\Sbar\ofst\ofk=\frac{1}{(2\pi)^2}\nffs
\sum_\yb W(\yb) \hsom\mcC\ofst\ofsy e^{-i \kb \cdot \yb}
\quad\mbox{where}\quad
W(\yb)=\sum_\xb w\ofx w(\xb+\yb)
,
\end{equation}
that is, we form $\Sbar\ofst\ofk$ as the Fourier transform of the isotropic
spatial covariance $\mcC\ofst\ofsy$ modified by the correlation of the sampling
window $W(\yb)$ computed over the relevant separation grid with lag distances
$\yb$. The blurred spectral density $\Sbar\ofst\ofk$ can be computed using a
multidimensional Fast Fourier Transform for sampling patterns that are
modulations of regular grids \cite[][their Lemma~2]{Guillaumin+2022}. The
explicit dependence of eqs~(\ref{eq:dwl}) and~(\ref{eq:blurredspec}) on spatial
wave vector $\kb$ rather than isotropic wavenumber $k=||\mbf{k}||$ is
intentional, as even for an isotropic model, the operation of blurring is not
radially symmetric. This is evident as the autocorrelation of the window $W\ofy$
is clearly not isotropic, even for a full rectangular grid. Compared to
eq.~(\ref{eq:blurredspec}), in the original formulation of \cite{Simons+2013},
the blurring of the spectral density by the sampling window~$w\ofx$ was inexact,
through numerically discretized convolution with the \textit{spectral} window
$|w\ofk|^2$ (the Fej\'er kernel in the unitary case), for which they used a
slightly different notation.

\begin{figure}\centering
  \includegraphics[width=0.8\textwidth]{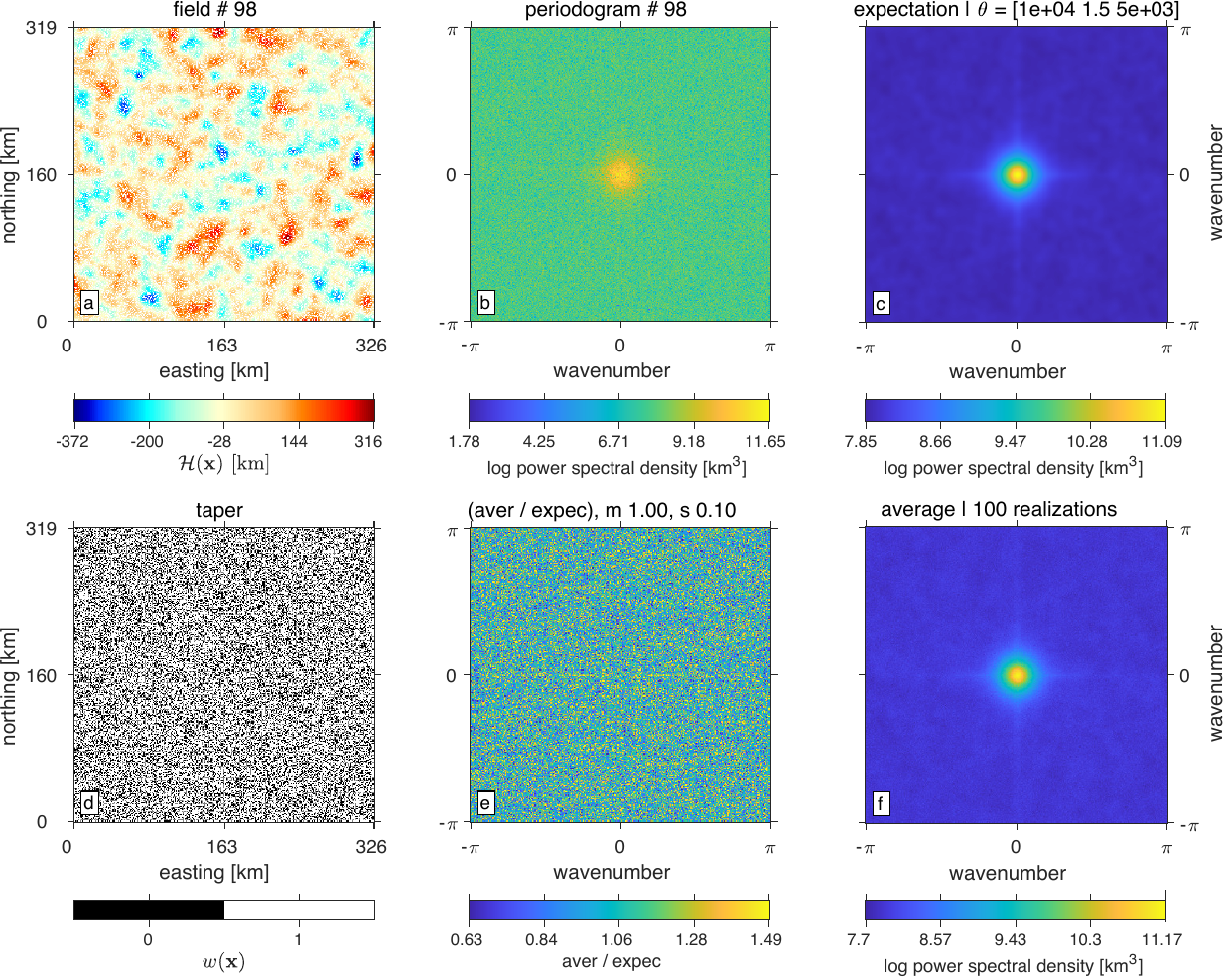}
  \caption{\label{fig:blurring} Maximum-likelihood estimation
    (Section~\ref{sec:mle}) on a randomly subsampled (66.7\% observed) grid: the
    effect of spectral blurring. \textit{(a)}~A single realization of a spatial
    random field $\mcH\ofx$ simulated through the circulant embedding of the
    Mat\'ern covariance with $\btheta= [10\,\km^2\; 1.5 \; 5\,\km]$ on a
    $319\times326$ grid with $1\,\km\times1\,\km$ spacing. \textit{(b)}~The
    empirical periodogram of the field $|H\ofk|^2$ in the Nyquist wavenumber plane $[-\pi \; \pi]\,\km^{-1}$. \textit{(c)}~The blurred
    spectral density $\Sbar\ofst\ofk$. \textit{(d)}~Spatial taper $w\ofx$
    indicating the presence or absence of data in white and black,
    respectively. \textit{(e)}~The ratio of the average empirical periodogram
    over 100 realizations to the blurred spectral density. \textit{(f)}~The
    average empirical periodogram. The empirical periodogram approaches the
    blurred spectral density in expectation. Boundary effects and sampling
    irregularities spoil the isotropy of both, while keeping their ratio
    unstructured and uninformative.}
\end{figure}

In Fig.~\ref{fig:blurring}, we illustrate the equivalence of
$\langle|H(\kb)|^2\rangle$ to $\Sbar\ofst\ofk$, along with several of the
quantities that appear in the debiased Whittle likelihood (eq.~\ref{eq:dwl})
which are significant to the spectral-domain method. One realization of a
spatial field $\mcH\ofx$ simulated from a known $\btheta_0$ and sampled on a
rectangular observation grid is shown in
Fig.~\ref{fig:blurring}(\textit{a}). Missingness in its sampling pattern is shown
as white in the colormap. In the data taper $w\ofx$ shown in
Fig.~\ref{fig:blurring}(\textit{d}) with the light color indicating observations
and dark color indicating missingness. This sampling pattern was created through
the uniformly random deletion of one-third of an otherwise evenly sampled
grid. The empirical periodogram $|H\ofk|^2$ of the spatial realization is shown
in Fig.~\ref{fig:blurring}(\textit{b}). The blurred spectral density
$\Sbar_{\btheta}\ofk$ evaluated at $\hbtheta$ is shown in
Fig.~\ref{fig:blurring}(\textit{c}). In contrast to the regularly sampled
rectangular fields shown by \cite{Simons+2026}, the blurred spectral density
displayed here demonstrates the loss of not only isotropy but mirror symmetry,
too, which arises from the sampling pattern, in the blurred model. From many
independent experimental realizations, we calculate the average periodogram
$\overline{|H\ofk|^2}$, shown in Fig.~\ref{fig:blurring}(\textit{f}), and
compare this in ratio to the blurred spectral density in
Fig.~\ref{fig:blurring}(\textit{e}), which displays very low perturbations that
lack structure. We interpret this as evidence that, in expectation, the periodogram approaches the
modeled blurred spectral density.

Maximizing the debiased Whittle likelihood (eq.~\ref{eq:dwl}) yields an
asymptotically unbiased, random estimator $\hbtheta$, such that its expectation
is the true underlying process model parameter set,
$\langle\hbtheta\rangle=\btruth$. Defining the blurred \textit{score} vector $\bbg\oft$
as the gradient of the blurred likelihood function~$\bar{\mcL}\ofth$, evaluation
at the local optimum $\hbtheta$ will in expectation yield the zero-vector
\begin{equation}\label{eq:bgth}
    \bnabla\hsomm\ofst\bar{\mcL}\ofth=\bbg\ofth=\bzero.
\end{equation}

We find that various iterative numerical methods are suitable for the
optimization of eq.~(\ref{eq:dwl}), including gradient-based (both constrained
and unconstrained) methods and non-gradient, simplex-based methods. In
practice, we choose to implement an unconstrained quasi-Newton algorithm for
ease of comparison between numerically calculated gradients and Hessians of the
likelihood function, evaluated at the numerical estimate $\hbtheta$, with our
evaluations of the analytical partial derivatives of eq.~(\ref{eq:dwl}). Our
initialization of the numerical optimizer randomly perturbs guesses of the
Mat\'ern parameters that we propose as follows: we choose an initial
$\sigma^2_0$ as the variance of the spatial field as observed within the
sampling window~$w\ofx$, we assign a guess for $\nu_0$ at a value we take to be
intermediate to the typical range found in practice ($\nu_0=2$), and we
arbitrarily calculate an initial $\rho_0$ relative to the grid spacing and the
area of the grid as $(1/20 \pi)\sqrt{\Dy \Dx \Ny\Nx}$. While practitioners may
provide initialization values for the parameters, or fix any of their values as
hyperparameters, all of the initialized guesses we use in practice follow this
automated strategy that is independent of prior knowledge of the truth.

Upon determining an estimator $\hbtheta$ for the random field $\mcH\ofx$ within
the observation window $w\ofx$, we must assess our confidence in the estimator
and how well the model, with all its inherent assumptions (i.e., local
stationarity, Gaussianity, isotropy, and a local non-varying mean), complies
with the observed data.

\subsection{Residual diagnostics and their interpretation}\label{sec:res}

In the spectral domain, our spectral residuals are the ratio of the periodogram to
the blurred spectral density parameterized by our estimator,
\begin{equation}\label{eq:chisq}
  X_{\btheta}(\kb) = \frac{|H(\kb)|^2}{\bar{S}_{\btheta}(\kb)}
  .
\end{equation}

In expectation, the periodogram~$|H\ofk|^2$ approximates the blurred spectral
density~$\Sbar\ofst\ofk$, and for a (Gaussian) process their ratio scales
(asymptotically) like a chi-squared distribution with two degrees of freedom,
except at 0 and the Nyquist wavenumbers \cite[][their Section~4.6]{Simons+2026},
hence, 
\begin{equation}\label{eq:stx}
  X_{\btheta}(\kb) \sim \frac{\chi^2_2}{2}\also  \stx = \frac{1}{K}\sum_\kb [X\ofst\ofk-1]^2 \inlaw \mcN(1,8/K)
  .
\end{equation}
The convergence of $\stx$ for data $H\ofk$ that abide by model assumptions forms a
basis for quantitative assessments. The appropriateness of the covariance model
as stationary, isotropic, and parametric, and the adequacy of the sample size
$K$, are addressed by posing the Mat\'ern estimator as the null hypothesis to be
evaluated based on the test statistic in eq.~(\ref{eq:stx}). 

While we report $\stx$ and the results of the test as such in our data examples,
these values are to be used for initial guidance only, as the relations in
eq.~(\ref{eq:stx}) are asymptotic, and sample sizes are often too small.  A much
more valuable qualitative assessment tool is the visual inspection of the spectral
residuals $X\ofst\ofk$ for unmodeled structure.

In the spatial domain, we employ simulation as a secondary qualitative tool that
allows us to compare by eye the appropriateness of a synthetic spatial field
that possesses the first- and second-order structure estimated from the observed
spatial field (e.g., Fig.~\ref{fig:bath2synth}). This is useful for real data
examples where we appraise the average variability of the data, rather than
unique or deterministic features.

\subsection{Analytical prediction of estimator covariance}\label{sec:cov}

Obtaining confidence intervals on our estimated parameters requires their
covariance for the given sampling pattern~$w\ofx$.  This is possible
analytically and exactly without direct reference to any particular data
realization, which provides us the opportunity to investigate the asymptotic
behavior of our estimator under various designed sampling plans.  Our exact and
analytical calculations are consistent with copious empirical estimates of model
covariance from simulation-derived estimator ensembles across the wide range of
sampling patterns we have generated. Here, we summarize the
parameter covariance calculation method that we find most intuitive and which
emphasizes the role of the sampling window~$w\ofx$ and its correlation $W\ofx$.

Following \cite{Guillaumin+2022} and \cite{Simons+2026}, the parameter
covariance for the estimated model is determined from~$\cov\{\bbg\oft\}$, the covariance
of the blurred score (the gradient of the log likelihood),
and~$\bar{\bmcF}^{-1}$, the blurred inverse Fisher information matrix (the
inverse of the curvature of the log likelihood), both
evaluated in practice at the estimator,
\begin{equation}\label{eq:magic}
  \cov\ofth\approx
  \bar{\bmcF}^{-1}\oftr\,\cov\{\bbg\oftr\}\,\bar{\bmcF}^{-1}\oftr
  .
\end{equation}
Asymptotic expressions often further approximate $\cov\ofth\approx
\bar{\bmcF}^{-1}$, but we have rarely found ourselves in the regime where that
holds comfortably, see Section~\ref{sec:asymp}. The blurred score vector, which we minimize per
eq.~(\ref{eq:bgth}), is
\begin{equation}\label{eq:bgth2}
  \bgth\oft=-\norml\bmth\ofk\left[1-\frac{|H\ofk|^2}{\Sbar\ofst\ofk}
\right] ,
\end{equation}
which depends on the sampling pattern $w\ofx$ through the exactly blurred
spectral density (eq.~\ref{eq:blurredspec}) and window-modified periodogram
(eq.~\ref{eq:dwl}), as well as the partial derivatives of the logarithmic
blurred spectral density presented as
\begin{equation}\label{eq:mthbar}
  \bmth\ofk=
  \Sbar\ofst^{-1}\hsomm\ofk\frac{\pl\Sbar\ofst\ofk}{\pl\theta}=
  \frac{\Sbar\ofst^{-1}\hsomm\ofk}{(2\pi)^2}\nffs
  \sum_\yb W(\yb)\hsom\frac{\pl\mcC\ofst\ofsy}{\pl\theta} e^{-i \kb \cdot \yb}.
\end{equation}
All of the partial derivatives $\partial_{\theta}\mcC_{\btheta}$ have
closed-form analytical expressions. The components of the Hessian, given by
\begin{equation}\label{hessian}
  F_{\theta\theta'}\oft
  =-\norml\Bigg[
\frac{\pl \bmthp\hsomm\ofk}{\pl\theta}
+\left\{\bmth\ofk \bmthp\hsomm\ofk
-\frac{\pl \bmthp\hsomm\ofk}{\pl\theta}\right\}
\frac{|H\ofk|^2}{\Sbar\ofst\ofk}
\Bigg]
,
\end{equation}
and its negative expectation, the Fisher information matrix, both depend on the
product of eq.~(\ref{eq:mthbar}) evaluated for parameter interactions,
\begin{equation}
   \label{eq:fisherinfo}
   \bar\mcF_{\theta\theta'}\oft=\norml\bmth\ofk\hsom\bmthp\hsomm\ofk.
\end{equation}

The covariance of the blurred score with full wave vector correlations,
\begin{equation}\label{eq:covscore}
  \cov\big\{ \bgth,\bgthp\big\}=
  \normltu \bmth\ofk \,\frac{ \cov\{|H\ofk|^2,|H\ofkp|^2\} }{ \Sbar\ofst\ofk\Sbar\ofstp\ofkp} \hsom\bmthp\hsomm\ofkp
  ,
\end{equation}
depends on values calculated from eqs~(\ref{eq:blurredspec}) and
(\ref{eq:mthbar}), as well as the covariance of the window-modified periodogram
and with correlated wave vectors $\cov\{|H\ofk|^2,|H\ofkp|^2\}$. We calculate
this term by applying the \cite{Isserlis1918} theorem for Gaussian processes
\cite[]{Percival+1993,Stein1995,Walden+1994,Simons+2026}, which we have assumed
holds for $\mcH\ofx$, so that
\begin{align}
  \cov\big\{|H\ofk|^2,|H\ofkp|^2\big\}
  &=\label{eq:percov}
  |\cov\{H\ofk,H^*\hsomm\ofkp\}|^2 + |\cov\{H\ofk,H\hsomm\ofkp\}|^2\\
  &=\label{eq:percov1}
  \left|\langle H\ofk H\ofkp \rangle \right|^2+
  \left|\langle H\ofk H^*\hsomm\ofkp \rangle \right|^2\\
  &=\label{eq:percov2}
  \big| \sQ\hsom \langle \mcH\ofx\mcH\ofxp \rangle \hsom\sQ' \big|^2+
  \big| \sQ\hsom \langle \mcH\ofx\mcH\ofxp \rangle \hsom{\sQ^*}' \big|^2\\
  &=\label{eq:percov3}
  \big| \sQ\hsom \tilde{\mcC}\ofst \hsom\sQ' \big|^2+
  \big| \sQ\hsom \tilde{\mcC}\ofst \hsom{\sQ^*}' \big|^2,
\end{align}
where $\sQ$ is the discrete Fourier transform matrix, and $'$ and $^{*}$$'$
denote the transpose and conjugate transpose. In our case of non-unitary
tapers, the elements of the spatial covariance in eq.~(\ref{eq:percov3}) are
weighted by the outer product of the sampling window $w\ofx$ with itself,
denoted $\tilde{\mcC}\ofst$.

Implementing eq.~(\ref{eq:percov3}) has proven to be computationally tractable for
problems under $K=10,000$ samples on a multi-CPU desktop computer (equipped with
Intel Xeon w7-3455 processor with 24 2.5--4.8~GHz cores with 64~GB memory) as
well as on a contemporary MacBook Pro. To handle larger sample sizes, as in
Fig.~\ref{fig:bath2synth}, we switch to a slower method that accesses the
wave vector correlation matrix diagonal per diagonal as described by
\cite{Simons+2026}, with identical results.

\begin{figure}
  \centering
  \includegraphics[width=0.95\textwidth]{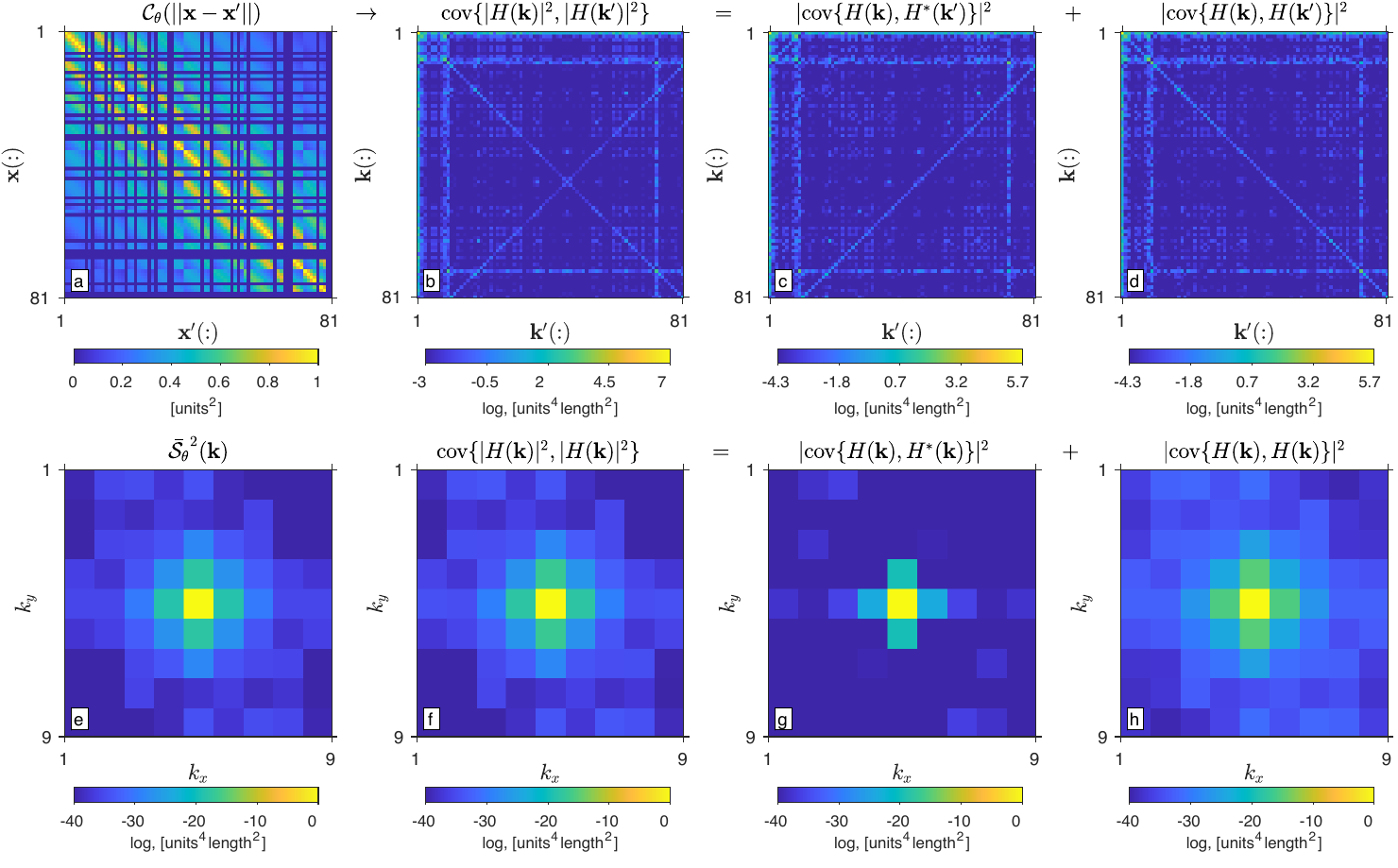}
  \caption{\label{fig:percov} Maximum-likelihood estimation
    (Section~\ref{sec:mle}) on a randomly subsampled (66.7\% observed) grid: the
    effect of wavenumber correlation. The covariance and pseudocovariance terms
    of the periodogram covariance (eq.~\ref{eq:percov}) for
    $\btheta=[1\,\mathrm{unit}^2\;0.5\;2\,\mathrm{length}]$. The product of the
    observed grid taper with the spatial autocovariance over all lag
    interactions \textit{(a)} reveals a tartan pattern of negligibly
    contributing lag pairs that manifests in the spectral-domain terms that
    include all wavenumber interactions \textit{(b--d)} as elongated hatches of
    diffuse power in contrast to the fully observed grid, concentric about the
    intersection of the diagonals of the covariance and pseudocovariance. When
    $\kb=\kbp$ \textit{(e--h)}, the random deletions of the observation grid
    display an oblique symmetry in the periodogram variance.}
\end{figure}

Fig.~\ref{fig:percov} illustrates the calculation of the periodogram
covariance~(eq.~\ref{eq:percov}) through the discrete Fourier transform matrix
implementation (eq.~\ref{eq:percov3}) for a small grid that is only $2/_{}3$
observed due to random deletions in its sampling pattern.
Fig.~\ref{fig:percov}, which, when compared to Fig.~6 in \cite{Simons+2026},
demonstrates how, for an irregularly sampled $w\ofx$, asymmetric anisotropy is
imparted in the periodogram's covariance (and thus the blurred spectral
density). The top row (Fig.~\ref{fig:percov}\emph{a--d}) visualizes each of the
terms that appear in eqs~(\ref{eq:percov})--(\ref{eq:percov3}). Panel
Fig.~\ref{fig:percov}(\emph{b}) in the top row labeled
$\cov\big\{|H\ofk|^2,|H\ofkp|^2\big\}$ includes all wavenumber correlation
terms. Panel Fig.~\ref{fig:percov}(\emph{f}) in the bottom row labeled
$\cov\big\{|H\ofk|^2,|H\ofk|^2\big\}$ only shows the diagonal ($\kb=\kbp$)
terms, i.e., the periodogram variance, where the diagonals of the second through
fourth columns have been wrapped. The nature of the anisotropy depends on the
sampling pattern, with the amplitude at $\kb,\kbp$ further depending on
$\btheta$. We find a near equivalence of $\cov\big\{|H\ofk|^2,|H\ofk|^2\big\}$
(Fig.~\ref{fig:percov}\emph{f}) with $\Sbar^2_{\btheta}\ofk$
(Fig.~\ref{fig:percov}\emph{e}).

Our ability to analytically calculate periodogram covariance
(eq.~\ref{eq:percov}), the covariance of the blurred score
(eq.~\ref{eq:covscore}), and ultimately estimator covariance
(eq.~\ref{eq:magic}) is a boon of our methodology. As the covariance between
the model parameter estimates only depends on the parameter values and the
geometry of the experiment, without reference to a particular data realization,
we are able to calculate confidence intervals for single data realizations
without the need for additional simulation, and, in addition, we can employ
eq.~(\ref{eq:magic}) for designing sampling plans that minimize estimation
errors or other criteria, which is of great applicability across the sciences.

\subsection{How to implement our method---and how to read this paper}

Section~\ref{sec:parametriccovariancemodels} stated the Mat\'ern model and
pointed to its two-parameter special cases.  Section~\ref{sec:sim} defined the
observation window~$w\ofx$, and formalized the approach to creating synthetics
that fully conform to all model assumptions, allowing for simulations and
scenario testing. In Sections~\ref{sec:mle} (formulating the Whittle likelihood
and maximizing it to obtain the Mat\'ern model parameters), \ref{sec:res}
(calculating residuals and interrogating them for model departures), and
\ref{sec:cov} (quantifying parameter uncertainties and inspecting their
correlations to enable model comparisons), we discussed the three key steps
involved in obtaining an estimate for the covariance parameters of a random
field~$\mcH\ofx$, as well as tools for their assessment.
Section~\ref{sec:missing} provides the full numerical validation of the
robustness of our strategy for irregularly bounded and incompletely sampled
fields with a series of visual indicators and a variety of metrics.  When
compared with their equivalents for the completely sampled, rectangular case
considered by \cite{Simons+2026}, they corroborate that Whittle likelihood
analysis of Mat\'ern covariance models need not suffer sampling.

To give the reader a clear picture of how to implement our methodology (and for
our computer code, see Section~\ref{code}) for their own science, we now outline
exploratory data analysis and preprocessing steps to help geospatial datasets
align with our modeling assumptions.

A basic requirement is that
finite-\textit{variance} field data should be collected at an interval and over
an extent that are broadly commensurate with domain-specific initial notions of
\textit{smoothness} (grid discretization needs to be attuned to the visible
variability) and \textit{correlation lengths} (grid size and spacing need to
capture characteristic scales of the process) over physical domains that are not
patently amalgamated (suggesting lack of \textit{stationarity}) or showing
preferred directionality (hinting at loss of \textit{isotropy}).
Next, observed data, histograms, and summary statistics can be checked for obvious
departures from \textit{Gaussianity}. Non-negligible excess kurtosis might be
treated by outlier removal, and areas marred by artifacts or
revealing a likely non-random nature should be excluded. The spatial
window~$w\ofx$ must capture data availability; it can be crafted to isolate
mapped regions of specific interest or provenance, or to select for good
behavior: it should be designed as a mask, taper, or weight with these purposes
in mind.
Lastly, zero-mean randomness is not guaranteed by simple mean removal: guarding
against locally varying means can be accomplished by detrending. In our
modeling, we consider polynomials up to order two as deterministic, and we
least-squares fit and remove them prior to analysis. Mat\'ern covariance
modeling using our procedure may be iterative: inspecting residuals for
unmodeled structure could motivate a reassessment of the data taken to
be~$\mcH\ofx$.

Not all readers will be interested in reading beyond the numerical
validation of statistical and computational efficiency presented in
Section~\ref{sec:missing}.  Those focused on how to conduct Mat\'ern
covariance parameter estimation may wish to skip to Section~\ref{code} for
our code availability.

Readers invested in utilizing specific models (e.g., exponential,
squared-exponential) or deciding why they should use the Mat\'ern class are
advised to read Section~\ref{sec:fixing}, which deals with model specification
by parameter fixing, e.g., to its special cases. Section~\ref{dunno} builds the
bridge between model selection and sampling patterns, and how these choices or
externalities interact. Realistic geoscience sampling scenarios are brought in
using synthetics and simulations.  Section~\ref{sec:separ} dwells in more detail
on the motivating example presented in our first real-data example of
Fig.~\ref{fig:bath2synth}, furthering the discussion on the implications of
analyzing non-stationary fields, with or without knowledge of the geographical
boundary separating them.

Section~\ref{sec:modeltest} takes us through a variety of real-data test cases
that complement the gallery of geoscientific examples that we presented as
\cite{Simons+2026}. They fit in this repertoire while being incompletely and
irregularly sampled because the debiased Whittle likelihood effectively erases
sampling-induced bias, while remaining computationally efficient in its
spectral-domain formulation.
These examples were not selected to prove that the world is governed only by
Gaussian processes belonging to the Mat\'ern covariance class; they were chosen
to convince the reader that \textit{if} they are, we can find their parameters
while being cognizant of the sampling function, we can find and interpret their
uncertainties, and we can synthesize simulated observations.  If they are not,
we can assess residuals for model departure: such diagnostics help alert us to
any and all of the aspects that our model cannot capture, helping identify
alternatives, and carrying information about the physical process under study
(e.g., the directionality of seafloor fractures). Not all datasets will be
Gaussian, stationary, isotropic, and Mat\'ern, but our analysis under those
assumptions as a null hypothesis will not be unduly influenced by how
specifically we collect or mask
(e.g., to remove clouds in remote sensing)
the data.
Thus, in Section~\ref{sec:modeltest}, we demonstrate the spatial isolation of a homogeneous background from
deterministic features, and how through our model evaluation tools, we can interpret the
random portion of the dataset to be isotropically Mat\'ern.
Additionally, we show how to extract
disjoint processes from within the same dataset in order to show
data departures from a model that may well turn out to reveal a feature of
physical interest (be it the phenomenon or how it is measured) as they can not
be readily attributed to artifacts and biases arising from spatial sampling
patterns.  Finally, we present a directly observed, presumably
single, statistical process, but under a sampling pattern so intricate that we
must make sure any inference based on it is not unduly influenced by it.  While
the sampling pattern explains much of the variability within the periodogram,
fine-scale, directional structure apparent within the model residuals shows that
some regions of seafloor are not isotropic.

The penultimate section is for the statistically minded: Section~\ref{sec:asymp}
helps understand large-sample asymptotic behavior of the estimators. Having
established that taking sampling pattern ``shape'' into account matters,
sampling ``size'' will matter, too. Increasing the \textit{number} of samples
taken within a specific ``template'' influences the behavior of the
estimators. Growing the domain, reducing data missingness, infilling within a
fixed domain: these are all strategies with consequences for parameter recovery
for which we developed the machinery to study. Extensions of our methodology
from two to $d$ dimensions are mathematically intrinsically contained in our
method and understanding them statistically will benefit from this section. Our
root-mean-squared error decay with sample size~$K$ as $1/\sqrt{K}$ is favorable
compared to methods where parameter estimation bias is expected to dominate over
variance with growing dimension \cite[]{Dahlhaus+1987}. The last section is for
aficionados and sampling strategy enthusiasts devoted to experimental
design. Section~\ref{sec:design} prepares the reader for the most practical
question of all: where should they be adding more data to better understand
their model, their science, their field of study? Designing experiments is
possible within the context of this paper because we are able to calculate
parameter uncertainties via dataless parameter covariance predictions
conditioned on the sampling pattern.

\section{U~N~B~I~A~S~E~D{\hsps}P~A~R~A~M~E~T~E~R{\hsps}E~S~T~I~M~A~T~E~S{\hsps}F~R~O~M{\hsps}I~N~C~O~M~P~L~E~T~E~L~Y{\hsps}S~A~M~P~L~E~D{\hsps}D~A~T~A}\label{sec:missing}

We illustrated the power of the debiased Whittle likelihood in
\cite{Simons+2026} for completely sampled, evenly spaced, rectangular grids. We
recall their and our ability to incorporate exactly, and fast, the effects of
sampling and bounding, into the likelihood~(eq.~\ref{eq:dwl}). In their Fig.~2,
they showed that the blurred spectral density~(eq.~\ref{eq:blurredspec}) matches
the average periodogram of the data $\overline{|H\ofk|^2}$, and in their Fig.~6,
they illustrated the geometrical structure of the covariance of the periodogram,
$\cov\{|H(\kb)|^2,|H(\kbp)|^2\}$, which controls the parameter
covariance~(eq.~\ref{eq:magic}). In the present paper, we have so far, in our
Fig.~\ref{fig:blurring}, documented the effect on the expected periodogram of
missingness through random grid subsampling, and we have, in our
Fig.~\ref{fig:percov}, shown the corresponding structure and geometry of the
periodogram covariance. In this section, we show how the debiased Whittle
likelihood continues to deliver asymptotically unbiased, Gaussian estimates of
Mat\'ern parameters for those kinds of sampling scenarios, including how well
our uncertainty quantification continues to capture the observed behavior. We
conclude this section by covering the behavior of the moments of the periodogram
for contiguously missing patches---and their complements.

\subsection{Randomly subsampled regular grids}

\begin{figure}\centering
  \includegraphics[width=0.75\textwidth,angle=0,trim=0cm 0cm 0cm 2.1cm,clip]{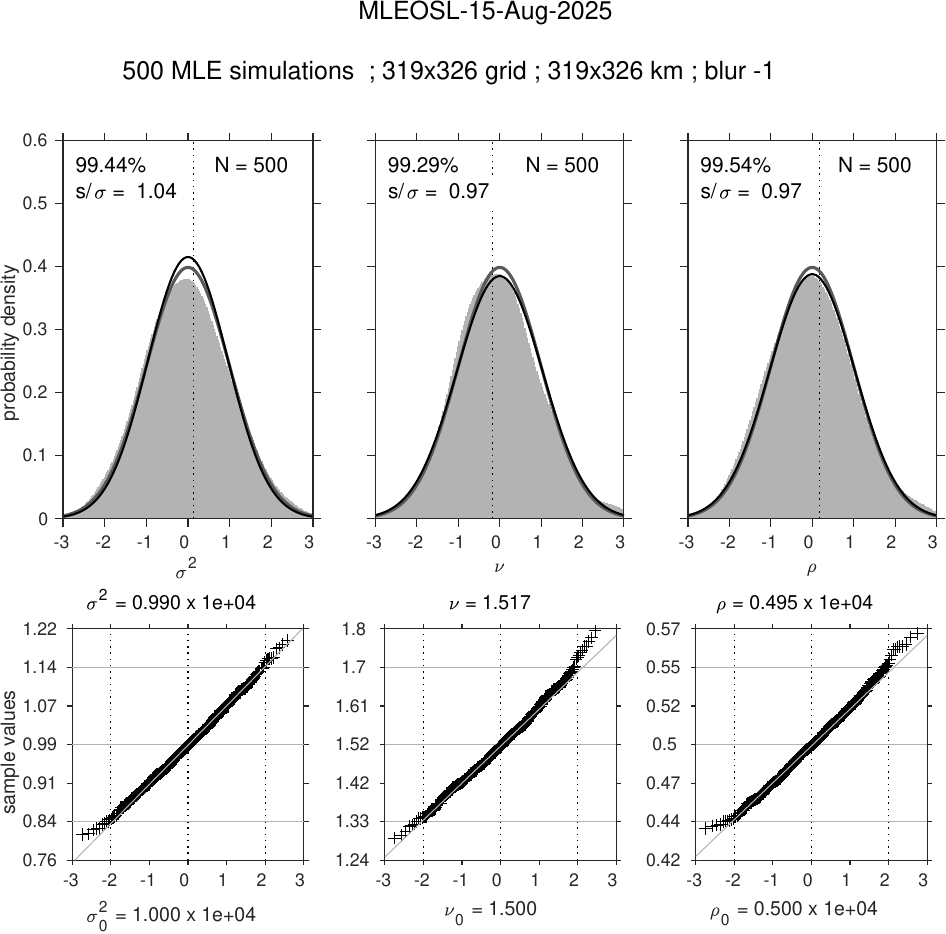}
  \caption{\label{fig:mlecurves}
    Maximum-likelihood estimation (Section~\ref{sec:mle}) on a randomly
    subsampled (66.7\% observed) grid as shown in
    Fig.~\ref{fig:blurring}\textit{(a)}: ensemble statistics. Mat\'ern parameter
    maximum-likelihood estimation statistics for 500 spatial-covariance
    embedding simulations carried out on a 319$\times$326 grid with $1~\km \times
    1~\km$ spacing exhibiting uniform random deletions (see
    Fig.~\ref{fig:blurring}) for true Mat\'ern parameter values of
    $\sigma^2_0=10$~km$^2$, $\nu_0=1.5$, and $\rho_0=5$~km, recovered via
    maximization of the exactly blurred, uncorrelated,
    likelihood~(eq.~\ref{eq:dwl}). The estimates average to
    $\st=9.90\pm0.75$~km$^2$, $\nu=1.52\pm0.09$, $\rho=4.95\pm0.26$~km, quoting
    one observed standard deviation. The thick gray line is derived from the
    covariance calculated from the ensemble of simulation and estimation
    outcomes. The thick black line is based on the covariance exactly calculated
    from eq.~(\ref{eq:magic}) shown and calculated at the truth, shown by the
    dotted vertical line in the top panels. The quantile-quantile plots of the
    bottom row show the ensemble distribution between its $[-3, 3]$ empirical
    standard deviations (abscissas) in relation to the theoretical normal
    distribution parameterized by the sample statistics (ordinates). }
\end{figure}

\begin{figure}\centering
  \includegraphics[width=0.95\textwidth,angle=0,trim=0cm 0cm 0cm 1.5cm,clip]{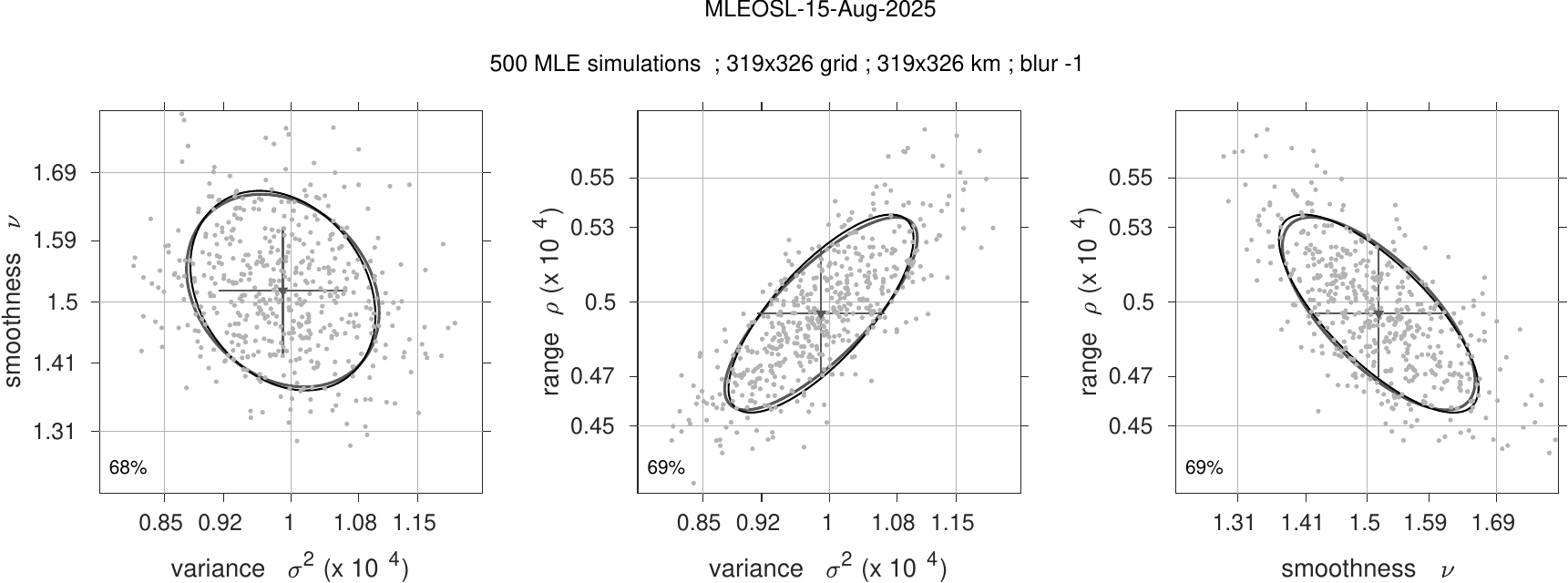}
  \caption{\label{fig:mleellipses}
    Maximum-likelihood estimation (Section~\ref{sec:mle}) on a randomly
    subsampled (66.7\% observed) grid as shown in
    Fig.~\ref{fig:blurring}\textit{(a)}: parameter covariance.
    Maximum-likelihood estimation statistics for the ensemble of 500 simulation
    and recovery experiments reported in Fig.~\ref{fig:mlecurves}. The mean
    estimate is highlighted by the gray triangle and two observed standard
    deviations marked by gray lines. The heavy gray ellipse is the 68\%
    confidence region based on the ensemble. The closely matching heavy black
    ellipse is the predicted 68\% confidence region based on the covariance
    predicted from eq.~(\ref{eq:magic}). Values quoted in the bottom left are
    the percentages of estimates falling within the 68.27\% predicted confidence
    interval.}
\end{figure}

\begin{figure}\centering
  \includegraphics[width=0.65\textwidth,angle=0,trim=0.0cm 0.75cm 0cm 0.5cm,clip]{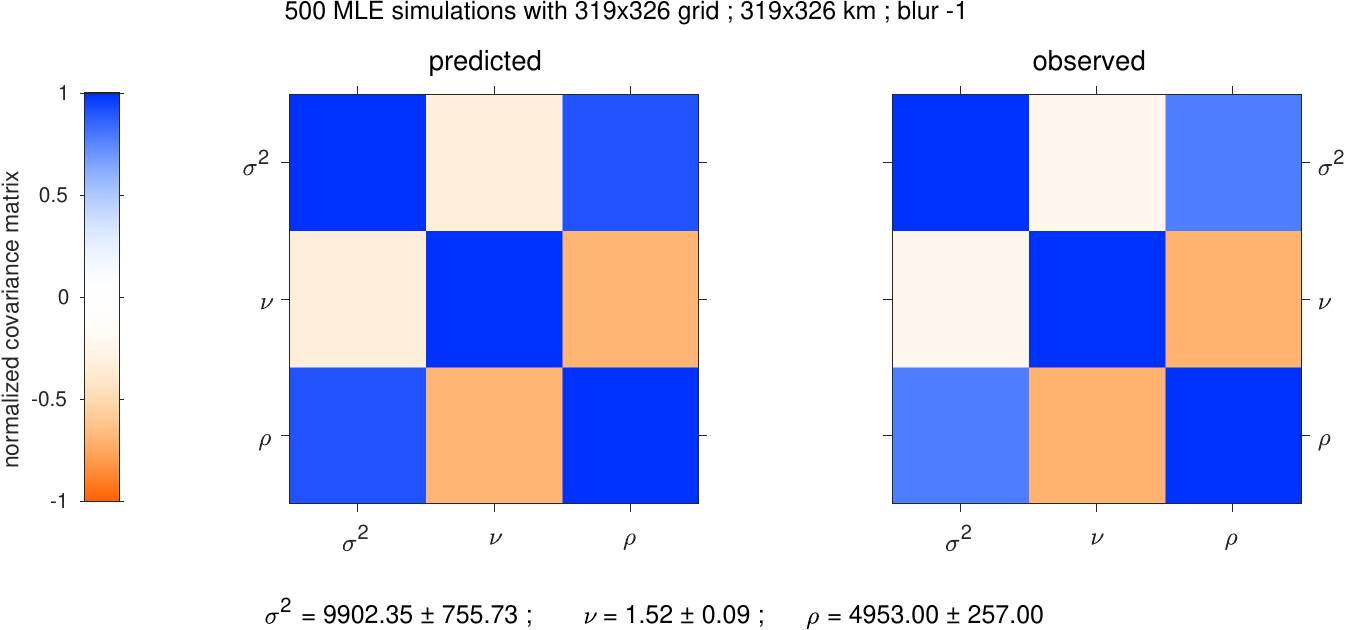}
  \caption{\label{fig:mlerubiks}
    Maximum-likelihood estimation (Section~\ref{sec:mle}) on a randomly
    subsampled (66.7\% observed) grid as shown in
    Fig.~\ref{fig:blurring}\textit{(a)}: ensemble correlations. Comparison of
    the covariance predicted via eq.~(\ref{eq:magic}) and the covariance
    observed on the set of experiments reported in Figs~\ref{fig:mlecurves}
    and~\ref{fig:mleellipses}. Shown are the relevant correlation matrices
    between the estimators for the three Mat\'ern parameters $\st$, $\nu$, and
    $\rho$, highlighting the relatively strong trade-off between $\nu$ and
    $\rho$.}
\end{figure}

\begin{figure}\centering
  \includegraphics[width=\textwidth]{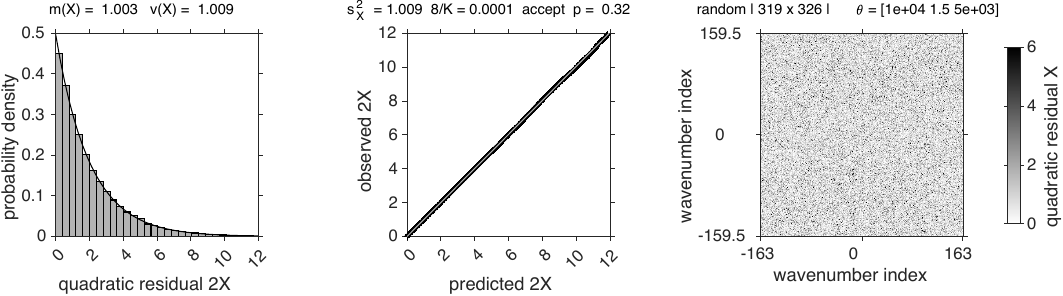}
  \caption{\label{fig:res}
    Maximum-likelihood estimation (Section~\ref{sec:mle}) on a randomly
    subsampled (66.7\% observed) grid as shown in
    Fig.~\ref{fig:blurring}\textit{(a)}: residual statistics. We report on a
    single, randomly selected simulation carried out in the experiment shown in
    Fig.~\ref{fig:blurring}\textit{(a)}. As in Fig.~11 of \cite{Simons+2026},
    the distribution of the variable~$X\ofst\ofk$ of eq.~(\ref{eq:chisq}) is
    shown as a histogram across all wavenumbers with the theoretical
    distribution superposed, as a quantile-quantile plot for the distribution in
    question, and as a spectral-domain map. Plot labels refer to (\emph{left})
    the mean, \textsf{m(X)}, and variance, \textsf{v(X)}, of $X\ofst\ofk$,
    (\emph{center}) the test-statistic ($\textsf{s}_\textsf{X}^2$,
    eq.~\ref{eq:stx}), variance of the convergence distribution given the number
    of samples $K$ for the observed grid, whether the rigorous test was
    rejected, and with what significance (\textsf{p}-value, \emph{right}), and
    the grid and model parameterization.}
\end{figure}

We demonstrate the robustness of our estimation strategy and quantitatively
evaluate the quality of our estimated models through numerical experiments with
stationary Mat\'ern random fields. Figs~\ref{fig:mlecurves}--\ref{fig:res}
report on an ensemble of simulations and their maximum-likelihood estimates over
many realizations from the same model~$\btheta_0$ for sampling operators~$w\ofx$
characterized as 66.7\% observed rectangular grids. The $w\ofx$ we implement
for these first experiments all exhibit uniformly random missingness so that we
can build intuition without the confounding factors of grid geometry, size, or
sampling pattern. Evaluating statistics from this ensemble and its realizations
provides us with information about how well we can hope to know our
parametrically modeled random fields, as a framework for interpretation when we
are confronted with real data, when not all of our modeling assumptions will be
met. We save the complications on estimators and their covariances related to
varying grid geometry, growing sample size, and (geographically) structured
sampling patterns for later in this manuscript.

We visualize ensembles of parameter estimates for an experiment of 500
realizations of the same model and grid parameters as in
Fig.~\ref{fig:blurring}. Fig.~\ref{fig:mlecurves} shows the distribution of the
marginals of the model estimates to be approximately normally distributed.
Shown are gray kernel density estimates (kdes) with gray overlayed normal
distribution fits to the sample mean and standard deviation in the top row, and
as quantile-quantile (Q-Q) plot comparisons of the observed estimates to a
theoretical normal distribution in the bottom row. From these diagnostics, we
see agreement of the numerical estimates with theory in that the centrality of
the distribution of estimates is aligned with $\btheta_0$ (i.e., $\hbtheta$ is
an unbiased estimator). The spread of the distribution of estimates is consistent with our
analytical calculation of parameter covariance, shown as the thick black line.
The analytically calculated parameter standard deviation follows
eq.~(\ref{eq:magic}), evaluated at the mean of the estimator for this ensemble.
The analytical calculation closely matches our numerical experiments for these
irregularly sampled fields, as well as it did for the complete sampling
scenarios presented by \cite{Simons+2026}, their Fig.~7. Optimizing the likelihood
in eq.~(\ref{eq:dwl}) correctly accounts for missingness; see Section~\ref{sec:asymp} for
a discussion of the asymptotics of this behavior.

Fig.~\ref{fig:mleellipses} shows our suite of estimates as parameter
cross-plots. Each of the estimates is plotted in the three permutations of the
two-dimensional parameter spaces. The 68\% confidence ellipses calculated are
drawn from the observed samples, in gray, and from the analytically calculated
covariance of eq.~(\ref{eq:magic}), in black. Here, for the 66.7\% observed
field with random deletions, we see a subtle negative correlation between
estimators of $\st$ and $\nu$, a strong positive correlation between estimators
of $\st$ and~$\rho$, and a strong negative correlation between those of $\nu$
and~$\rho$. The empirical values are $\{\st,\nu\}=-0.2254$,
$\{\st,\rho\}=0.7830$, and $\{\nu,\rho\}=-0.7006$, respectively. The
correlations theoretically calculated via eq.~(\ref{eq:magic}) are
$\{\st,\nu\}=-0.2569$, $\{\st,\rho\}=0.7637$, and $\{\nu,\rho\}=-0.7441$, which
are good predictors for those observed. The results for these incompletely
sampled fields remain in qualitative agreement with the complete sampling
scenarios presented by \cite{Simons+2026}, their Fig.~9. As we will see later,
the strength of the parameter correlation is dependent on the sampling geometry,
structure, and size.

Fig.~\ref{fig:mlerubiks} gives a second means of assessing the correlation of
the estimates. We render the normalized parameter covariance as a heat map. The
correlation structure observed through experiment is consistent with our
prediction, as it was for \cite{Simons+2026} in their Fig.~10.

Fig.~\ref{fig:res} evaluates the model via a graphical representation of the
distribution of the spectral residuals $X_{\btheta}(\kb)$ of
eq.~(\ref{eq:chisq}) for the same model, grid, and taper as previously detailed
in Fig.~\ref{fig:blurring}. The histogram and Q-Q plot follow the $\chi^2_2$
theoretical distribution very closely, and the wave vector plot retains no
structure. This behavior mirrors the completely sampled case shown by
\cite{Simons+2026} in their Fig.~11. In scenarios where we evaluate models for
real data and the model is rejected, an inspection of such statistics and
visualizations allows us to glean information regarding how the statistics of
our dataset depart from our model assumptions.

\subsection{Irregularly bounded patches}

Repeating the experiments as illustrated in
Figs~\ref{fig:mlecurves}--\ref{fig:res} for different non-random sampling
scenarios, in which patches with particular boundary geometries are either
present or absent, only corroborates our assertion that the debiased Whittle
likelihood~(eq.~\ref{eq:dwl}) produces unbiased Gaussian estimators with fully
characterizable uncertainties with powerful diagnostics for evaluating model
fitness. Hence, while we have validated this performance extensively, to save
space, we are not showing the full results here. Our openly available code (see
Section~\ref{code}) allows the reader to explore all manner of sampling
scenarios.

\begin{figure}
  \centering
  \includegraphics[width=0.8\textwidth,trim=1.65cm 3.2cm 4.75cm 1.65cm,clip]{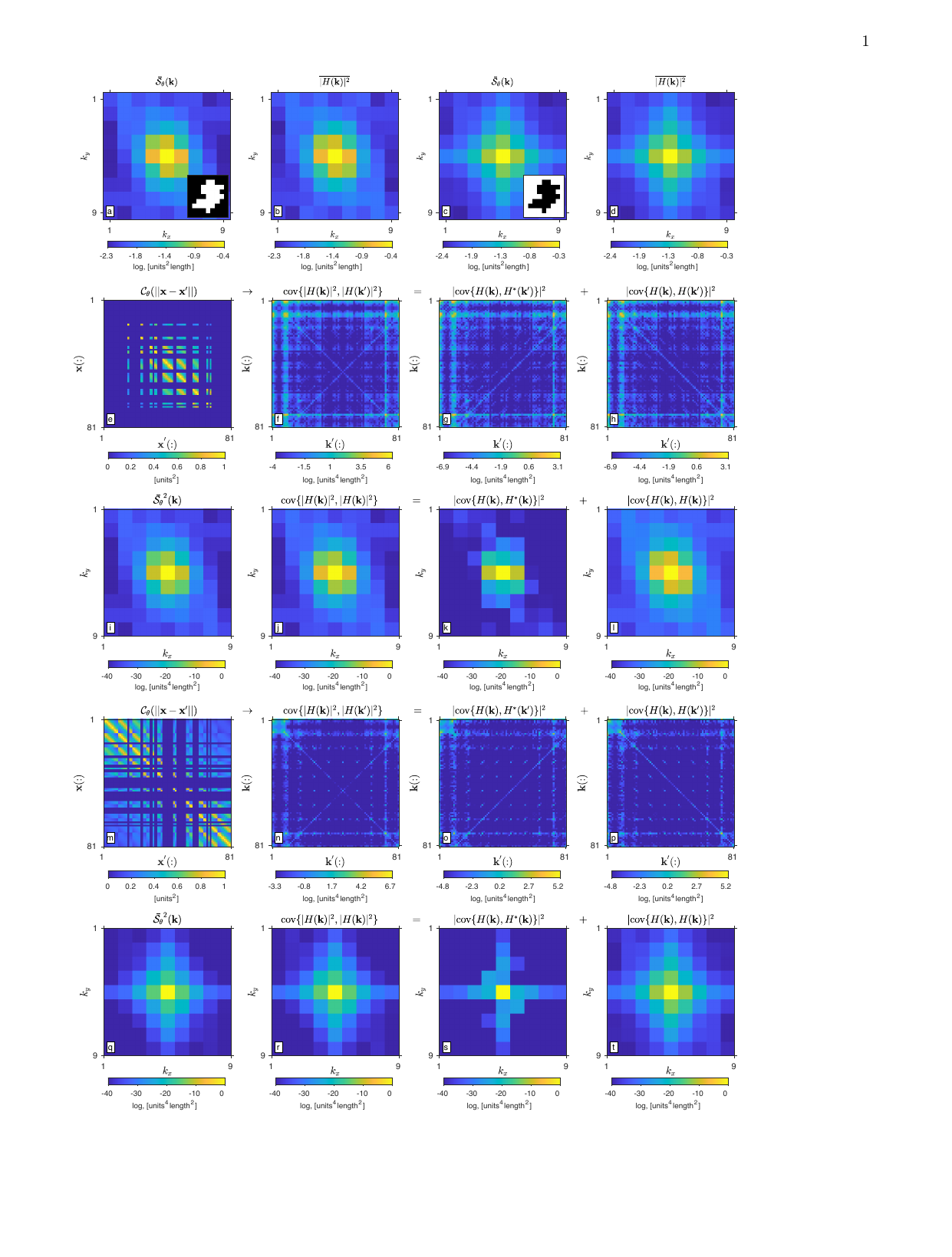}
  \caption{\label{fig:percovnj}
    Expected and averaged periodograms with their computed covariances for
    contiguous, bounded fields. All panels and models are for a $9\times 9$
    perimeter with $1\,\m$ even spacing and
    $\btheta=[1\,\m^2\;0.5\;2\,\m]$. Top, \emph{(a--d)}: the expected
    periodogram $\Sbar\ofst\ofk$ and the average periodogram over 100
    realizations $\overline{|H\ofk|^2}$ for a field sampled within the interior
    of an irregular boundary comprising 68\% of the complete grid, and
    $\Sbar\ofst\ofk$ and $\langle|H\ofk|^2\rangle$ for a field sampled from the
    complementary exterior of the irregular boundary as 32\% of the complete
    grid. $w\ofx$ for the interior and exterior sampling patterns are shown in
    the subset of the $\Sbar\ofst\ofk$ subfigures. The covariance and
    pseudocovariance terms of the periodogram covariance (eq.~\ref{eq:percov})
    for the interior (\emph{e--l}) and exterior (\emph{m--t}) sampling patterns. Layouts of the periodogram covariance
    panel sets are as in Fig.~\ref{fig:percov}. }
\end{figure}

The behavior of the estimates made via eq.~(\ref{eq:bgth}) is controlled by the
shape of the sampling kernels $w\ofx$, their correlations $W\ofx$, and the
blurring effect of the spectral window $|w\ofk|^2$, where $w\ofk$ is the Fourier
transform of $w\ofx$, on the spectral densities $\Sbar\ofst\ofk$ in
eq.~(\ref{eq:blurredspec}), which is equal to the expected periodogram of the
data $\langle |H\ofk|^2\rangle$ as could be seen from the average
$\overline{|H\ofk|^2}$ shown in Fig.~\ref{fig:blurring}. For a sampling scenario
where a New Jersey shaped domain is the region of interest or omission,
Fig.~\ref{fig:percovnj} shows the analytically blurred spectral densities
$\Sbar\ofst\ofk$ and the average periodogram $\overline{|H\ofk|^2}$ over 100
realizations for a Mat\'ern model noted in the caption.

The parameter uncertainties and covariances obtained with
eqs~(\ref{eq:magic})--(\ref{eq:percov3}) are to be understood geometrically
through the covariance of the periodogram $\langle |H\ofk|^2,|H\ofkp|^2\rangle$
as shown in Fig.~\ref{fig:percov} for random sampling, and now illustrated by
Fig.~\ref{fig:percovnj} for New Jersey and its complement.
Sections~\ref{sec:fixing}--\ref{sec:modeltest} elaborate on this behavior for cases
that carry specific geophysical relevance. After discussing the statistics of
asymptotics in Section~\ref{sec:asymp}, we take a final look in
Section~\ref{sec:design} at the effects of aliasing and blurring, sampling and
bounding, across a variety of data observation scenarios that arise in the Earth
sciences, specifically through the lens of the spectral window.

\section{S~I~M~P~L~I~F~Y~I~N~G{\hsps}T~H~E{\hsps}M~A~T~\'E~R~N{\hsps}D~E~N~S~I~T~Y{\hspst}:{\hsps}F~I~X~I~N~G{\hsps}P~A~R~A~M~E~T~E~R~S}\label{sec:fixing}

Models frequently assumed within the geophysical community
\cite[e.g.,][]{Tarantola+84,Frankel+1986,Goff+88,North+2011,Muir+2023} are
special cases of the general Mat\'ern form. In Table~\ref{tab:MaternSpecial}, we
provide analytic simplifications of the Mat\'ern spatial covariance
(eq.~\ref{eq:spatmatern}) and two-dimensional spectral density
(eq.~\ref{eq:specmatern2d}) by specifying the smoothness parameter $\nu$ to
fixed values that correspond to covariance functions that are commonly selected
within the physical sciences. Many of these are motivated by the mean-squared
behavior of the solutions of differential equations
\cite[e.g.,][]{vonKarman+1938,Bui-Thanh+2013,CardenasAvendano+2023}. We provide
these simplifications for the $d$-dimensional spectral density
(eq.~\ref{eq:specmaterndd}) in Table~\ref{tab:MaternSpecialAll}.

In relating our forward and inverse modeling of the general three-parameter
Mat\'ern spatial and spectral covariance functions to that of a nested model
where one or more parameters are fixed \textit{a priori} (e.g., a two-parameter
special case), we adjust the parameterization from
\begin{equation}\label{eq:111}
  \btheta=[\st,\,\nu,\,\rho]^{T}
\end{equation}
to the set of nested parameters that will be allowed to vary, e.g.,
\begin{equation}\label{eq:101}
    \btheta^{101}=[\st,\,\cdot\,,\rho]^{T},
\end{equation}
where the superscript indicates the parameters actually inverted for (as 1), or
held fixed (as 0).

Beyond taking a specialized covariance function to be representative of the
statistics of a dataset due to precedent or assumption, prior knowledge of its
character may make working with a nested covariance model desirable. One can
determine the potential consequences of adopting a special-case parametric
covariance model rather than the general form through model selection tests
\cite[e.g.,][]{Vuong1989} or through simulations conducted to assess estimator
ensemble bias and uncertainties.
We include the option for estimating nested models with parameters specified
as in eq.~(\ref{eq:101}) within the implementation of our maximum-likelihood
estimation strategy. Any (or all) parameters may be held fixed to a supplied
initial value and the inversion calculations will only optimize the likelihood
(eq.~\ref{eq:dwl}) by varying the free parameters. In the case of the smoothness
parameter $\nu$ held fixed to a value that corresponds to one of the special
cases (Tables~\ref{tab:MaternSpecial} and~\ref{tab:MaternSpecialAll}), one may invert for the simplified
analytic function rather than eq.~(\ref{eq:specmatern2d}). In such special
nested model cases, we optimize only over variance $\st$ and range $\rho$ (as in
eq.~\ref{eq:101}), reducing the estimation problem to characterizing a
covariance structure based upon the amplitude of the random field and a scale
factor.
Verification calculations of the gradient vector (eq.~\ref{eq:bgth2}), Hessian
(eq.~\ref{hessian}), and Fisher (eq.~\ref{eq:fisherinfo}) matrix to compare with
the observed numerical behavior and to evaluate eq.~(\ref{eq:magic}) at the
estimate $\hbtheta$, rely on first and second partial derivatives of the
simplified nested models, for which we will provide
analytic expressions in forthcoming work.

\begin{figure*}
  \centering
  \includegraphics[height=10cm,trim={2.3cm 16.2cm 2.75cm 1.6cm},clip]{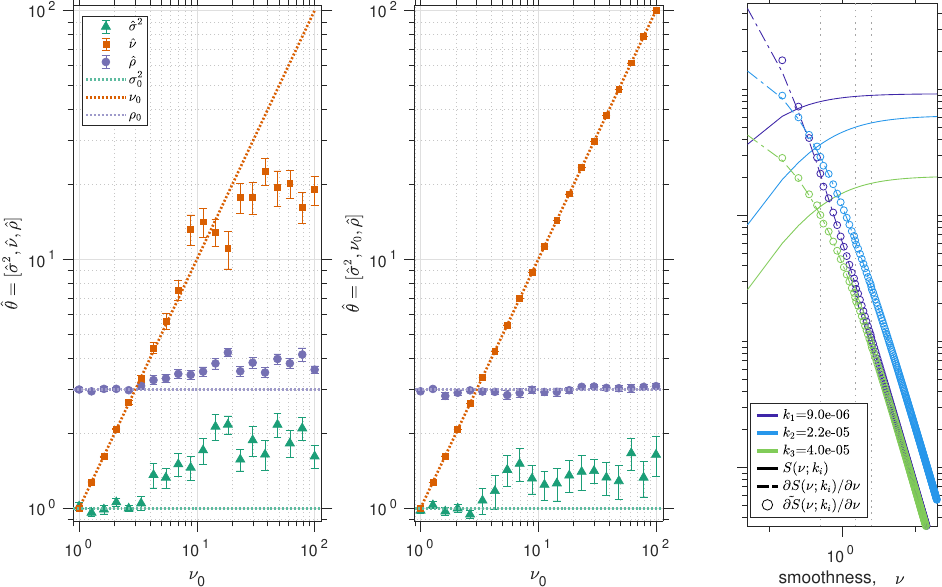}  
  \caption{\label{fig:estincnu}
    Dependence of Mat\'ern parameter estimates on smoothness. (\emph{Left})
    Three-parameter estimates. (\emph{Center}) Two-parameter estimates with
    correctly fixed $\nu_0$. The $128\,\m\times128\,\m$ sampling grid is
    completely, regular, and evenly spaced. Markers show mean estimates
    $\hat{\theta}=\{\hat{\sigma}^2, \hat{\nu}, \hat{\rho}\}$ for 48 realizations
    with their empirical standard errors. True values $\theta_0=\{\sigma_0^2,
    \nu_0, \rho_0\}$ are displayed as colored, dotted lines for each trial model
    where the variance $\sigma_0^2=1~\m^2$ and range $\rho_0=3~\m$ are held
    constant while the smoothness $\nu_0$ increases from $1$ to $100$.
    (\emph{Right}) At three selected wavenumbers $k_i$, $i=1,2,3$, values of the
    normalized radial spectra (solid), and their derivatives with respect to the
    smoothness (exact, dashed), (numerically approximated, open circles) are
    displayed.}
\end{figure*}

In Fig.~\ref{fig:estincnu}, we study the influence of increasingly large
smoothness parameter $\nu$ on the reliability of our estimation strategy through
two numerical experiments. In both experiments, we set a common variance
$\sigma^2_0=1\,\m^2$ and range $\rho_0= 3\,\m$, and select 20 logarithmically
spaced values of $\nu_0\in [1, 100]$. We simulate 48 realizations for each of
these 20 target models on a common, fully sampled, evenly spaced $128\,\m \times
128\,\m$ grid. In the first experiment (Fig.~\ref{fig:estincnu},
\textit{left}), we estimate all three parameters of the Mat\'ern model, and in
the second (Fig.~\ref{fig:estincnu}, \textit{center}), we fix the smoothness
correctly to $\nu_0$ and estimate the variance $\st$ and range $\rho$ as
two-parameter nested models.
In the three-parameter inversion experiment, we observe that as $\nu_0$
increases, all three parameters are well estimated until a point when they begin
to accumulate bias which grows until it saturates. The one-standard error
(1~se) empirical confidence interval for the variance parameter $\st$ excludes
$\sigma^2_0$ when $\nu_0>4$, while the 1~se interval for the smoothness $\nu$
increases until about $\nu_0>9$, after which $\hat{\nu}$ saturates and hovers
around 10--20. The 1~se empirical confidence interval for the range parameter
$\rho$ fails to contain $\rho_0$ when $\nu_0>4$. In the second experiment, the
1~se interval for the variance $\st$ exceeds $\sigma^2_0$ for $\nu_0>5$ and its
bias plateaus at about half that of the three-parameter inversion. Not being
estimated, $\nu$ is plotted along the 1:1 line. The confidence intervals of
$\hat{\rho}$ overlap with $\rho_0$ for all trial models.

The difficulty in accurately estimating all three Mat\'ern parameters for large
$\nu$ should be expected due to problems with identifiability in ``sloppy''
models \cite[]{MonsalveBravo+2022}, a trait that can be observed, e.g., in the
parameter correlation structure of Figs~\ref{fig:mleellipses}
and~\ref{fig:mlerubiks}. The strong trade-off between $\nu$ and $\rho$ is
apparent; when we know $\nu$ and fix it, $\rho$ is more adeptly estimated, and
along with it, $\sigma^2$. The failure of $\hat{\nu}$ to approximate large
$\nu_0$ suggests there is an upper bound in being able to distinguish variation
in $\nu$ for smooth fields, depending on sampling density $\Delta y,\, \Delta
x$. As a function of $\nu$ the Mat\'ern spectral
density~(eq.~\ref{eq:specmatern2d}) begins to plateau first for small
wavenumbers, followed by intermediate and large wavenumbers. In
Fig.~\ref{fig:estincnu} (\textit{right}), we explore how the spectral density
and its first derivative with respect to the smoothness varies with increasing
smoothness for small, medium, and large wavenumbers. In the presence of very
smooth, large $\nu$, fields, fixing the smoothness parameter should be
considered. However, the squared-exponential ($\nu\rightarrow\infty$) model is
still not an advisable choice due to its accelerated decay with wavenumber,
which is numerically problematic without additional precautions taken in the
analysis (e.g., including a nugget effect). Analytically, the limiting behavior
$\nu\rightarrow\infty$ of the spectral domain illuminates its asymptotically
Gaussian nature, which depends on $\st$, $\rho$, and wavenumber~$k$, as in
Table~\ref{tab:MaternSpecial},
\begin{equation}\label{eq:SkNuLimInf}
\lim_{\nu\to\infty} S(k) = \sigma^2 \left( \frac{\pi\rho^2}{4}
    \right) \exp \left(- \frac{ \pi^{2} \rho^{2} k^2 }{ 4 } \right)
    \quad \for \quad k\in\mathbb{R}^+
    .
\end{equation} 

The experiments shown in Fig.~\ref{fig:estincnu} pertained to complete
rectangular samplings. We next illustrate the nested special cases of the
Mat\'ern form for realistic synthetics comprising three geophysically motivated
spatial sampling cases subject to irregular boundaries, structured tracks, and
random missingness. For each of those we will discuss both correctly and
incorrectly selected nested models.

\section{M~O~D~E~L{\hsps}S~E~L~E~C~T~I~O~N{\hsps}A~N~D{\hsps}S~A~M~P~L~I~N~G{\hsps}P~A~T~T~E~R~N~S}\label{sec:specialsamples}\label{dunno}

In this section, we create realistic synthetics to further elucidate two of the
problems we introduced that are regularly encountered in modeling geophysical
data: selecting appropriate covariance models (see Section~\ref{sec:fixing}) and
handling discrete, missing data (see Section~\ref{sec:missing}).

The flexible attributes of the Mat\'ern density are the scale, shape, and size
parameters. The comparison of three-parameter and two-parameter inversions shown
in Fig.~\ref{fig:estincnu} illustrated how in the case of large smoothness, the
option to fix the smoothness parameter can be advantageous. When the fixed
hyperparameter(s) are selected correctly, the behavior of the estimators is
approximately equivalent with that of the general Mat\'ern model. If the value
of the fixed parameter is chosen incorrectly, the model is effectively unable to
characterize the others, and accuracy in the estimation of the parameters
inverted for will suffer. The magnitude of this compensatory parameter bias
depends on the spatial sampling of the data.

We now inspect the applicability of two-parameter special cases for three types
of realistic sampling patterns: a geographic patch, sweeps of sampling tracks,
and uniform random missingness. We then address the issue of separability, both
in space and in model parameters.  We conduct numerical experiments for a
variety of sampling patterns (contiguous, structured, and dispersed) on the
partial grid where we compare summary statistics of estimating the three
Mat\'ern parameters (eq.~\ref{eq:111}) versus two
(eq.~\ref{eq:101}). Specifically, we estimate the $\sigma^2$ and $\rho$ that
capture the amplitude and length scale of the covariance structure while setting
the smoothness $\nu$ as a hyperparameter.

\begin{figure*}
    \centering
    \includegraphics[height=0.85\textheight,angle=0,trim=4.2cm 0.2cm 1.5cm 1.3cm,clip]{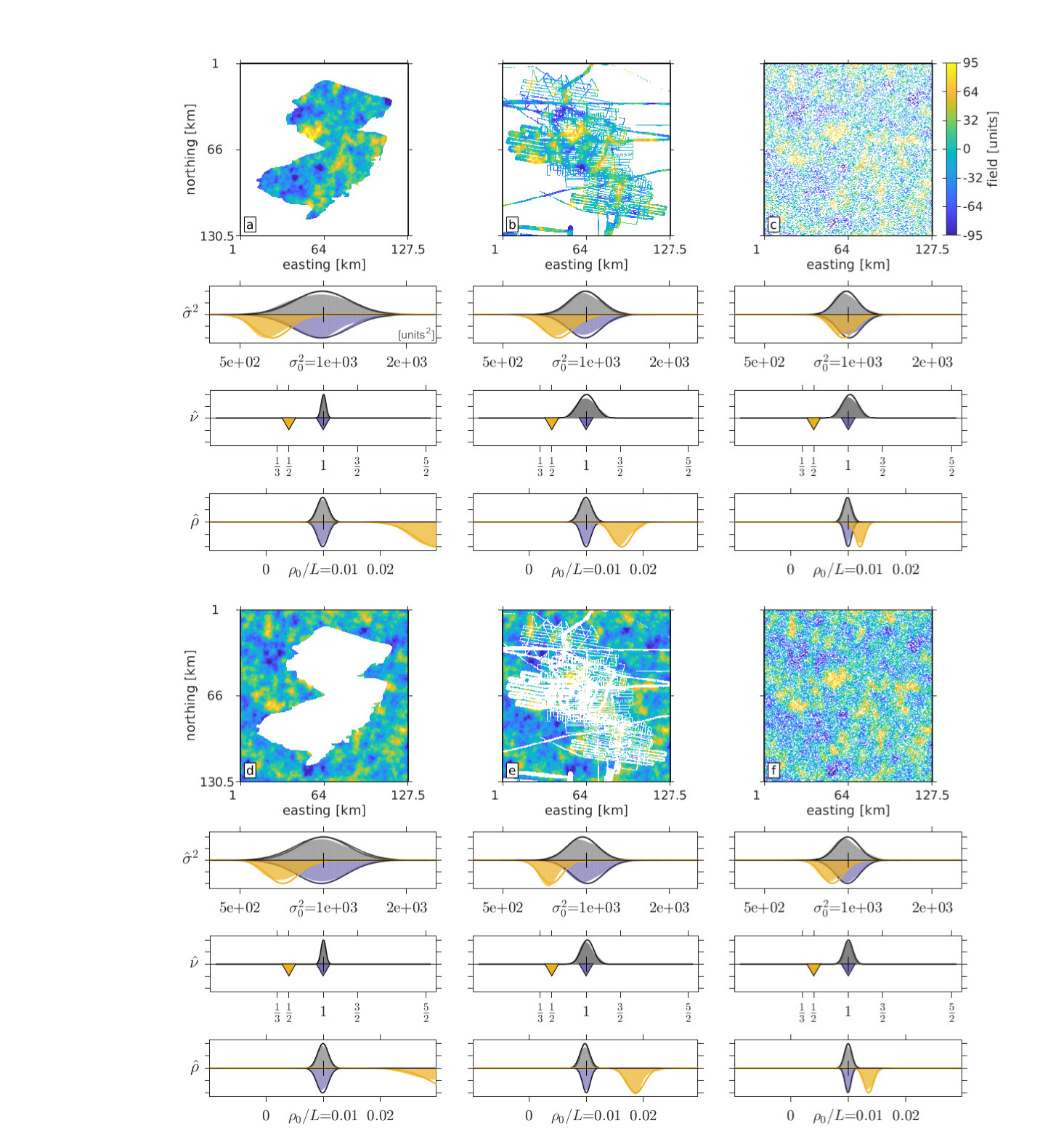}
    \caption{\label{fig:specialsamples}
      Maximum-likelihood estimation statistics for 100 spatial-covariance
      circulant embeddings for irregularly sampled fields (\emph{top}: 33.6\%
      observed, \emph{bottom}: complementary 66.4\% observed) on a
      $261\times255$ grid with spacings $\Delta y=\Delta x=0.5~\km$ for true
      values (black vertical ticks) $\sigma^2_0=1000$~units$^2$, $\nu_0=1.0$, and
      $\rho_0=20$~km. Single realizations for each sampling pattern are shown,
      \emph{a, d}: irregular boundary, \emph{b, e}: structured, \emph{c, f}:
      random deletions. Parameter estimates for the general Mat\'ern model are
      displayed in the top, up-right smoothed ensemble distributions (kdes;
      gray) with curves indicating the normal distribution fit for the sample
      mean and the ensemble observed (gray) and analytically calculated (black)
      standard deviations. Parameter estimates for 2-parameter special case
      covariance models where the smoothness $\nu$ is correctly fixed to the
      true value (purple; ``Whittle'', $\nu=1$) and where it is incorrectly
      fixed (light yellow; ``Exponential'', $\nu=1/2$) are shown by the
      downward, mirrored kdes with normal distribution fits for the sample mean
      and ensemble observed (respective colors) and analytically calculated
      (black; only for the correctly fixed model) standard deviations in
      estimated variance $\st$ and range $\rho$. The fixed value associated with
      the special case $\nu$ is shown by the inverted color-coded triangles.
      Tick labels in estimated smoothness $\nu$ represent additional popular
      special cases of the Mat\'ern class mentioned in
      Table~\ref{tab:MaternSpecial}: the von K\'arm\'an case for $\nu=1/{}3$,
      exponential $\nu=1/{}2$, 2nd-order autoregressive $\nu=3/{}2$, Whittle
      $\nu=1$, and 3rd-order autoregressive $\nu=5/{}2$. Estimated range $\rho$
      is normalized relative to the diagonal length of the grid, $L=[(N_y\Delta
        y)^2+(N_x\Delta x)^2]^{1/2}$.}
\end{figure*}

Fig.~\ref{fig:specialsamples} shows these cases where we recover approximately
Gaussian distributed parameters centered at their true value for both three- and
two-parameter inversions, but only for the latter if we have fixed $\nu$ to its
true value. We demonstrate and discuss the repercussions of incorrectly selecting a
two-parameter special case in terms of parameter bias and uncertainty for the
variance $\sigma^2$ and range $\rho$ estimates.
We consider two overarching synthetic cases for this set of experiments where
data availability comprises a third of the encompassing, regular, rectangular
grid (Fig.~\ref{fig:specialsamples}\emph{a--c}), and its two-thirds
complement (Fig.~\ref{fig:specialsamples}\emph{d--f}). For both cases, we show three sampling
scenarios: a geographically-motivated, bounded patch in the shape of New Jersey
(left column, \emph{a, d}), a sweep of the seafloor along bathymetric
sounding tracks in the South Atlantic \cite[][center column, \emph{b, e}]{GEBCO2024},
and another realization of uniformly random subsampling (right column, \emph{c, f}).

For each of these sampling scenarios, Fig.~\ref{fig:specialsamples} shows three subplots with the
distributions of the estimators. Within the subplots, we display right-side-up,
gray histograms for the three-parameter inversion. These are presented as
they were in Fig.~\ref{fig:mlecurves}, as a kde of the estimator ensemble with
curves displaying Gaussian distributions parameterized by the ensemble mean and
the empirical (dark gray) and analytically calculated (black) standard
deviations. The upside-down histograms are two cases of inverting for two
parameters, one where the smoothness is held fixed $\nu=\nu_0=1$ at the truth
(darker, cool color), and the other held fixed at an incorrect value $\nu=1/2$
(lighter, warm color).
As expected, the three-parameter inversion, regardless of whether the region of
data availability is a New Jersey patch or an Atlantic hatch or a uniform
stipple, yields unbiased (visibly centered at the truth $\btruth$) and (to the eye) Gaussian
distributed ensembles, whose empirical parameter uncertainties are
well predicted by the analytical calculation of eq.~(\ref{eq:magic}).
Fig.~\ref{fig:specialsamples} is our first demonstration of the effect of
sampling patterns on the estimators, which so far we had only illustrated
geometrically (in Figs~\ref{fig:percov} and~\ref{fig:percovnj}) for the specific
regions of interest. As the columns in the top and bottom panels each have the
same number of data points, the visible distinctions in parameter deviations
between the three sets of distributions can be attributed to the geometry of the
acquired samples. These ensembles allow us to interpret the effects of boundary
structure and regularity in the missingness. The numerical results of the
experiments in Fig.~\ref{fig:specialsamples} are summarized in
Table~\ref{tab:specialsamples}, from which we will point out a few key
observations.

The three-parameter inversions (upward-facing histograms in Fig.~\ref{fig:specialsamples}) result in unbiased
estimators regardless of the sampling window~$w\ofx$. The spread of the
ensembles, however, does show a dependence on~$w\ofx$. Beginning with the results
from the top panel of Fig.~\ref{fig:specialsamples}\emph{a--c} for 33.4\% of the
encompassing grid sampled, and progressing from left to right, we will report on
the standard deviations of the three Mat\'ern parameters, noting the empirical
versus analytical values as $(~\cdot~\%~|~\cdot~\%)$. For the process variance~$\st$, 
the relative standard deviations for the irregular boundary case
(Fig.~\ref{fig:specialsamples}\emph{a}) of the New Jersey patch
(16.07\%~$|$~17.09\%), the structured bathymetric tracks
(Fig.~\ref{fig:specialsamples}\emph{b}) from the Atlantic (8.91\%~$|$~9.17\%),
and the uniformly random subsampling (Fig.~\ref{fig:specialsamples}\emph{c}) of
the grid (6.67\%~$|$~6.83\%) reveal a decrease in uncertainty with increased
dispersion in sampling, which we interpret to be due to the subsampling
effectively removing the confounding effects of spatial correlation. We
initially observed this effect in the growing-domain experiments in
\cite{Simons+2026}, their Figs~4 and 5, and we will return to this in
Section~\ref{sec:growing}. In contrast, for the smoothness parameter~$\nu$, we see from
the New Jersey (3.28\%~$|$~4.06\%), Atlantic (13.03\%~$|$~13.62\%), and uniform
(11.10\%~$|$~11.61\%) sampling experiments that more contiguous sampling
patterns increase the precision on $\nu$, revealing greater process
discrimination with denser sampling. We will be studying this effect in
Sections~\ref{sec:infill}--\ref{sec:infillirr} through the perspective of infill asymptotics for regular
and irregular allocations of additional samples. For the correlation length~$\rho$, the
New Jersey (9.27\%~$|$~10.12\%), Atlantic (9.96\%~$|$~10.53\%), and uniform
(6.34\%~$|$~7.37\%) samplings reduce the spread for the uniform random
subsampling pattern only.

The two-parameter inversions (downward-facing, light colored histograms in Fig.~\ref{fig:specialsamples}) display
a pronounced bias in estimated variance $\sigma^2$ and range $\rho$ when the
smoothness $\nu$ is fixed to an incorrect value (in other words, when we chose
the wrong model). For $\st$, the bias (sample mean of the estimators,
$\overline{\st}$ vs truth $\sigma^2_0$) for the New Jersey (695~vs~1000~units$^2$),
Atlantic (833~vs~1000~units$^2$), and uniform (972~vs~1000~units$^2$) sampling reveals a decrease in
bias as the sampling becomes more random because the boundary terms no longer
dominate the estimation procedure. Needless to say, the bias in $\nu$ is
intentional because we (knowingly) fixed it to the incorrect value. For $\rho$,
the bias for the New Jersey (58.1~vs~20.0~km), Atlantic (32.5~vs~20.0~km), and
uniform (24.1~vs~20.0~km) sampling decreases as we further randomize sampling, which
we interpret, too, as being due to reduced boundary effects. When the
smoothness parameter is fixed correctly, however, the estimates of the
two-parameter inversion are unbiased with correct model specification, as they
were for the three-parameter inversion. For $\st$, differences in parameter
uncertainty are negligible between the three-parameter and correctly fixed,
two-parameter model. For $\rho$, we observe a slight decrease in the relative
standard deviation between the three-parameter versus the two-parameter model in
the cases of the New Jersey (9.27\%~vs~8.60\%), Atlantic (9.96\%~vs~8.05\%), and
uniform (6.34\%~vs~5.77\%) sampling patterns.

The complementary sampling scenarios in the bottom panel of
Fig.~\ref{fig:specialsamples}(\emph{d--f}) substantiate all these behaviors, but
we see the effects of having more data available. For example, comparing the
right-side-up histograms for the three-parameter inversions from the top panel
with the bottom panel (33\% grid observations versus 66\%), the parameter
standard deviations obtained from the ensemble are similar or subtly
reduced. The standard deviations of the variance $\st$ for the New Jersey
(16.07\%~vs~14.91\%), the Atlantic (8.91\%~vs~8.88\%), and the uniform 
(6.67\%~vs~7.17\%) sampling scenarios remain relatively constant with increased data
observations. For $\nu$, in the case of the New Jersey (3.28\%~vs~3.46\%),
Atlantic (13.03\%~vs~9.29\%), and uniform (11.10\%~vs~5.79\%) sampling, the
empirical standard deviations remain similar in the bounded case and decrease in
the distributed sampling cases for increased data size. For $\rho$, in the case
of the New Jersey (9.27\%~vs~8.90\%), Atlantic (9.96\%~vs~7.52\%), and uniform
(6.34\%~vs~6.17\%) sampling, as expected, the empirical standard deviations
decrease with increased sample size. These observed changes in the spread of the
ensemble between the three sets of models for increased data size are
small. However, inspecting the upside-down histograms where the smoothness
parameter $\nu$ has been fixed incorrectly, bias grows in the presence of more
data. The relative bias in $\st$ (33\% vs 66\% vs the truth) for the Atlantic
(833 vs 790 vs 1000~units$^2$) and uniform (972 vs 906 vs 1000~units$^2$) sampling follows this
behavior, while in the New Jersey (696 vs 759 vs 1000~units$^2$) case, which has the
greatest bias across the sampling scenarios, we see an improvement with sample
size. The smoothness $\nu$ is again intentionally wrong. For $\rho$, the case of
New Jersey (58.1 vs 68.9 vs 20.0~km), the Atlantic (32.5 vs 37.3 vs 20.0~km), and the
uniform (24.1 vs 27.1 vs 20.0~km) sampling all show larger bias with sample size. We
will further discuss the imprint of the sampling window on estimator behavior in
Section~\ref{sec:design}.

We recommend using the general three-parameter Mat\'ern covariance in modeling
\cite[]{Stein99}. There are no improvements to be made in terms of bias
reduction when selecting the correct simplified form over the general, and
parameter uncertainty decreases only subtly---for the range parameter $\rho$.
In practice, the true value of $\nu$ can rarely, if ever, be assumed to be
known. If one must, a preliminary estimation of the three parameters can be used
to inform which two-parameter special case is most appropriate. Spatial
sampling patterns should be considered in fixing the smoothness, as they tightly
connect to estimator bias. Bias bred by an errant choice is reduced when samples
are randomly dispersed and amplified when contiguous and dense.

As to the effects of having (for reasons pragmatic) or needing (for adjacent
processes) a fixed (e.g., geographic) boundary that separates domains with
potentially different Mat\'ern parameters, additional considerations come into
play, which we explore next.

\begin{figure*}\centering
    \includegraphics[height=0.76\textheight,angle=0,trim=3.23cm 9.7cm 2.2cm 1.5cm,clip]{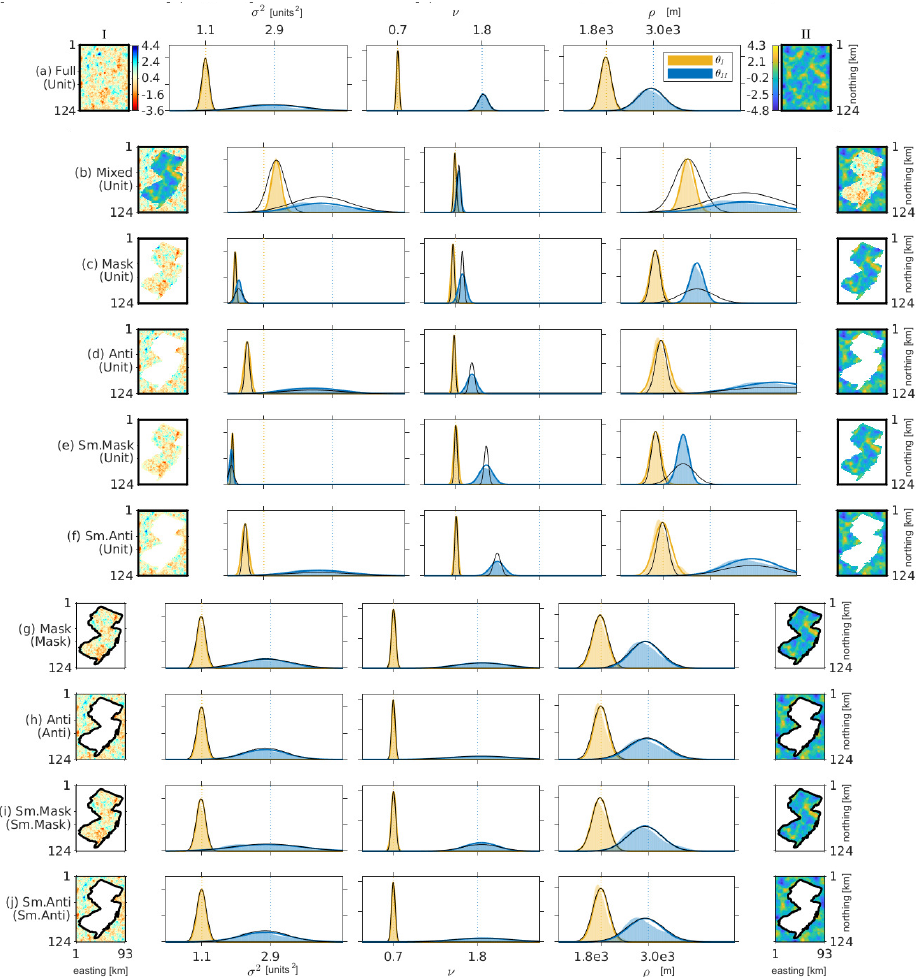}
  \caption{\label{fig:merged}
   Recovery of spatially merged fields with different Mat\'ern covariance
   structures: the role of the analysis window in the debiased Whittle
   likelihood. Synthetic experimental results conducted on a common rectangular
   grid, $\Ny=124$, $\Nx=93$, and $\Dy=\Dx=1$~km, with different processes on an
   exterior (59.83\% of the total area) and interior (40.17\%) subdomain
   delineated by an irregular boundary. The far left and right columns display
   realizations for Mat\'ern models $\btheta_I=[1.15\,\mathrm{units}^2\,\,
     0.65\,\, 1.8~\km]$ (light, yellow-toned colormap) and
   $\btheta_{I\!I}=[2.90\,\mathrm{units}^2\,\, 1.75\,\, 3~\km]$ (deep, blue-toned
   colormap), respectively, for each of the 10 experimental scenarios. In each
   row, the observation window and analysis taper are varied, which we denote
   along the rows with labeling and the display for a single realization of
   fields $I$ (left) and $I\!I$ (right). Order of labeling: experiment (a--j),
   observation window of the field (\emph{Full}, \emph{Mixed} at the boundary,
   \emph{Mask} of interior, \emph{Anti}-mask of exterior, \emph{Sm(ooth) Mask},
   \emph{Sm(ooth)~Anti}-mask), taper used in analysis for the blurred likelihood
   (the $w\ofx$ in eq.~(\ref{eq:dwl}); \emph{Unit}, \emph{Mask},
   \emph{Anti}-mask, \emph{Sm.~Mask}, \emph{Sm.~Anti}-mask). Estimate ensembles
   are displayed as kernel density estimates for $\hat{\theta}_I$ (left, yellow) and
   $\hat{\theta}_{I\!I}$ (right, blue), overlain with normal fits in matching
   colors, and with, in black, analytically predicted parameter standard
   deviations (eq.~\ref{eq:magic}), evaluated at the sample means, for the
   analysis taper used in the experiment, which is plotted by the thick black
   line on the spatial renderings. The colored empirical and black analytically
   predicted distributions are unbiased and agree when the correct taper is used
   in analysis, that is, for experiments (a) and (g--j), offset horizontally for
   distinction.}
\end{figure*}

\section{S~E~P~A~R~A~B~I~L~I~T~Y{\hsps}O~F{\hsps}S~P~A~T~I~A~L~L~Y{\hsps}M~E~R~G~E~D{\hsps}P~R~O~C~E~S~S~E~S}\label{sec:separ}

We explore a series of numerical experiments to demonstrate the ramifications of
analyzing multiple Mat\'ern processes that are spatially merged along an
irregular boundary where we are either ignorant or aware of that boundary and
analyze the observation as a single or multiple fields, respectively. In the
case of the former, the random field departs from our model assumption of local
stationarity.

Fig.~\ref{fig:bath2synth} showed a first such example. In it, we showed
15-arcsecond resolution seafloor bathymetry comprised of direct soundings acquired
by single and multibeam sonars along ship sampling tracks filling 38\% of the
bounding grid with the complementary 62\% predicted from bathymetry data
augmented by indirectly obtained sea surface satellite altimetry. We fit and
removed a plane over the entire domain, split the data into its non-overlapping
components, demeaned them, and estimated the maximum-likelihood Mat\'ern
covariance parameters from the separate and merged datasets. The synthetic
realizations simulated on the partial and complete grids were shown with the
removed trends added back in. The parameter estimates for $(\st, \nu, \rho)$ are $(38871,1.46,2.841)$ for
the merged dataset (``\textsf{mer}''), which is effectively nonstationary, is
not a simple average of the estimates made on the constituent components
$(45452, 0.96, 4.322)$ for ``\textsf{dir}'', and $(47764,1.45,4.881)$ for
``\textsf{ind}''. The apparently concentrated uncertainty of the merged estimate
$(\pm 6657,\pm 0.02,\pm 0.208)$ versus $(\pm 10852,\pm 0.18,\pm 0.895)$ and
$(\pm 7857,\pm 0.24,\pm 0.907)$, is illustrative of how estimable such a process
is within its observation grid rather than how well it corresponds to the data,
which, other than delivering the parameters, have no part in shaping the
uncertainty.

For the case of bounded domains with an interior and exterior portion, the
experiments in this section address three queries: to ascertain (1)~how the
debiased Whittle likelihood (eq.~\ref{eq:dwl}) accounts for spatial sampling
patterns and produces an unbiased estimator $\langle\hbtheta\rangle=\btruth$
whose covariance can be exactly predicted via eq.~(\ref{eq:magic}), and (2)~how
the appearance of multiple processes within a single window of observation
affects the estimator when their distinctions are not indicated by the taper
$w\ofx$ employed in the analysis, and (3)~how the acknowledgment of the spatial
extent of each of the merged processes by the sampling taper allows for the
robust estimation of the processes individually.

In Fig.~\ref{fig:merged} we consider a rectangular grid with an interior region
(the ``\textit{mask}'') that comprises 40.17\% of the total sample size, and an
exterior (the ``\textit{anti-mask}'') possessing the complementary 59.83\%. The
interior is separated from the exterior by an irregular, sock-like, closed
boundary. Fig.~\ref{fig:merged} shows a suite of 10 experiments for two
processes that we organized by permuting the parametric process (two sets of
Mat\'ern parameters, $I$ and $I\!I$), the extent of the process within the
rectangular observation window, and whether the sampling window applied in the
analysis $w\ofx$, smoothed or not, has acknowledged the region within it. The
set covers the following 10 scenarios: (a)~a single process observed and
analyzed on a rectangular grid; (b)~two processes merged along an irregular interior boundary but
analyzed on the rectangular grid; a single process in the interior without (c)
and with (e) smoothing along the boundary analyzed on the rectangular grid; a
single process in the exterior without (d) and with (f) smoothing along the
boundary analyzed on the rectangular grid. In all those cases (a--f), the
analysis window $w\ofx$ is a rectangle, rendered as a thick black line. The
next four rows show a single process in the interior without (g) and with (i)
smoothing along the boundary analyzed in the interior only and in
the exterior without (h) and with (j) smoothing along the boundary analyzed in
the exterior only. In those cases (g--j) the analysis window $w\ofx$ is the
boundary between the processes, rendered as a thick black line. If applied, the
smoothing process used along the boundary is a cosine squared taper applied along
each of the segments of the boundary, iteratively convolved to smooth the
corners.

We simulate 200 realizations of all permutations, estimate the corresponding
Mat\'ern parameters, and report on the empirical and analytical distribution
(calculated from eq.~\ref{eq:magic} evaluated at the sample mean) of the
ensemble in the inner panels of Fig.~\ref{fig:merged}. In
Table~\ref{tab:merged}, we summarize, for each experiment and for both processes the
taper used, the sample means, and the standard deviations (relative,
analytical and empirical) of the variance, smoothness, and range of the process. In
addition, Table~\ref{tab:merged} lists the correlations between the Mat\'ern
parameter estimates, in per cent, first as predicted and then as empirically
observed.

In Fig.~\ref{fig:merged} and Table~\ref{tab:merged}, we begin with the base
case~(a) of two rectangular, stationary sampled random fields. The first field,
$I$, possesses the Mat\'ern parameterization
$\btheta_I=[1.15~\mathrm{units}^2\quad0.65\quad1800~\km]$, and the second, $I\!I$, with
$\btheta_{I\!I}=[2.90~\mathrm{units}^2\quad1.75\quad3000~\km]$, is larger in magnitude for all
three parameters. For both field $I$ and $I\!I$, we simulate realizations via
the circulant embedding of the Mat\'ern spatial covariance for $\btheta_{I}$ and
$\btheta_{I\!I}$, both on a common grid size of $124~\km \times 93$~km with even
$1$~km spacing in each direction.

Case~(a) in the top left of Fig.~\ref{fig:merged} displays a single realization
for process $I$ and in the top right for process $I\!I$. For each simulation,
we apply our maximum-likelihood estimation strategy to determine estimators
$\hbtheta_I$ and $\hbtheta_{I\!I}$, taking the sampling window for the analysis,
$w\ofx$ in eqs~(\ref{eq:dwl}) and~(\ref{eq:blurredspec}) to be a unit taper
(i.e., fully observed). The ensembles of estimates for the three Mat\'ern
parameters are displayed in the center columns as smooth distributions, with
yellow kdes distinguishing the ensemble of $\hbtheta_I$, yellow overlying curves
marking the normal distribution fit to the $\hbtheta_I$ sample mean and sample
standard deviation, and black overlying curves showing the normal distribution
fit to the sample mean and analytically calculated parameter standard deviation
(eq.~\ref{eq:magic}) evaluated at the ensemble sample mean for the unit taper
sampling window on the rectangular grid. Similarly, the blue kdes, matching
blue curves, and overlying black curves with the shared centers correspond to
the distribution of the $\hbtheta_{I\!I}$ and its empirical and analytically
calculated standard deviations. The true parameter values for $\btheta_{I}$ and
$\btheta_{I\!I}$ are labeled on the horizontal axis, displayed as dotted
vertical lines in yellow and blue, respectively, with $\btheta_{I\!I}$ falling
to the right as its true parameter values were selected to always be greater
than $\btheta_{I}$. The experimental outcomes of (a) include two ensembles
centered on their truths with predictable parameter uncertainties that
correspond to the observed ensembles. The spread of the ensembles scales with
the magnitude of the parameter value: the larger the $\theta_i$,
the larger the $\var\{\theta_i\}$. We retain the setup of the fields and
organization of results throughout the experiments that follow.

Case~(b) considers the two processes occurring within the same unit window,
either with process $I\!I$ (coded by the deeper color, blue-toned colormap
within the realizations shown in Fig.~\ref{fig:merged}) within the interior of
process $I$ (shown in the lighter, yellow-toned colormap of
Fig.~\ref{fig:merged}) in the left hand-side ``$I$'' column, or the opposite
(process $I$ within the interior of process $I\!I$) in the right hand-side ``$I\!I$''
column. In this example, we are (knowingly incorrectly) attempting to estimate
a stationary Mat\'ern process from a non-stationary field. The estimate
ensembles and normal approximations reveal estimates that are intermediate to
$I$ and $I\!I$ in comparison to case~(a). Additionally, our analytical prediction
of uncertainty evaluated at the ensemble sample mean for a unit taper window
(black curve) reveals a larger anticipated spread than we observe from the
ensemble.

Cases~(c) through (f) consider a single process confined to the same bounded
area (denoted \emph{mask} in Fig.~\ref{fig:merged} and Table~\ref{tab:merged})
or its complement (\emph{anti-mask}) observed and analyzed with the \emph{unit}
taper, with or without smoothing (\emph{Sm.}) applied at the edges of the
field. These experiments are deemed unsuccessful due to the bias imparted on the
ensemble of estimates and the discrepancy between the observed and predicted
parameter uncertainty.

Cases~(g) through (j) are formatted as (c--f), however, they are analyzed
according to the proper observation window, either tapered by the \emph{mask} or
\emph{anti-mask} with possible smoothing at the edges. In all of these
experimental cases, the estimates are unbiased and the empirical and analytical
parameter uncertainties are well matched, a testament to the debiased Whittle
likelihood of eq.~(\ref{eq:dwl}). Smoothing the outer edges of the field does
not appear to improve the ensemble mean in these cases, though we do see a
subtle increase in parameter uncertainty when smoothing is applied.

The outcome of our suite of experiments reveals what one might expect for our
first stated query~(1): estimation of single processes whose sampling pattern is
taken as the analysis tapering window ($w\ofx$ in eq.~\ref{eq:dwl}; experiments
a, g--j) results in an unbiased estimator with estimator uncertainty that is
inversely proportional to the degrees of freedom. Deviations in $w\ofx$
introduce increased bias and uncertainty, whose magnitudes depend on the field
observation window and the magnitude of the process parameters. The 68.27\%
confidence intervals for the variance $\st$ and smoothness $\nu$ do not overlap
for $I$ and $I\!I$ in the experiments conducted on the homogeneous subdomains, while the
distributions of the range $\rho$ parameters display overlap due in part to the
increased parameter variance from the reduced degrees of freedom in the
\emph{mask} and \emph{anti-mask} tapered fields. In the experiments for
\emph{(anti-)masked} single processes analyzed with a \emph{unit} taper (c--f),
overlap is only a problem for $\st$ in (c) and (e) for the masked field with and
without smoothing applied to the boundary of the mask which is subject to large,
negative bias. Bias is observed for all parameters in (c--f), with a greater
effect displayed by the larger-magnitude parameter model $I\!I$. Smoothing
along the boundaries yields similar ensemble results to analyzing fields with a
hard indicator boundary.

The resolution to our query~(2) is highlighted by the outcome of the
mixed-process experiment (b), which is an example of multiple, spatially
separable processes being observed and analyzed within the same window. This is a
realistic scenario for geophysical sampled random fields where multiple,
spatially confined, physical processes have imprinted neighboring regions. The
results of this experiment for process~$I\!I$ within the interior of process~$I$
in light yellow (left column of Fig.~\ref{fig:merged}) and the inverse in blue
(right), show ensemble estimates that are intermediate to model $I$ and $I\!I$
in $\st$, estimates that favor the smaller $\nu$ of $I$, and estimates that
favor the larger value $\rho$ of $I\!I$, with overlap in the confidence
intervals for both scenarios.
In terms of model identifiability, \cite{Guillaumin+2017,Guillaumin+2022}
discuss how the interactions between the geometry of the spectral density and
the geometry of the sampling process can be understood via what they term the
\emph{significant correlation contribution}.

What we wish to emphasize in our last query~(3) for merged processes is that
unbiased estimates with well predicted parameter covariances can be determined
from a single dataset if it is possible to spatially delineate the processes
within the domain and analyze them separately for an observation window defined
by the detected boundary (as in cases g through j). This strategy will allow
practitioners to study the covariance structure of spatial geophysical datasets
that develop under the action of multiple, spatially jigsawed phenomena, if their heterogeneity
can be demarcated first. We utilize this strategy of demarcation for our real-data
applications where we bring in outside (e.g., geological) information. In
the absence of such guiding information, our framework and analysis will help
evaluate the success of alternative data-driven (e.g., segmentation,
classification, and clustering) approaches.

\section{G~E~O~S~C~I~E~N~C~E{\hsps}A~P~P~L~I~C~A~T~I~O~N~S{\hspst}:{\hsps}R~E~A~L{\hsps}D~A~T~A}
\label{sec:modeltest}

Real, sampled spatial data are unlikely to be inherently and unquestionably
Gaussian, stationary, and isotropic. However, from our synthetic studies, we
are heedful of the bias that observations of multiple merged,
nonstationary, processes may impart on model estimates
(Section~\ref{sec:separ}) and of the bias and increased estimator uncertainty (namely
of the smoothness parameter, $\nu$) for irregularly, incompletely sampled fields
(in comparison to those on a regular grid; Section~\ref{sec:specialsamples}). 
In our previous work \cite[Section~5 of][]{Simons+2026}, we encountered 
the effect of departures from these assumptions on the residuals for
covariance estimates of real-data examples sampled on a complete regular grid.

\begin{figure*}\centering
  \includegraphics[width=\textwidth,angle=-0,trim=0cm 0cm 0cm 0cm,clip]{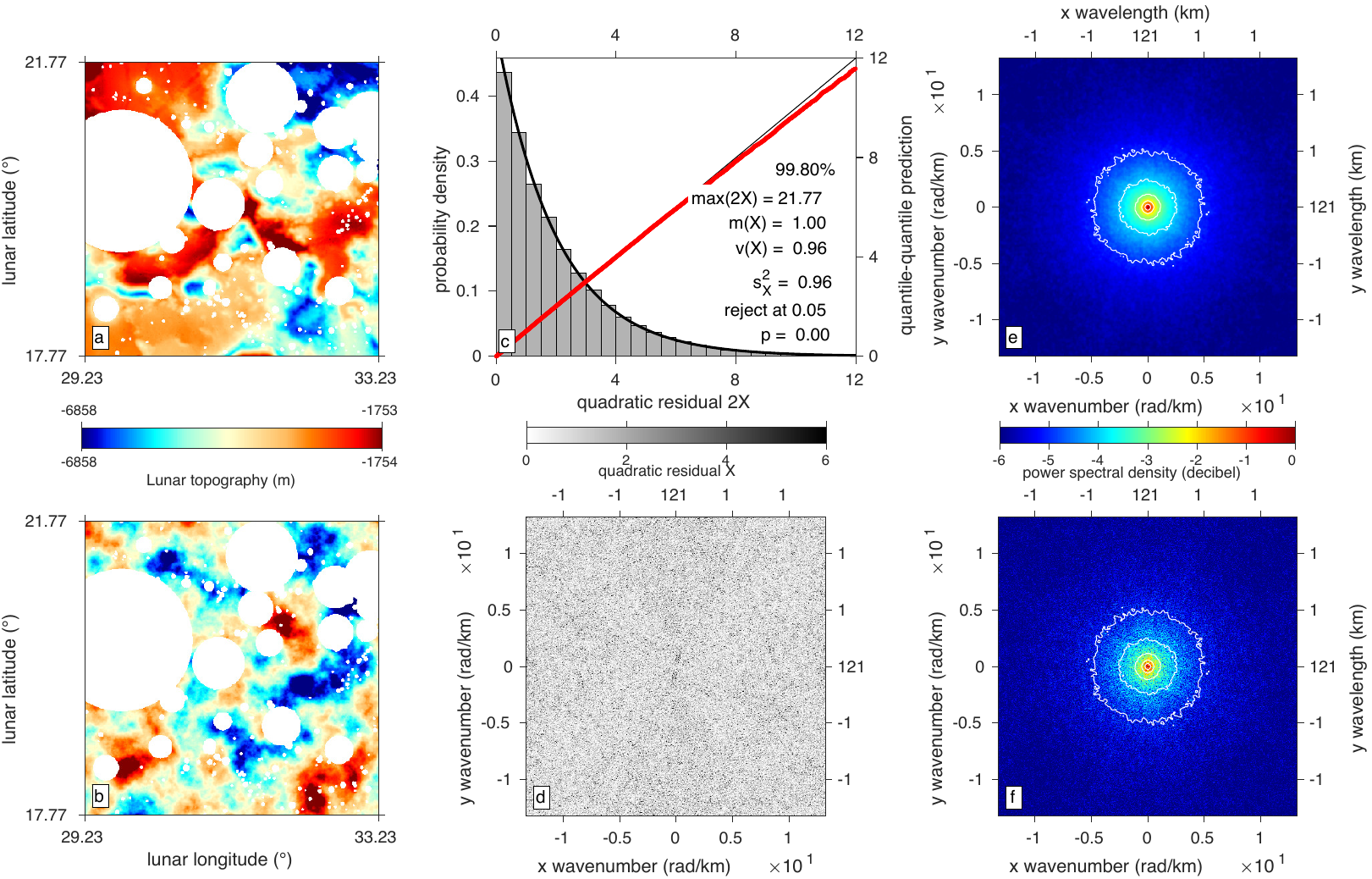}
  \caption{\label{fig:sldem} Maximum-likelihood analysis for the Mat\'ern
    covariance structure of lunar topography \cite[]{Barker+2016} without
    craters \cite[]{Robbins2018} near Taurus-Littrow Valley.  Shown are the
    observed data (\textit{a}) and a synthetic generated through the circulant
    embedding of the Mat\'ern spatial covariance parameterized by our estimate
    (\textit{b}; Table~\ref{tab:applications}), a histogram and a
    quantile-quantile (Q-Q) plot of the quadratic residual $2X\ofst$
    (\textit{c}), with $X\ofst\ofk$ being rendered and inspected for patterns in
    reciprocal-space (\textit{d}; Section~\ref{sec:res}). Also shown are the
    expected (\textit{e}) and observed (\textit{f}) periodograms,
    $\bar{\mcS}\ofst\ofk$ and $|H\ofk|^2$, respectively, with contour lines at
    each integer decibel of the former overlain on the latter.  This example
    demonstrates the analysis of a spatially contiguous process \textit{(a)}
    that is geomorphically distinct from the union of its bounded impact
    craters.  The synthetic simulation \textit{(b)} is visually comparable to
    the random features within the observed data.  The model for the
    non-cratered lunar surface fits very well \textit{(c)}, though not quite
    passing the rigorous statistical test (eq.~\ref{eq:stx}), with no evident
    structure within the model residuals \textit{(d)}.  The modeled spectral
    density \textit{(e)} fits the data periodogram \textit{(f)} very well.  }
\end{figure*}

\begin{figure*}\centering
  \includegraphics[width=\textwidth,angle=-0,trim=0cm 0cm 0cm 0cm,clip]{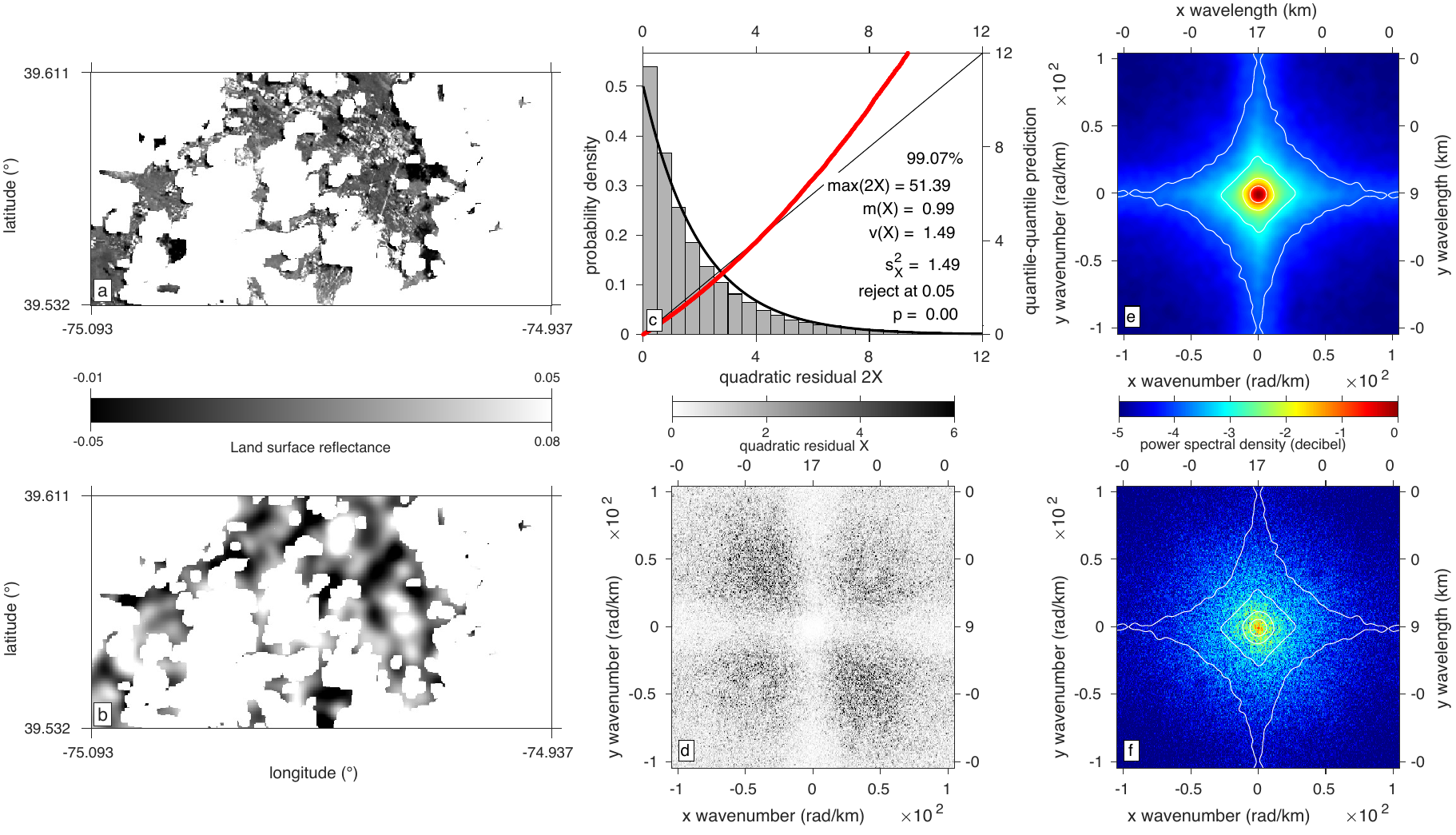}
  \caption{\label{fig:landsatland} Maximum-likelihood analysis for the Mat\'ern
    covariance structure of LANDSAT 8 Collection 2, Level-2 surface reflectance
    data for land observations. Red, green, and blue spectral bands converted to
    a grayscale composite with pixels labeled as ``land'' in the accompanying
    quality analysis file selected \textit{(a)}, which comprise 29.71\% of the
    encompassing rectangular grid; (diffuse) clouds, cloud shadows, and minor
    water bodies make up the remaining masked pixels.  The estimated covariance
    structure has a large smoothness (Table~\ref{tab:applications}).  The
    synthetic \textit{(b)} shares some resemblance to the data.  The $2X\st\ofk$
    model residuals \textit{(c)} for the land surface reflectance data follows
    the $\chi_2^2$ distribution well through the bulk of the distribution, with
    higher than expected small residuals and a long tail, as shown in the
    departures of the Q-Q-plot.  The residual plot in reciprocal-space
    \textit{(d)} possesses a windowpane-like structure that our model is unable
    to capture.  The modeled spectral density \textit{(e)} captures the general
    structure of the noisy data periodogram \textit{(f)}.  Layout and annotation
    are as in the caption of Fig.~\ref{fig:sldem}.}
\end{figure*}

\begin{figure*}\centering
  \includegraphics[width=\textwidth,angle=-0,trim=0cm 0cm 0cm 0cm,clip]{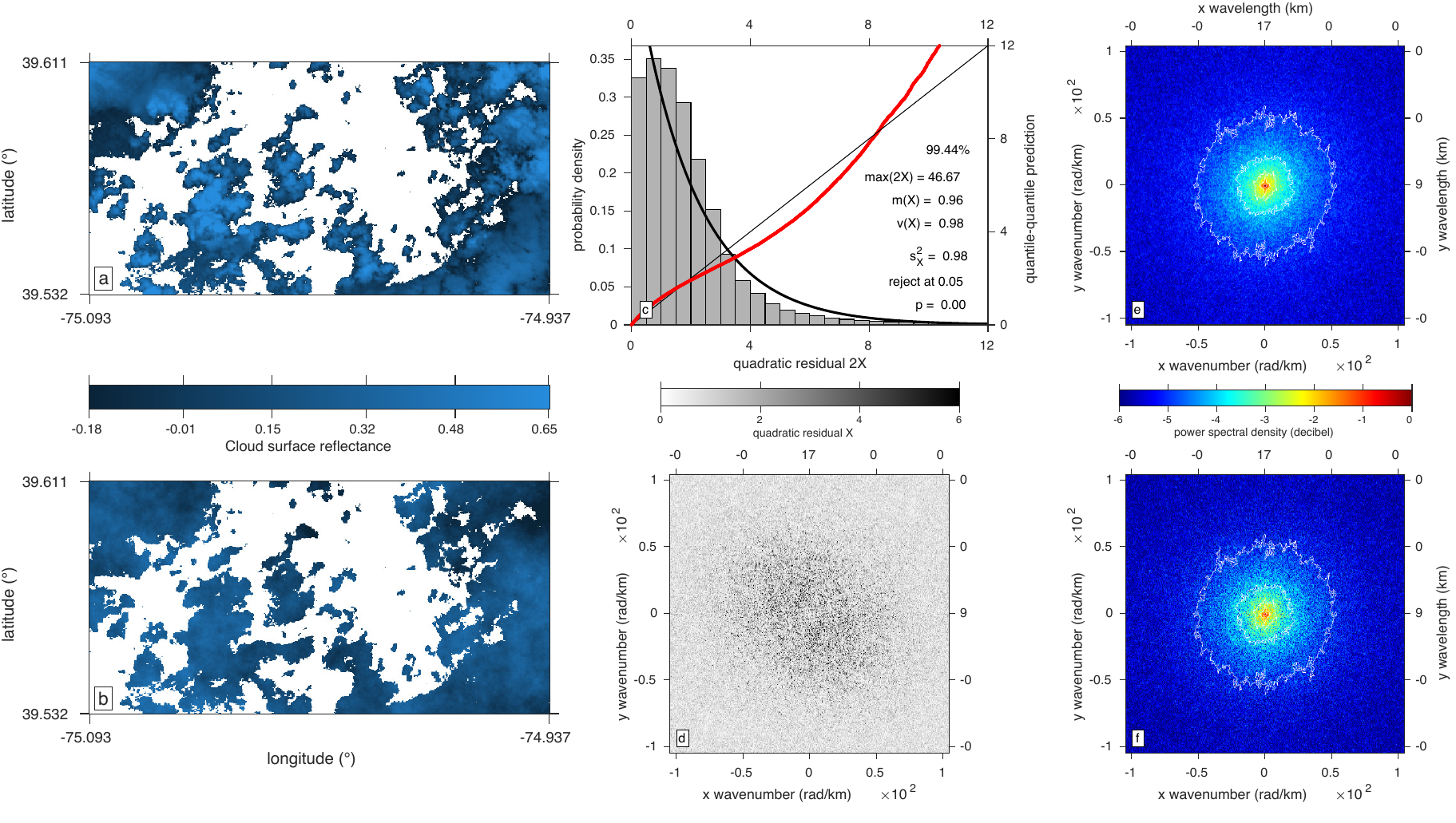}
  \caption{\label{fig:landsatcloud} Maximum-likelihood analysis for the Mat\'ern
    covariance structure of LANDSAT~8 Collection~2, Level-2 surface reflectance
    data for cloud observations \textit{(a)}, complementary to the dataset in
    Fig.~\ref{fig:landsatland}, comprising 54.23\% of the encompassing grid. The
    spatial synthetic \textit{(b)} generated from the model is appropriate by
    eye.  The model for the observed cloud data (Table~\ref{tab:applications})
    does not fit well in terms of the histogram of the residuals \textit{(c)} as
    they do not conform to the theoretical $\chi_2^2$ distribution of
    eq.~(\ref{eq:chisq}) as would be anticipated for the ratio of uncorrelated,
    Gaussian random variables.  The structure of the residuals in reciprocal-space \textit{(d)} appears isotropic, albeit with higher than expected
    structure remaining for low wavenumbers.  The modeled spectral density
    \textit{(e)} fits the data periodogram \textit{(f)}.  Layout and annotation
    are as in the caption of Fig.~\ref{fig:sldem}.}
\end{figure*}

We apply our maximum-likelihood estimation strategy to four geophysical data
examples that exhibit irregularity in their sampling to characterize their
covariance structures with quantified uncertainties and a rigorous assessment of
the model residuals. As in the synthetic example in Fig.~\ref{fig:res} and the
real-data examples of \cite{Simons+2026}, we estimate the covariance structure
of a geophysical field and assess the model by a qualitative assessment of
whether a synthetic realization simulated from the estimate corresponds to the
behavior of the original data and whether the model residuals $X\ofst\ofk$
(eq.~\ref{eq:chisq}) retain any structures from the periodogram $|H(\kb)|^2$
(eq.~\ref{eq:dwl}) that have been left unresolved by the modeled blurred
spectral density $\bar{S}\ofst\ofk$ (eq.~\ref{eq:blurredspec}). Quantitatively,
we evaluate the goodness-of-fit of our model and how well it scales with the
theoretical distribution $\chi^2_2$ (eq.~\ref{eq:chisq}) by way of a Q-Q plot
and the test statistic $s_X^2$ (eq.~\ref{eq:stx}). Our analysis results for
these four data applications are displayed in
Figs~\ref{fig:sldem}--\ref{fig:landsatcloud} and Table~\ref{tab:applications}.
For the partially sampled datasets we consider here, we create an observation
window that masks either indirect sources of data, secondary processes, or
observed phenomena that we seek to omit, thus performing our analysis on the
data type or feature of interest only, with exact blurring of the likelihood and
covariance calculation. We preprocess the data by detrending up to a
second-order polynomial fit when a locally varying mean is evident in the
original data. A 5\%-cosine window taper is applied to the encompassing
rectangular perimeter of the observation window, only smoothing the outermost
samples to reduce edge effects that appear in the spectral residuals and which
obfuscate the remaining signal we wish to interpret.

In the first example (Fig.~\ref{fig:sldem}), we demonstrate a procedure for
assessing spatial processes where an irregular boundary can be drawn to
delineate features with geomorphic traits or histories that are distinct from
their surroundings, for which we create a mask for the observation window of the
process under analysis.  Our data source is SLDEM2015, a lunar digital elevation
model (DEM) developed by \cite{Barker+2016} from observations gathered by the
Lunar Orbiter Laser Altimeter (LOLA) onboard NASA's Lunar Reconnaissance Orbiter
(LRO) spacecraft that were corrected by co-registered stereo-derived DEMs from
the SELENE Terrain Camera.  We isolate a rectangular patch of lunar topography
encompassing Taurus-Littrow Valley (centered at approximately 20$^\circ$N,
31$^\circ$E).  One application of parametric covariance estimation is as a
metric for surface roughness. For this dataset, the historic landing site of
Apollo~17, our analysis might provide a useful and interpretable surface
roughness metric for comparative site planning in autonomous navigation
\cite[e.g., for landers or rovers;][]{Thrun+2006,Haselich+2013,Pardo-Igúzquiza+2022}.  To exclude
deterministic geologic features from their homogeneous background, we create a
mask that fits cataloged lunar impact craters with diameters greater than
$1~\km$ \cite[]{Robbins2018} as circles. The non-cratered lunar surface
comprises 62.8\% of the bounding rectangle. The grid spacing is 0.24~km.  Our
analysis of the Mat\'ern model for the non-cratered topography reveals a near
$\chi^2_2$ distribution of the spectral residuals $X\ofst\ofk$, a strong fit to
eq.~(\ref{eq:stx}), though not quite sufficient to fail to reject the hypothesis
test. We see no structure within the spectral map of the residuals. In the
visual comparison between the DEM and the simulation, subtle along-latitude
streaking appears in the DEM, most likely arising from a data artifact, which our
model for random processes \emph{is not} meant to replicate.

In our next two companion examples (Figs.~\ref{fig:landsatland}
and~\ref{fig:landsatcloud}), we select a dataset that exhibits missingness
distributed throughout the observation window as pixel-labeled `land' and
`cloud' features corresponding to surface reflectance data captured by LANDSAT 8
on July 20, 2025 provided as a Collection 2, Level-2 data product for the tile
aligned on path 014, row 032 with 30 m resolution \cite[]{LANDSAT8}. From the
red, green, and blue surface reflectance bands, we convert the digital recorded
values to physical measurements of surface reflectance ratios, masking pixels
that exceed the specified valid ranges \cite[]{LANDSATHandbook}, and converting
the three bands into a grayscale composite for univariate analysis. We select a
region within the tile between 39.53$^\circ$N--39.61$^\circ$N and
74.94$^\circ$W--75.09$^\circ$W where 54\% of the pixels are labeled in the
quality assessment band file as cloud coverage, 30\% as land, and of the pixels
we do not consider within the tile, 18\% are labeled as cloud shadows
(possessing considerable overlap with other labels), 16\% dilated clouds, and,
negligibly, 0.02\% water bodies. We estimate the covariance structure for the
land-tagged data (Fig.~\ref{fig:landsatland}) and the cloud-tagged data
(Fig.~\ref{fig:landsatcloud}) as separate experiments. For the land surface
reflectance data, the estimated smoothness is notably large, as are its
parameter uncertainty bounds. Due to the (expected) strong, negative trade-off
between the smoothness and range parameters in the Mat\'ern model (see the
correlations listed in Table~\ref{tab:applications}) it is possible that, at
this resolution, we are unable to estimate the covariance structure of this
particular dataset well as the true range parameter is smaller than the grid
spacing in our observation window. We include this example to show how a
``too-smooth'' physical dataset may fall outside of our sampling grid
constraints, which we will discuss in Section~\ref{sec:asymp}. The cloud dataset
fits its estimated Mat\'ern model better, but not perfectly. We observe
isotropic structure within the reciprocal-space plot of the residuals that is
not captured by the model. The histogram of the residuals does not sufficiently
conform to a $\chi_2^2$ distribution. This deviation of the data from their most
likely estimated model and the residuals from their expectation could be indicative
of non-Gaussianity.

\begin{figure*}\centering
  \includegraphics[width=\textwidth,angle=-0,trim=0cm 0cm 0cm 0cm,clip]{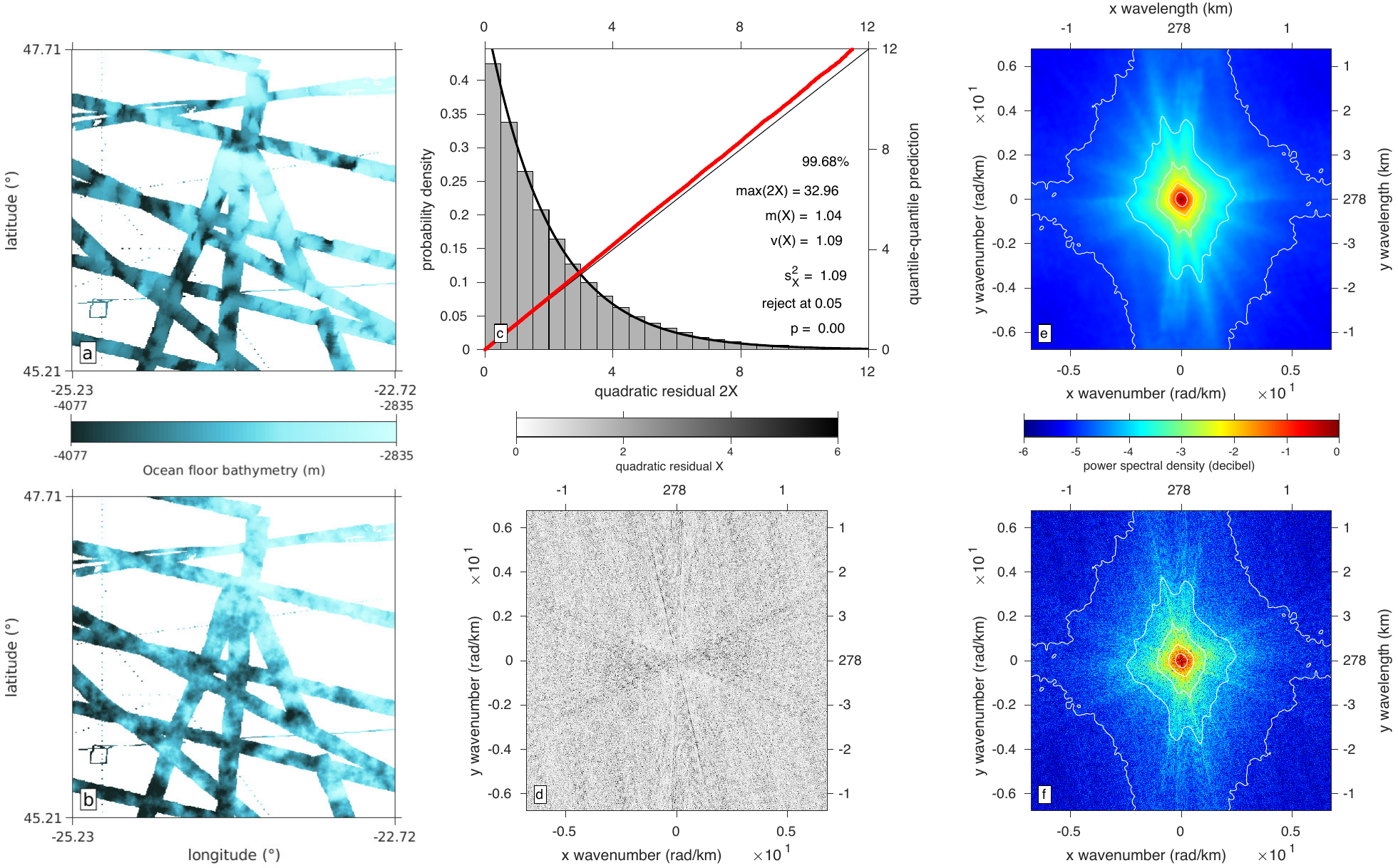}
  \caption{\label{fig:gebco} Maximum-likelihood analysis for the Mat\'ern
    covariance structure of partially observed Atlantic seafloor
    bathymetry. This dataset provides an example of structured geophysical data
    collected along sampling tracks by shipboard direct single-beam and multibeam
    measurements \cite[\textit{a}; ][]{GEBCO2024}.  The synthetic \textit{(b)}
    is very similar to the data.  The spectral domain residual histogram and
    Q-Q-plot \textit{(c)} show the model to be a good fit, not quite passing the
    rigorous statistical test (eq.~\ref{eq:stx}), with subtle, fine-scale
    radiating structure retained within the reciprocal-space residual plot
    \textit{(d)}, which we interpret as multi-directional geometric anisotropy
    within the data.  The modeled spectral density \textit{(e)} overlays the
    data periodogram \textit{(f)} very well.  Layout and annotation are as in
    the caption of Fig.~\ref{fig:sldem}. }
\end{figure*}

Our final example (Fig.~\ref{fig:gebco}) is of the directly sampled, structured
tracks with 0.46~km resolution of ocean floor bathymetry \cite[]{GEBCO2024}
centered at $46.45^\circ$N, $23.97^\circ$W in the Atlantic from
Fig.~\ref{fig:bath2synth}.  The directly measured data are collected shipboard
as predominantly multibeam (98.9\%) with few single beam (1.1\%) soundings that
collectively make up 38.0\% of the enclosing rectangular window. As we stated in
the bathymetric full patch example of \cite{Simons+2026}, we do not anticipate
that ocean bathymetry in general will exhibit isotropic spatial behavior due to
the anisotropy of its plate tectonic formative processes. For this patch with
only direct measurements considered, we do however find that the isotropic model
fits very well. The residuals retain a minor, radial pattern that we take to
represent multi-directional, geometric anisotropy, in contrast to the more
diffuse directional anisotropy that we observed in Fig.~16 of
\cite{Simons+2026}.

\begin{table}
    \centering
    \begin{tabular}{rcrrrrrrrrrrrr}
       Example                    & Fig.                  &  $K$  & \% Obs  &  $\st/s^2$ & $\pm$\% & $\nu$ & $\pm$\% & $\rho$ & $\pi\rho/r$ & $\pm$\% & $\{\st,\nu\}$ & $\{\st,\rho\}$ & $\{\nu,\rho\}$ \\\hline
       Land surface reflectance   & \ref{fig:landsatland}  & 50061 & 29.7   &  6.7290    & 12.3    & 30.53 & 533.7   & 0.18           & 0.0296      &  12.9  & -29.5        & 45.5          & -93.0         \\
       Cloud surface reflectance  & \ref{fig:landsatcloud} & 91371 & 54.2    &  2.1084    & 109.9   &  0.46 & 41.6    & 7.39           & 1.1969      & 190.3  & -10.9        & 74.5          & -72.6         \\
      Ocean bathymetry           & \ref{fig:gebco}        & 137251 & 38.0  &  2.3073  &  8.99   &  0.96   &   7.19        &   4.32    &  0.0110  &    7.89      &     21.1      &   49.0   &  -69.0  \\
      Lunar topography           & \ref{fig:sldem}        & 165255 & 62.8   &   1.0727   &  21.4   & 0.98  &   25.9   &     5.12       &   0.0936   &  38.8  &    1.89     &    32.66       &   -91.54        
       \end{tabular}
    \caption{\label{tab:applications}
    Results from experiments with irregularly sampled geophysical data, sorted
    by degrees of freedom of the analysis for the tapered patch, $K=\sum w\ofx$.
    The grid percentage observed is reported relative to the area of the
    encompassing regular grid as $K/(\Ny\Nx)$. The estimate for the variance
    $\st$ is quoted as a fraction of the sample variance of the observed
    dataset, $s^2$. In the presence of significant range, $s^2$ is small
    relative to $\st$. We present the estimate of the range in units of field
    distance [km], see Fig.~\ref{fig:specialsdfcov}, and as
    a fraction of $r$, the enclosing grid diagonal, scaled by $\pi$. For all three parameters, the
    one-standard-deviation estimation uncertainty is listed to the nearest per
    cent of the parameter. The final three columns contain the correlation
    between estimates, in per cent.}
\end{table}

\section{A~S~Y~M~P~T~O~T~I~C~S{\hsps}O~F{\hsps}S~P~A~T~I~A~L{\hsps}S~A~M~P~L~I~N~G}\label{sec:asymp}

\begin{figure*}
 \centering
  \includegraphics[width=0.95\textwidth,trim=1.6cm 20.5cm 2cm 1.6cm,clip]{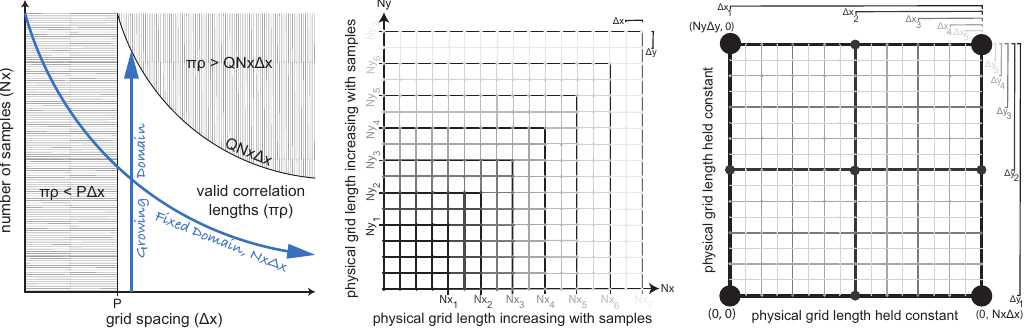}
  \caption{\label{fig:asymcartoon}
  Cartoons illustrating sampling grid constraints (\textit{left}, simplified to
  1-D) and increasing sample size within a growing-domain (\textit{center}) and
  a fixed-domain (\textit{right}) scheme. A Mat\'ern model is estimable for a
  sufficiently large grid where multiple observations of the correlation length
  ($\pi\rho$) may be made and the spacing is small in relation. Growing-domain
  asymptotics assume a fixed grid spacing ($\Dy$, $\Dx$) where samples are
  increased by extending the physical grid length ($\Ny\Dy\times\Nx\Dx$). In
  contrast, the fixed-domain or infill approach increases samples through
  densifying sampling on a fixed physical grid length by reducing the spacing
  between existing samples on a fixed physical grid size
  ($\Ny\Dy\times\Nx\Dx$).}
\end{figure*}

Our debiased Whittle maximum-likelihood estimation strategy provides an
(asymptotically) unbiased estimator by incorporating the effects of finite data,
possessing edges either hard or smooth in transition, regular or irregular in
boundary, and sampling patterns coarse or fine, discretized by spaces even or
agape. The results from the experiments discussed in Section~\ref{dunno} indicated
that sample size and spatial geometry bleed into estimator covariance, as seen
both from empirical observation and through our analytical prediction
(eq.~\ref{eq:magic}). In the case of real data, as in Section~\ref{sec:modeltest},
that are limited to a single spatial realization, understanding the (well
predicted) distance at which a lone estimate may fall from the truth, and how
increasing sample size according to particular modes of data collection can
reduce this distance, is valuable. Here, we design synthetic experiments that
demonstrate the asymptotic behavior of the estimator bias and its covariance
when sample size increases through three mechanisms: growing domain, infill
(fixed domain), and reduced missingness.

In Fig.~\ref{fig:asymcartoon}, we illustrate in a series of cartoons the
allocation of samples in growing and fixed domains for fully observed, regular
grids. Robust estimation of the underlying Mat\'ern model from a realization
requires that the sampling grid is crafted or observed such that it is
sufficiently large in terms of how the correlation length $\pi\rho$ relates to
the physical grid size and its sample spacing (Fig.~\ref{fig:asymcartoon},
\emph{left}). Growing-domain \cite[][]{Zhang2004} asymptotics
(Fig.~\ref{fig:asymcartoon}, \emph{center}) are concerned with how increasing
sample sizes by increasing the physical grid size with sample spacing ($\Dy$,
$\Dx$) held constant affects the estimator. Infill
\cite[]{Cressie1993,Zhang+2005}, or fixed-domain \cite[]{Stein1995,Stein99},
asymptotics (Fig.~\ref{fig:asymcartoon}, \emph{right}), on the other hand,
explore increases in sample sizes through densifying the sampling grid by
decreasing the spacing between samples such that the physical grid size ($\Ny\Dy
\times \Nx\Dx$) is held constant. Mixed-domain asymptotics, as the name
suggests, can capture the interplay between these end-member studies through
sensitivity to local and large-scale features of estimated parametric covariance
structures \cite[]{Chu2023}. Another mechanism for increasing the sample size
of a field is by replacing missing samples with new observations, which is
effectively a variation on fixed-domain asymptotics for an irregular grid with
an irregular filling strategy \cite[]{Deb+2017}. In the growing-domain and
infill asymptotic studies, we will consider observation windows of simulated
fields that are full, regularly sampled rectangular grids, and windows with
66.7\% of the grid observed, such that 33.3\% of the samples are missing at
random. For the missingness-reduction study, we will explore broad proportions
of random deletions and our results will emphasize the dependence of the
estimator's behavior on the true range parameter $\rho_0$ of the simulated model.

In each of the experiments that follow, the model is constant and only a single
feature of the grid is adjusted in each trial. We therefore attribute any
sensitivity in parameter bias or covariance between trials to the design of the
sampling grids. The behavior of $\bar{\mcS}_{\btheta}\ofk$ and its partial
derivatives $\pl\bar{\mcS}_{\btheta}\ofk/\pl\theta$ and their interactions with
the spectral window $|w\ofk|^2$ as dependent on data size $\Ny,\Nx$, sampling
size $\Dy,\Dx$, completeness, and arrangement of the sampling, collectively
control whether the chi-squaredness of the ordinates of the periodogram and the
asymptotic normality of the estimators can be reached
\cite[][]{Olhede+2004,Sykulski+2019,Guillaumin+2022}. Rather than discussing
these effects theoretically, we opt for a suite of practical experiments
conducted in regimes that geoscientists will recognize as relevant to their
experience, if only to guard against the unrealistic optimism that large-sample
behavior will be un-intentionally met or realized. Section~\ref{sec:design} will
discuss means for arriving at desired regimes through the design of experiments.

\subsection{Growing-domain asymptotics}\label{sec:growing}

\begin{figure*}
  \centering
  \includegraphics[width=0.93\textwidth,angle=0,trim=0cm 0cm 0cm 0.7cm,clip]{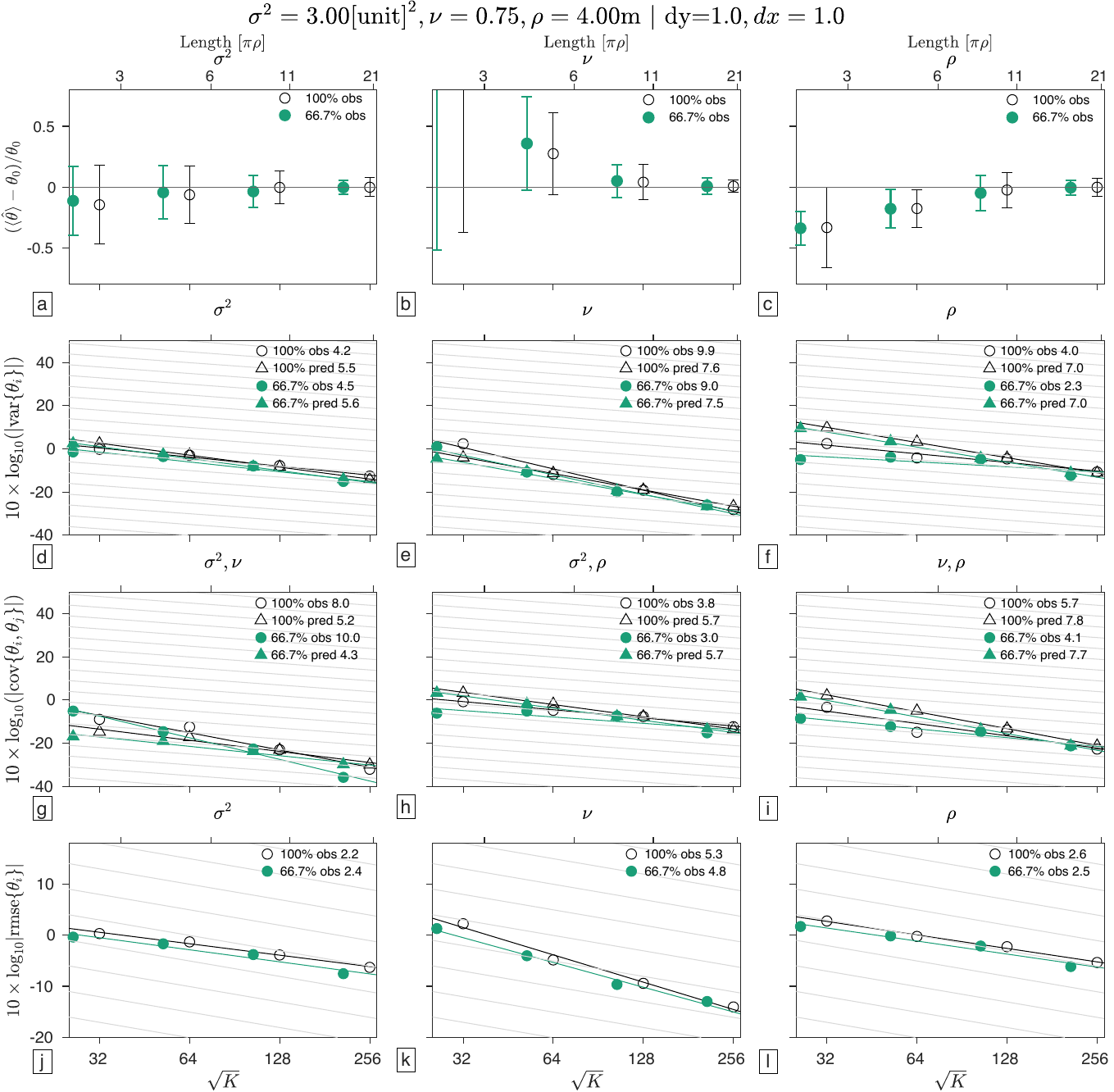}
  \caption{\label{fig:growingdomain} Growing-domain (i.e., constant $\Dy=\Dx$)
    asymptotic behavior of observed and predicted estimator covariance for
    increasing sample size through physical grid length growth of a fully
    (\textit{black}) and partially (66.7\%; \textit{green}) observed random
    field. One hundred realizations are simulated for the Mat\'ern model
    $\btruth = [ 3\, \mathrm{unit}^2\; 0.75\; 4\, \m ]$ with $1\,\m$ spacing in
    both spatial directions of each trial grid. Relative ensemble estimator bias
    $(\langle\hat{\theta}\rangle-\theta_0)/_{}\theta_0$ 
    (\textit{a}, $\sigma^2$; \textit{b}, $\nu$; \textit{c},~$\rho$) with
    empirical standard deviation error bars plotted against the square root
    of the sample size, $\sqrt{K}$. Estimator variance $\var\{\theta_i\}$
    (\textit{d}, $\sigma^2$; \textit{e}, $\nu$; \textit{f}, $\rho$) and absolute
    covariance $|\cov\{\theta_i,\theta_j\}|$ (\textit{g}, \{$\sigma^2,\nu$\};
    \textit{h}, \{$\sigma^2,\rho$\}; \textit{i}, \{$\nu,\rho$\}) from the
    observed ensemble (\textit{circles}) and predicted from eq.~(\ref{eq:magic})
    (\textit{triangles}) scaled as $10\log_{10}(\cdot)$. Root-mean-squared error
    $|\mathrm{rmse}\{\theta_i\}|$ (\textit{j}, $\sigma^2$; \textit{k}, $\nu$;
    \textit{l}, $\rho$) calculated from the ensemble. Absolute best-fit linear
    slopes in the $10\log_{10}-\log_2$ space for the empirical and predicted
    (co)variances and RMSE are reported in the subplot legends for each
    experiment; gray guide lines show $1/\sqrt{K}$ slopes. The number of
    approximate one-third decorrelation lengths ($\pi \rho$) for this grid are quoted on the upper horizontal axis.}
\end{figure*}

In our growing-domain experiments, we perform four simulation trials on evenly
spaced, square grids for which we generate 100 realizations parameterized by the
Mat\'ern model $\btruth=[ 3\, \mathrm{unit}^2\; 0.75\; 4\, \m ]$. Each trial
corresponds to a sample size that increases by a power of two from $32\times 32$
to $256\times 256$, which, for a grid with $1~\m$ spacing between samples,
equates to a physical grid length of approximately $5\pi\rho$ to $20\pi\rho$.
We perform this numerical experiment on the full, regularly sampled grid, and on
a grid 66.7\% observed, with the complementary 33.3\% of samples randomly
deleted over the grid.

In Fig.~\ref{fig:growingdomain}, we present the estimate ensembles yielded by
the simulation trials by comparing their sample mean relative to the
truth~$\btruth$, and their sample parameter (co)variances to their analytical
prediction (eq.~\ref{eq:magic}). For both fully and partially observed grids,
increasing sample size by increasing the size of the physical grid for a fixed
spacing $\Dy,\Dx$ reduces the relative estimator bias
(Fig.~\ref{fig:growingdomain}, \emph{first row}), parameter variance
(Fig.~\ref{fig:growingdomain}, \emph{second row}), parameter covariance
(Fig.~\ref{fig:growingdomain}, \emph{third row}), and the root-mean-squared
error, RMSE (Fig.~\ref{fig:growingdomain}, \emph{fourth row}),
\begin{equation}\label{eq:rmse}
\mathrm{rmse}(\hat{\theta})=\sqrt{\mathrm{var}(\hat{\theta})+\mathrm{bias}^2(\hat{\theta})}
\end{equation}
for all parameter and cross-parameter terms. The rate of decay of each scenario
is similar between the full and partial grid experiments.

\begin{figure*}
  \centering
  \includegraphics[width=0.93\textwidth,angle=0,trim=0cm 0cm 0cm 0cm,clip]{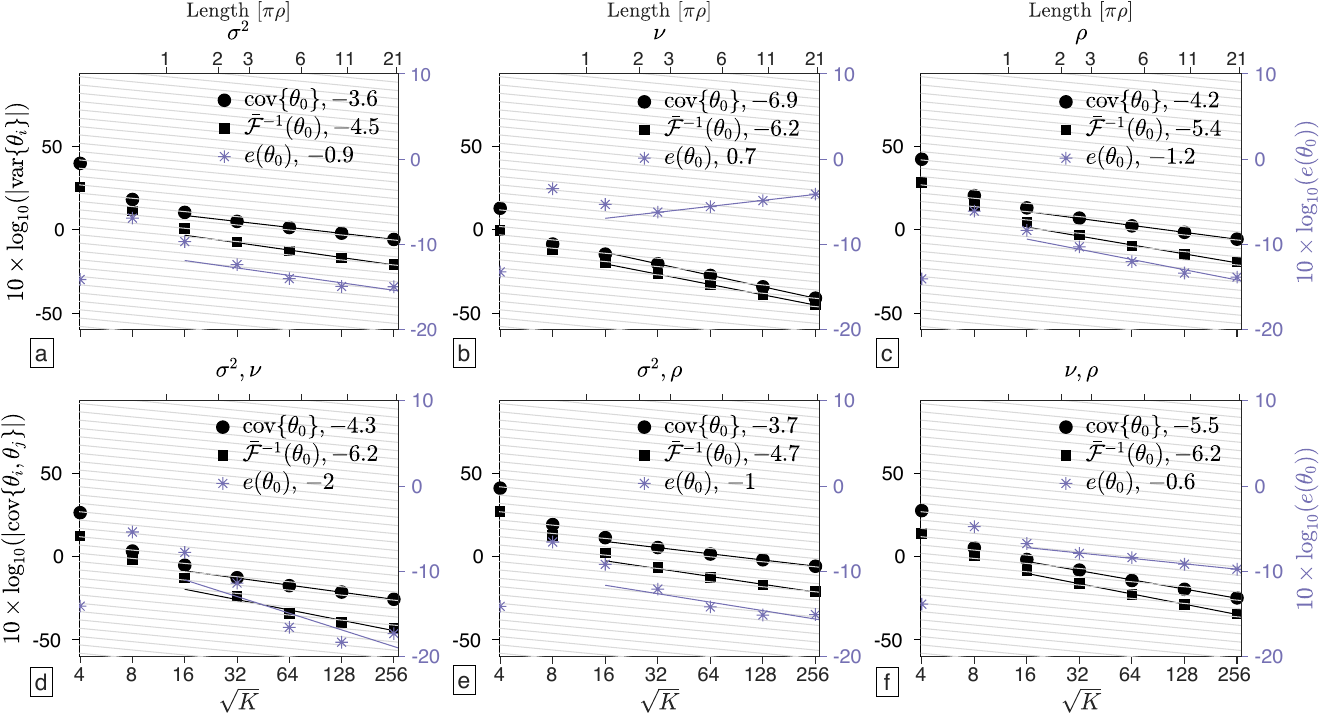}
  \caption{\label{fig:eff} Growing-domain asymptotic behavior of the elements of
    the absolute value of the blurred inverse Fisher information matrix
    (\emph{black squares, left axis}), the absolute value of the analytically
    calculated estimator covariance (\emph{black circles, left axis}), and their
    ratio $e(\btruth)$ (\emph{purple stars, right axis}) as a function of the
    sample size via the grid-length increasing evenly in both spatial directions
    for a fully observed grid. Plot organization and annotation details are as
    in the second and third rows of Fig.~\ref{fig:growingdomain}, with
    horizontal axes in multiples of the correlation length $\pi\rho$ and in
    terms of number of samples $\sqrt{K}$.}
\end{figure*}

The common approximation $\cov\ofth\approx\bar{\bmcF}^{-1}\oftr$ requires
comparison to the full expression of eq.~(\ref{eq:magic}): how well does the
inverse Fisher information matrix approximate the full theoretical calculation
of parameter covariance for a Mat\'ern model in terms of increasing window
sizes?  Inspired by the formulation of the Cram\'er-Rao
\cite[]{Rao1945,Cramer1946} lower bound efficiency, we define
\begin{equation}\label{eq:eff}
  e(\btruth)=\frac{\bar{\mcF}^{-1}(\btruth)}{\cov\{\btruth\}},
\end{equation}
for each parameter (pair), i.e., including the parameter interactions in the
ratio of simulation scheme cross-terms $\bar{\mcF}^{-1}(\theta_{i},\theta_{j})$
and $\cov\{\theta_{i},\theta_{j}\}$.

In Fig.~\ref{fig:eff}, we display these results within the context of a
growing domain for the full, regularly sampled experimental grid and the
Mat\'ern parameters used in the previous experiment
(Fig.~\ref{fig:growingdomain}). Both $\bar{\mcF}^{-1}(\btruth)$ and
$\cov\{\btruth\}$ decrease linearly in the $\log_{10}-\log_2$ space for all
parameter interactions, with the exception of $\{ \sigma^2,\, \nu \}$. With
increasing sample size, $\bar{\mcF}^{-1}(\btruth)$ and $\cov\{\btruth\}$
converge in $\{ \nu, \, \nu \}$, remain constant in $\{ \nu,\, \rho \}$, and
diverge in all other parameter pairings. Repeating this experiment for
additional models, we find that the blurred inverse Fisher information
approximation inconsistently approaches the parameter uncertainty in only some
of the Mat\'ern parameters for some models with small smoothness ($\nu_0< 3$)
parameters observed for grid sizes with lengths of $\sqrt{K}\leq2^{10.5}$. In
comparison to the analytically predicted parameter covariance calculated at the
truth $\cov\{\btruth\}$, we find that $\bar{\mcF}^{-1}(\btruth)$ is an overly
confident estimate of parameter uncertainty. The rate of growth of the ratio for
many of these parameters, at least for the grid sizes we inspect, indicates that
the blurred inverse Fisher information matrix does not consistently approach
parameter uncertainty values and is thus not a sufficient approximation for
sample window sizes that are realistic for much of the geophysical sciences.

\subsection{Infill (fixed-domain)  asymptotics}\label{sec:infill}

\begin{figure}
  \centering
  \includegraphics[width=0.93\textwidth,angle=0,trim=0cm 0cm 0cm 0.7cm,clip]{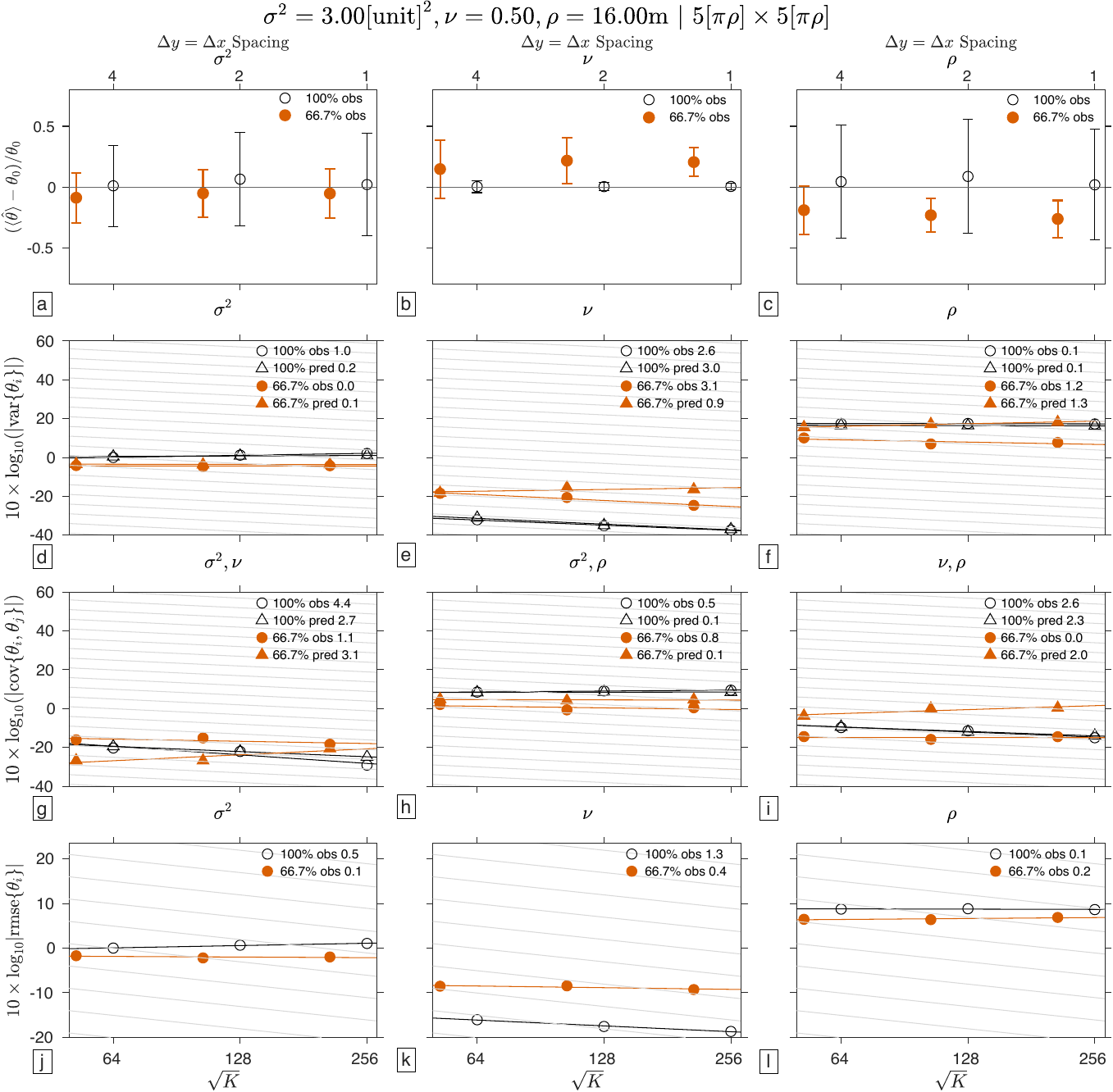}
  \caption{\label{fig:fixeddomain} Fixed-domain (i.e., constant $\Ny\Dy=\Nx\Dx$)
    asymptotic behavior of observed and predicted estimator covariance for
    increasing sample size through grid densification of a fully and partially
    (66.7\%) observed field for the Mat\'ern model $\btruth = [ 3\,
      \mathrm{unit}^2\; 0.5\; 16\, \m ]$ with fixed physical grid-length of $5
    \pi\rho \times 5 \pi\rho$ for 100 realizations of each trial grid. Plot
    organization and annotation are as in Fig.~\ref{fig:growingdomain}.}
\end{figure}

For our infill asymptotics experiments, we design square, evenly spaced grids
with a common physical length dimension of $5\pi\rho$, i.e., relative to the
Mat\'ern range parameter $\rho_0=16\,\m$, that we maintain by decreasing the
sample spacing $\Dy=\Dx$ from $4\,\m$ to $1\,\m$ for sample sizes that increase
by powers of two from $64\times64$ to $256\times256$. We simulate 100
realizations for the Mat\'ern model $\btruth=[ 3\, \mathrm{unit}^2\; 0.5\; 16\,
  \m ]$ on each trial grid for the case of full, regular sampling, and the case
of the partially (66.7\%) observed grid whose complement was randomly
deleted. As we did in the growing-domain experiment (Section~\ref{sec:growing}), we
calculate the relative estimator bias of the ensembles, parameter variance and
covariance from the ensemble and the analytic prediction (eq.~\ref{eq:magic}),
and the RMSE (eq.~\ref{eq:rmse}) of the ensemble.

In Fig.~\ref{fig:fixeddomain}, we show that densifying the full, regularly
sampled grid only reduces the estimation variance of the smoothness parameter,
denoted here as $\var\{\nu\}$, and the estimated parameter covariances that the
smoothness pairs into, which we denote again here as $\cov\{\sigma^2,\nu\}$ and
$\cov\{\nu,\rho\}$. Relative to the full, regularly sampled grid, the partially
observed grid experiment reveals bias in the estimated smoothness $\nu$ and
range $\rho$ parameters that is roughly constant with increasing sample size,
reduced parameter variance in the variance $\st$ parameter, larger parameter
variance in the smoothness $\nu$, and some disagreement in magnitude between the
ensemble and analytically calculated covariances. As increasing the number of
samples within the interior of the fixed domain improves the resolution of the
high-frequency content within the periodogram, to which the smoothness $\nu$
parameter is sensitive, the improvement in parameter (co)variance for the
smoothness $\nu$ should be expected within the infill scenario
\cite[]{Zhang2004}.

\subsection{Reducing missingness}\label{sec:infillirr}

\begin{figure}
  \centering
  \includegraphics[width=0.9\textwidth,angle=0,trim=0cm 0cm 0cm 0.7cm,clip]{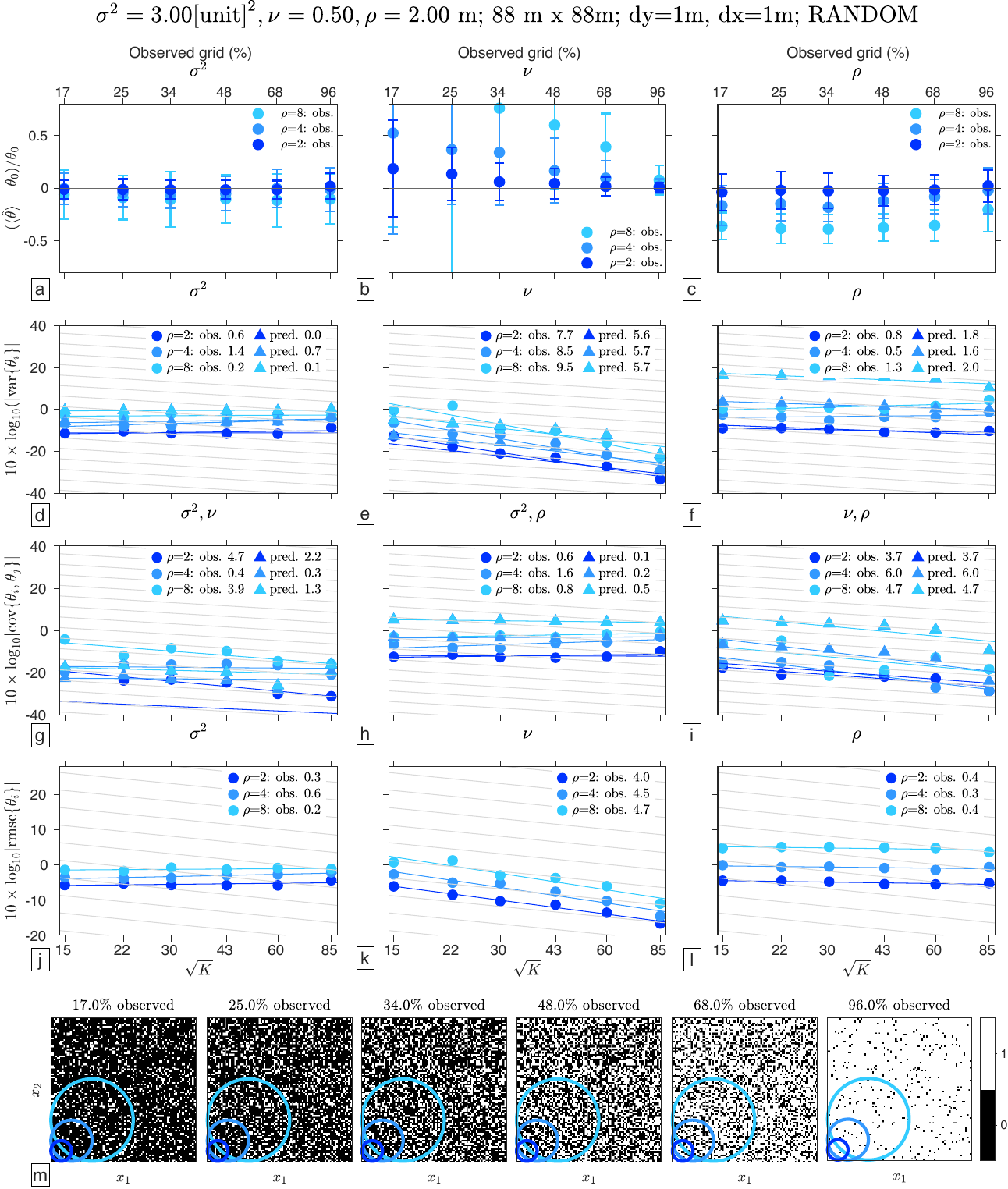}
  \caption{\label{fig:incmissrandom}
   Estimator relative bias (\emph{first row}), variance (\emph{second}),
   covariance (\emph{third}), and RMSE (\emph{fourth}) as a function of
   increasing observations within a fixed domain. Plot organization and
   annotation details for top four rows are as in Fig.~\ref{fig:growingdomain}.
   Experiments are performed for three Mat\'ern models where
   $\st_0=3\,\mathrm{unit}^2$ and $\nu_0=0.5$ for all, with values of
   $\rho_0=2\m$ (light blue; 0.0160 of the grid diagonal), $\rho=4\m$ (medium
   blue, 0.0321 of the grid diagonal), and $\rho=8\m$ (dark blue; 0.0642 of the
   grid diagonal) on a $88\times88$ grid with even 1~m spacing. The spatial
   taper accumulates random samples after each trial of 100 realizations for the
   three models considered where the percent observed is distributed from 17\%
   to 96\%. Spatial sampling patterns for each percent observation (omissions
   in black) are shown in the bottom row \textit{(m)}; circle radii correspond to the
   color-coded value of $\pi\rho$ for visual context of the correlation
   lengths.}
\end{figure}

We consider how reducing missingness in the sampling of an observed random field
affects the estimator and the behavior of its uncertainty. For increasing
sample sizes, we replace irregularly omitted samples with data randomly
distributed across the observation window. This experiment is akin to infill
asymptotics in that we are (irregularly) densifying the sampling of an irregular
grid as a function of sample size. 

Fig.~\ref{fig:incmissrandom} displays the results of three experiments
estimating the Mat\'ern parameters for 100 realizations of a field with
accumulating observations to simulate an increasing percentage of observation
for three Mat\'ern models with a common sampling grid and variance $\sigma^2_0$
and smoothness $\nu_0$ parameters, and with three range parameters $\rho_0=\{2,
4, 8\}$ that correspond to 1.6\%, 3.2\%, and 6.4\% of the diagonal length of the
sampling grid. The relative estimator bias (Fig.~\ref{fig:incmissrandom},
\emph{first row}) reveals a positive dependence scaling with the magnitude of the range
parameter~$\rho_0$. For small enough~$\rho_0$, the average parameter estimate is
within 1 standard deviation (s.d.)  of $\btheta_0$ for each trial grid
size. However, for large~$\rho_0$, the parameter bias exceeds 1 s.d. for
$\nu$ and $\rho$. Parameter variance (Fig.~\ref{fig:incmissrandom},
\emph{second row}) increases at a slow linear rate for $\sigma^2$,
decreases rapidly in $\nu$, and decreases slowly in $\rho$ as the sample
size and percent observation increase. Parameter covariance
(Fig.~\ref{fig:incmissrandom}, \emph{third row}) increases rapidly in estimated
$\{\sigma^2,\nu\}$ and $\{\nu,\rho\}$, but negligibly for
$\{\sigma^2,\rho\}$. As in the infill experiment, the largest improvements
are observed for the smoothness parameter, such as in the high rate of decay in
its RMSE (\emph{fourth}) for all three $\rho_0$ models.

\section{{\hsps}D~E~S~I~G~N~I~N~G{\hsps}S~P~A~T~I~A~L{\hsps}S~A~M~P~L~I~N~G{\hsps}E~X~P~E~R~I~M~E~N~T~S}\label{sec:design}

Sampling in the spatial domain shapes information in the spectral domain, which
affects our statistical knowledge of the underlying process. As we illustrated
in our realistic synthetic experiments in
Figs~\ref{fig:specialsamples}--\ref{fig:merged}, the geometry of the observation
window contributes to parameter estimate covariance, in terms of both the spread
of estimates and the strength of the parameter correlation. The spatial window
$w\ofx$ appears in the calculation of both the blurred inverse Fisher
information matrix (eq.~\ref{eq:fisherinfo}) and the covariance of the blurred
score (eq.~\ref{eq:covscore}), which itself depends on the sampling-window
informed covariance of the periodogram (eqs~\ref{eq:percov}--\ref{eq:percov3}).
As our calculation of predicted parameter covariance (eq.~\ref{eq:magic})
depends on these terms, the expected behavior of the model estimates ultimately
hinges on the sampling window $w\ofx$. As we wrote in Section~\ref{sec:mle}, the
debiasing effect of the sampling pattern blurring the spectral density can be
understood through the spectral window $|w\ofk|^2$ and its interplay with
$\bar{\mcS}_{\btheta}\ofk$ and $\pl\bar{\mcS}_{\btheta}\ofk/\pl\theta$ and their
asymptotics already discussed in Section~\ref{sec:asymp}.

Gaining an intuitive appreciation for how sampling geometry influences estimator
biases and uncertainties is relevant for a variety of signal processing tasks,
e.g., spectral estimation methods for irregularly sampled processes
\cite[e.g.,][]{Bronez88,Abreu+2016}, function approximation on bounded domains
\cite[e.g.,][]{Simons+2011a,Matthysen+2018}, and basis representations of
covariance spaces \cite[e.g.,][]{Buell1975,Buell1979,Richman1986}. The imprints
of the sampling domain need to be accounted for lest they manifest and be
misinterpreted as signal \cite[e.g.,][]{Lehr+2025}. Other disciplines that offer
valuable perspectives for viewing the structure imparted by spatial sampling
patterns include topology \cite[]{Ibanez+1996} and graph signal processing
\cite[]{Chen+2016}. In the analog setting of (Fourier) optics, a given aperture
$w\ofx$ can be shaped like a lens, and then the spectral window $|w\ofk|^2$ is a
far-field diffraction pattern.

Moreover, if we could design a sampling geometry for a spatial field from
scratch, or if we had the opportunity to modify a given sampling scenario by the
addition of more sampling points, where should they be allotted? In field
campaigns or remote experiments, pragmatics and logistics and a prudent use of
resources may be additional considerations \cite[]{Evans+2022,Zhang+2025} but
even in the least restricted of experimental settings where samples are cheap
and the region-of-interest is broad, appealing to sampling designs that
prioritize information gain will reduce redundancy in samples collected
and retained \cite[]{Zhang+2024}, alleviating the downstream burdens of large
data in terms of storage and modeling.

For parametric covariance modeling, the first concern of data collection should
be adequately sampling over an extent that is broad enough to capture several
repetitions of the correlation length yet dense enough to prevent its aliasing
(Fig.~\ref{fig:asymcartoon}, \emph{left}). Ideally, the field or remote-sensing
scientist will be aware of the spatial bounds at which the assumption of local
stationarity holds during their sample acquisition. Subsequent data collection
campaigns can consider whether employing a growing-, fixed- or mixed-domain
sampling plan would be most useful (e.g., in terms of parameter resolution) and
whether these approaches are viable (e.g., whether the region-of-interest is
geographically confined). From our studies of growing- and fixed-domain
asymptotics for full and partial, regular and irregular grids
(Figs~\ref{fig:growingdomain}--\ref{fig:incmissrandom}), we observed that
increasing the number of samples generally reduces parameter bias and
(co)variance. An exception to this behavior was observed for estimates of the
variance $\st$ and range $\rho$ in the infill experiments
(Figs~\ref{fig:fixeddomain} and \ref{fig:incmissrandom}), whose performance
either remained unchanged or negligibly degraded, worsening for experiments that
included random missingness, and/or large correlation lengths~$\rho_0$. While
more samples will improve the behavior of spatial covariance model estimates, we
have shown that the rate at which they do is determined by the sampling
pattern. For the regularly sampled grid, the sampling pattern can be modified to
tune the recovery of our estimator by densifying the discretization or by
physically expanding its size. For the partial grid, the geometry of sample
distribution across a domain is an additional factor that must be considered in
the design of spatial sampling experiments.

Optimal spatial sampling \cite[]{Smith1918,Fedorov1971} can be designed to
determine sample size and locations by which to minimize parameter uncertainties
or maximize information gain. Many optimality criteria \cite[][]{Cressie1993}
can be implemented, and for $K$ samples in the plane, one could consider as many
as $\sum^{K}_{n=1} { }_{n}C_{K}$ combinatorial sampling patterns. To quell this
mass of possibilities, the intuition and understanding built up so far can
provide general guidance, and our theory and software for estimation,
simulation, and (data-less) uncertainty quantification for (ir)regularly sampled
fields, fully equips us for a variety of constructions. 
Optimizing a sampling pattern for parameter uncertainty
reduction can be geometrically understood as designing an aperture mask that
concentrates and centers signal in the spectral domain, and reduces destructive
interference by apodization. Such a mask would reduce irregularities, in
particular those from random stippling, in the sampling pattern, which act like
optical aberrations, producing clutter within the spectral window that diffuses
signal.

\begin{figure}
\centering
  \includegraphics[width=0.87\textwidth,trim=0cm 45.2cm 0cm 0cm,clip]{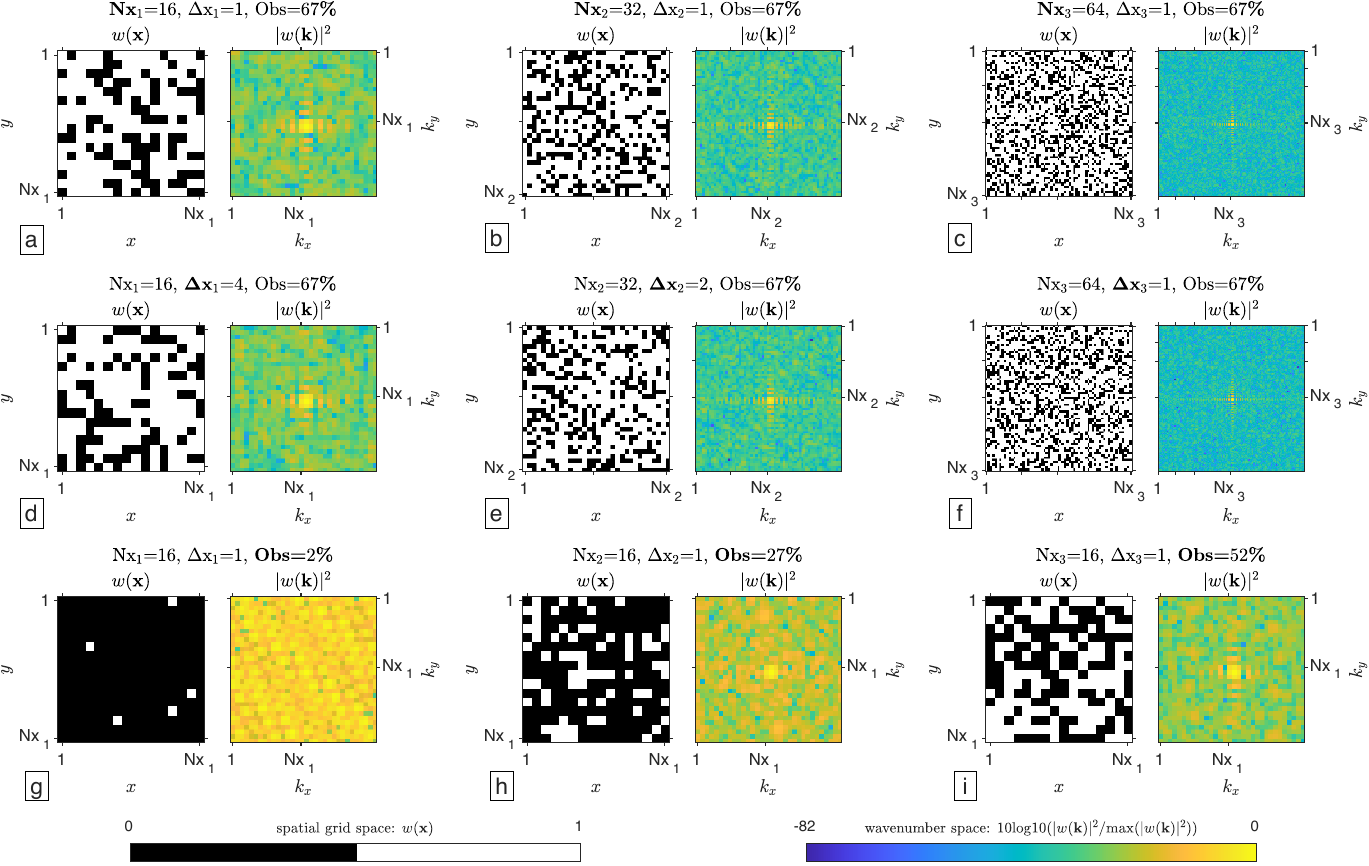}
  \caption{\label{fig:wofk1}
  Spatial $w\ofx$ (\emph{left-most subfigure of each column pair}) and spectral
  $|w\ofk|^2$ (\emph{right-most of each column pair}) windows computed with a
  factor of two zero-padding in both directions for incompletely sampled grids.
  Sampling patterns organized as panels for missingness distributed as a
  uniformly random pattern. The three rows in each panel show allocation
  strategies for increasing sample size such that the physical grid increases in a growing domain
  (\emph{a--c}), spacing decreases in a fixed
  domain (\emph{d--f}), and missingness is reduced in an irregularly filled fixed domain
  (\emph{g--i}). Grid parameters for the number of samples $\Nx_{i}$, spacing
  $\Dx_{i}$, and the percent observed `Obs' of the regular grid are noted in the
  subfigure titles. The subfigures share a common colormap.}
\end{figure}

Figs~\ref{fig:wofk1}--\ref{fig:wofk4} illustrate the appearance of several
sampling geometries as spatial windows and their corresponding spectral
windows. We configure the spatial and spectral window pairs from left to right
for three scenarios of increasing sample size following the three mechanisms we
studied in Section~\ref{sec:asymp}, that is, for a growing domain, fixed domain,
and reduced missingness. We show two styles of dispersed sampling as uniform
random missingness (Fig.~\ref{fig:wofk1}) and checkerboard
(Fig.~\ref{fig:wofk2}) patterns. For the random missingness pattern, increasing
the number of samples by any mechanism results in the amplitude of the spectral
window focusing near the zero wavenumber. This is particularly impactful in the
sparse sampling pattern shown in the bottom left for a 2\% observed $16\times16$
grid whose spectral window has transitioned to a lack of structure that will
efface any modeled spectrum entirely. For the checkerboard pattern, the
orientation of the longest sampled diagonal, which might be purely a matter of
arbitrary availability, generates orthogonal diagonals in the spectral window,
which will impart an indelible signature on the recovered spectrum. We revisit
sampling contiguous patches, including within the interior
(Fig.~\ref{fig:wofk3}) and the exterior (Fig.~\ref{fig:wofk4}) of an irregular
boundary. In general, we have observed parameter (co)variance to be reduced for
sampling patterns that follow contiguous patches rather than dispersive
stippling (e.g., Fig.~\ref{fig:specialsamples}). Signal is concentrated over a
broader range of low wavenumbers compared to the dispersed sampling examples,
with clutter at higher wavenumbers reduced in magnitude for larger sample
sizes. For both end-member (dispersed and contiguous) irregular sampling
scenarios, the imprint of the sampling window $w\ofx$ induces a diffraction
pattern, plainly visible in the spectral window $w\ofk$, and mathematically
carried by the very terms that control the estimator and shape its uncertainty,
that is, the blurred likelihood and its various derivatives.

\begin{figure}
\centering
  \includegraphics[width=0.87\textwidth,trim=0cm 30.5cm 0cm 15cm,clip]{figure23through26.pdf}
  \caption{\label{fig:wofk2}
  Spatial $w\ofx$ (\emph{left-most subfigure of each column pair}) and spectral
  $|w\ofk|^2$ (\emph{right-most of each column pair}) windows computed with a
  factor of two zero-padding in both directions for incompletely sampled grids
  following a checkerboard pattern. Organization of rows within the panels is
  as in Fig.~\ref{fig:wofk1} with grid parameters listed in the subfigure
  titles.}
\end{figure}

\begin{figure}
  \centering
  \includegraphics[width=0.87\textwidth,trim=0cm 15cm 0cm 30.25cm,clip]{figure23through26.pdf}
  \caption{\label{fig:wofk3}
  Spatial $w\ofx$ (\emph{left-most subfigure of each column pair}) and spectral
  $|w\ofk|^2$ (\emph{right-most of each column pair}) windows computed with a
  factor of two zero-padding in both directions for incompletely sampled grids
  constrained to the interior of the boundary of contiguous geographic patches.
  Organization of rows within the panels is as in Fig.~\ref{fig:wofk1} with grid
  parameters listed in the subfigure titles.}
\end{figure}

\begin{figure}
\centering
  \unboxed{\includegraphics[width=0.87\textwidth,trim=0cm 0cm 0cm 45.25cm,clip]{figure23through26.pdf}}
  \caption{\label{fig:wofk4}
  Spatial $w\ofx$ (\emph{left-most subfigure of each column pair}) and spectral
  $|w\ofk|^2$ (\emph{right-most of each column pair}) windows computed with a
  factor of two zero-padding in both directions for incompletely sampled grids
  constrained to the exterior of the boundary of contiguous geographic patches.
  Organization of rows within the panels is as in Fig.~\ref{fig:wofk1} with grid
  parameters listed in the subfigure titles.}
\end{figure}

\section{C~O~N~C~L~U~S~I~O~N~S}

Irregularly sampled spatial data are pervasive in the geophysical sciences.
Treating such observations as random fields presents the opportunity to
characterize them as single realizations of a statistical process. We have
demonstrated how to properly account for sampling patterns, edges and omissions,
by incorporating a (weighted) window of observation into a spectral-domain
maximum-likelihood analysis. Given incomplete and irregularly spaced
observations we estimate the covariance structure of a locally non-varying-mean
field as a stationary, isotropic, parametric Mat\'ern form. Our debiased Whittle
maximum-likelihood estimation relies on the exact blurring of the modeled
spectral density as the expected periodogram to balance the finite-field effects
that affect the (windowed) periodogram of the data. While we show that we indeed
can exclude wave-vector correlations for efficient parameter estimation, we also
did demonstrate that these correlations can be completely calculated and
correctly incorporated through the periodogram covariance for an uncertainty
assessment that fully matches empirically observed behaviors. In realistic
simulation scenarios and by analysis of real geophysical data, we substantiate
that Mat\'ern parameter estimates, their aleatoric uncertainty, and epistemic
model goodness-of-fit are fully and reliably quantifiable. Correctly fixing the
smoothness~($\nu$) to special familiar models that only parameterize
variance~($\st$) and range~($\rho$) yields estimate ensembles and parameter
covariances comparable to when all three parameters of the general Mat\'ern
class are inverted for. Crucially, incorrect \textit{a priori} assumptions
impart considerable bias. Little is gained beyond functional simplicity in
selecting a special case over the general Mat\'ern form: we should fully
estimate it in real-data scenarios when we are without knowledge of the
underlying process. We provide the numerical tools for implementation,
estimation, and simulation, and the analytical expressions for connecting the
general Mat\'ern with its special cases in both spatial and spectral
representations to make this suggestion practical. Spatially merged processes
are separable when their boundaries, however irregular, are known \textit{a
  priori} and accounted for, but their analysis on an encompassing rectangular
grid, violating the modeling assumption of stationarity, leads to strong
parameter bias. Theory and simulation reveal that sample size and geometry play
a considerable role in parameter (co)variance, which we understand through the
spectral window. Simulations under growing-domain and infill asymptotic regimes
for full or partial observations, with sampling irregularities in the way of
boundaries, structure, or missingness, reveal the rates at which estimator
performance improve with sample size, setting the stage for experimental design
and (optimal) sample allocation. Analysis and synthesis of planetary-scale
geophysical data, tapestries of time and terrain, as random fields, is efficient
under our framework, and our ability to critically assess it will enable
relaxing our current null hypotheses that such data are Gaussian, stationary,
isotropic, random, Mat\'ern, fields.

\section{D~A~T~A{\hsps}A~V~A~I~L~A~B~I~L~I~T~Y}\label{code}

All the code used to conduct the calculations and produce the figures in this
paper made in \textsc{Matlab} by the authors is openly available as
Release~2.0.0 from \url{https://github.com/csdms-contrib/slepian_juliet}, doi:
\texttt{10.5281/zenodo.4085253}. For an alternative code base in Python, see
also \cite{Guillaumin+2026}. Also compare with the \textit{R} package by
\cite{Paciorek2007}.  The seafloor bathymetry model \texttt{GEBCO\_2024} and its
Type Identifier Grid are available from
\url{https://www.gebco.net/data-products/gridded-bathymetry-data/gebco2024-grid};
both data products are described by \cite{GEBCO2024}.  The lunar digital
elevation model \texttt{SLDEM2015} is available from
\url{https://pgda.gsfc.nasa.gov/products/54} and is described by
\cite{Barker+2016}. The crater catalog of \cite{Robbins2018} is available from
\url{https://astrogeology.usgs.gov/search/map/moon_crater_database_v1_robbins}. The
remote sensing data are available from \url{https://earthexplorer.usgs.gov} for
the Collection 2, Level-2 data products of the path 014, row 032 tile captured
on July 20, 2025 and are further described by \cite{LANDSAT8} and
\cite{LANDSATHandbook}.

\section{A~C~K~N~O~W~L~E~D~G~E~M~E~N~T~S}

OLW gratefully acknowledges financial support from the Schmidt DataX Fund, grant
number 22-008, at Princeton University, made possible through a major gift from
the Schmidt Futures Foundation, and NSF grants EAR-2341811 and EAR-2422649, and
Gordon and Betty Moore Foundation grant 13614 to FJS. We thank the Associate
Editor, Carl Tape, and the reviewers, Andrew Valentine and John Goff, for very
constructive reviews that helped improve the manuscript.

\bibliographystyle{gji}
\bibliography{olw.bib}

\clearpage
\section*{A~P~P~E~N~D~I~X}

\renewcommand{\thesubsection}{\Alph{subsection}}
\renewcommand{\thetable}
{\thesubsection\arabic{table}}

\subsection{Tables}
\label{app:tabs}
\setcounter{table}{0}

\begin{table}
    \centering
    \small
    \begin{tabular}{| c c c c |}
    \hline
 Smoothness~$\nu$ & Normalized spatial covariance~$\mcC_{\btheta}(r)/\st$ & Normalized spectral density~$\mcS^2_{\btheta}(k)/\st$ & Street name \\
    \hline   
    $\hspace{0.25em} {1}/{3}\hspace{0.25em}$ & $
        \frac{4^{1/3}}{\Gamma(\frac{1}{3})} \left( \frac{2\sqrt{3}}{3\pi\rho} r
        \right)^{1/3} K_{1/3}\left( \frac{2\sqrt{3}}{3\pi\rho} r \right)$ &
        $\frac{4^{1/3}\pi \rho^2 }{(4 + 3
        \pi^2 \rho^2 k^2)^{4/3}}$ 
        & von K\'arm\'an \\
    $1$ & $ \left( \frac{2}{\pi\rho} r \right) K_1\left( \frac{2}{\pi\rho} r
        \right)$ & 
        $\frac{4\pi \rho^2}{(4+ \pi^2 \rho^2 k^2)^2}$ 
        & Whittle \\
    ${1}/{2}$ & $ \exp{\left( - \frac{\sqrt{2}}{\pi\rho} r \right)}$ &
        $\frac{2\pi \rho^2 }{(4 +
        2 \pi^2 \rho^2 k^2)^{3/2}}$ 
        & Exponential \\
    ${3}/{2}$ & $ \exp{\left( - \frac{\sqrt{6}}{\pi\rho} r \right)} \left(1 +
        \frac{\sqrt{6}}{\pi\rho} r \right)$ & 
        $\frac{9\sqrt{6}\pi \rho^2 }
        {(6 + \pi^2 \rho^2 k^2)^{5/2}}$ 
        &  Autoregressive
        (2nd order) \\
    ${5}/{2}$ & $ \exp{\left( - \frac{\sqrt{10}}{\pi\rho} r \right)} \left(1 +
        \frac{\sqrt{10}}{\pi\rho} r + \frac{10}{3 \pi^2 \rho^2} r^2 \right)$ &
        $\frac{250\sqrt{10}\pi \rho^2 }{(10 +
        \pi^2 \rho^2 k^2)^{7/2}}$ 
        & Autoregressive (3rd order) \\
      $\infty $ & $ \exp{\left(-\frac{1}{\pi^2
        \rho^2}r^2\right)} $ & $ \left(\frac{\pi \rho^2}{4}\right)
        \exp{\left(-\frac{\pi^2 \rho^2 k^2}{4}\right)}$ & Squared exponential\\
    \hline
    \end{tabular}
    \caption{\label{tab:MaternSpecial}
      Analytic forms of the isotropic Mat\'ern covariance and two-dimensional
      spectral density for special values of $\nu$. The $d$-dimensional spectral
      density expressions are provided in Table~\ref{tab:MaternSpecialAll}. All
      simplifications are trivial, with $\lim_{\nu\rightarrow \infty}$ requiring
      the application of L'Hospital's rule in triplicate.}
\end{table}

\begin{table}
    \centering
    \small
    \begin{tabular}{| c c c |}
    \hline
    Smoothness~$\nu$ & Normalized spatial covariance~$\mcC_{\btheta}(r)/\st$ & Normalized spectral density~$\mcS^d_{\btheta}(k)/\st$\\
    \hline   
    $\hspace{0.25em} {1}/{3}\hspace{0.25em}$ & $
        \frac{2^{2/3}}{\Gamma(\frac{1}{3})} \left( \frac{2\sqrt{3}}{3\pi\rho} r
        \right)^{1/3} K_{1/3}\left( \frac{2\sqrt{3}}{3\pi\rho} r \right)$ 
        & $\frac{\Gamma(1/3+d/2)}{\Gamma(1/3)}\frac{4^{1/3}(3 \pi \rho^2)^{d/2} }{(4 + 3
        \pi^2 \rho^2 k^2)^{1/3+d/2}}$
\\
    $1$ & $ \left( \frac{2}{\pi\rho} r \right) K_1\left( \frac{2}{\pi\rho} r
        \right)$ 
        & $\Gamma(1+d/2)\frac{4(\pi \rho^2)^{d/2}}{(4+ \pi^2 \rho^2 k^2)^{1+d/2}}$
\\
    ${1}/{2}$ 
        & $ \exp{\left( - \frac{\sqrt{2}}{\pi\rho} r \right)}$ 
        & $\frac{\Gamma\left(1/2+d/2\right)}{\sqrt{\pi}}\frac{\sqrt{2}(\pi\rho^2)^{d/2}}{(2 +  \pi^2 \rho^2 k^2)^{(1+d)/2}}$ 
\\
    ${3}/{2}$ & $ \exp{\left( - \frac{\sqrt{6}}{\pi\rho} r \right)} \left(1 +
        \frac{\sqrt{6}}{\pi\rho} r \right)$ 
        & $\frac{2\,\Gamma\left(3/2+d/2\right)}{\sqrt{\pi}}\frac{6\sqrt{6}(\pi\rho^2)^{d/2}}{(6 +  \pi^2 \rho^2 k^2)^{(3+d)/2}}$
\\
    ${5}/{2}$ 
        & $ \exp{\left( - \frac{\sqrt{10}}{\pi\rho} r \right)} \left(1 +
        \frac{\sqrt{10}}{\pi\rho} r + \frac{10}{3 \pi^2 \rho^2} r^2 \right)$ 
        & $\frac{4\,\Gamma\left(5/2+d/2\right)}{3\,\sqrt{\pi}}\frac{100\sqrt{10}(\pi\rho^2)^{d/2}}{(10 +  \pi^2 \rho^2 k^2)^{(5+d)/2}}$
\\
        $(n+1)/2$ &
        $\exp{\left(
    -\frac{2\sqrt{n+1/2}}{\pi\rho}r \right)} \frac{n!}{(2n)!}
    \sum^n_{k=0} \frac{(n+k)!}{k!(n-k)!}\left(
    \frac{4\sqrt{n+1/2}}{\pi\rho} r \right)^{n-k}$ 
        & $\frac{\Gamma\left(n+(1+d)/2\right)}{\Gamma\left(n+1/2\right)}
        \left(\frac{(4(n+1/2))^{n+1/2}(\pi\rho^2)^{d/2}}{4(n+1/2)+\pi^2\rho^2k^2}\right)^{n+(1+d)/2}$
 \\
      $\infty $ & $ \exp{\left(-\frac{1}{\pi^2
        \rho^2} r^2\right)} $ 
        & $\left( \frac{\pi\rho^2}{4} \right)^{d/2} \exp \left(- \frac{ \pi^{2} \rho^{2} k^2 }{ 4 } \right)$
\\
    \hline
    \end{tabular}
    \caption{\label{tab:MaternSpecialAll}
     Simplified analytic forms of the isotropic Mat\'ern covariance and
     $d$-dimensional spectral density for special values of $\nu$ for comparison
     with the two-dimensional spectral density Table~\ref{tab:MaternSpecial}, with the addition of the general half-integer smoothness case, $\nu=(n+1)/2$.}
\end{table}

\begin{table}
    \centering
    \begin{tabular}{rcccccccccccc}
         Taper    & $\btheta$          & \% Obs & $\overline{\st}\,\mathrm{units}^2$ & $\pm$\% & $\overline{\nu}$ & $\pm$\% & $\overline{\rho}\times10\,\km$ & $\pm$\% & $\{\st,\nu\}$ & $\{\st,\rho\}$ & $\{\nu,\rho\}$ \\\hline
         Full &3p           & 100      &   999.79    & 19.7 \V 19.8 &     1.001 &  1.3 \V  1.5 & 1.986 &   11.4 \V 12.1 &  -50.5 \V -58.7 & 97.6 \V 97.7 & -67.2 \V -74.3 \\
         Full &2c           & 100      &   999.95    & 19.8 \V 18.7 &   {\it 1} &     - \V     - & 1.986 &   10.1 \V 9.6\p &      - \V      - &  - \V - & - \V - \\
         Full &2i           & 100      &  1028.08    & 32.6 \V 11.3 & {\it 0.5} &     - \V     - & 10.404 &  165.3 \V 11.4\p &      - \V      - &  - \V - & - \V - \\
        \hline
         Bound &3p        & 33.6    &   994.57    & 17.1 \V 16.1 &     1.006 &  4.1 \V  3.3 & 1.986 & 10.1 \V 9.3\p&  10.2 \V \p2.3 & 94.2 \V 95.2 & -18.3 \V -19.5 \\
         Bound &2c        & 33.6    &   988.62    & 17.1 \V 15.1 &   {\it 1} &     - \V     - & 1.980 &  9.7 \V  8.6 &      - \V      - & 96.4 \V 95.9 &      - \V      - \\
         Bound &2i        & 33.6    &   695.62    &10.0 \V 10.4 & {\it 0.5} &     - \V     - & 5.806 & 10.6 \V 38.6 &      - \V      - & 98.4 \V 93.4 &   \hp - \V      - \vspace{0.1cm}\\
         Struct &3p      & 33.6    &   996.77    &  9.2 \V  8.9 &     1.005 & 13.6 \V 13.0 & 1.991 & 10.5 \V 10.0&  -0.92 \V-8.4\p & 65.9 \V 61.8 & -62.5 \V -69.5 \\
         Struct &2c      & 33.6    &   991.95    & 10.1 \V9.5\p &   {\it 1} &     - \V     - & 1.997 &  8.7 \V  8.1 &      - \V      - & 83.2 \V 81.8 &      - \V      - \\
         Struct &2i      & 33.6    &   833.08    &  8.6 \V  9.2 & {\it 0.5} &     - \V     - & 3.246 & 11.8 \V 16.5 &      - \V      - & 89.9 \V 81.4 &   \hp - \V      - \vspace{0.1cm}\\
         Rand &3p          & 33.6    &   987.82    &  6.8 \V  6.7 &     1.025 & 11.6 \V 11.1 & 1.974 &  7.4 \V  6.3 & -18.6 \V -9.5\p& 70.0 \V 73.7 & -68.6 \V -53.4 \\
         Rand &2c          & 33.6    &  1006.43    &  7.0 \V  7.0 &   {\it 1} &     - \V     - & 1.997 &  5.5 \V  5.8 &      - \V      - & 82.5 \V 80.4 &      - \V      - \\
         Rand &2i          & 33.6    &   971.84    &  6.3 \V  6.8 & {\it 0.5} &     - \V     - & 2.410 &  8.9 \V  7.8 &      - \V      - & 85.3 \V 83.6 &      - \V      - \\
         \hline
         Anti-Bnd &3p   & 66.4    &   999.24    & 17.1 \V 14.9 &     1.003 &  3.6 \V  3.5 & 1.991 & 10.2 \V 8.9\p&  35.5 \V  33.5 & 92.1 \V 90.6 &   2.7 \V   1.1 \\
         Anti-Bnd &2c   & 66.4    &   989.04    & 15.7 \V 14.4 &   {\it 1} &     - \V     - & 1.980 & 10.7 \V 9.8\p&      - \V      - & 95.7 \V 95.1 &      - \V      - \\
         Anti-Bnd &2i   & 66.4    &   759.05    & 12.9 \V9.3\p & {\it 0.5} &     - \V     - & 6.892 & 13.8 \V 49.5 &      - \V      - & 99.6 \V 93.0 &   \hp - \V      - \vspace{0.1cm}\\
         Anti-Str &3p & 66.4    &   978.19    &  8.9 \V  8.9 &     1.020 &  8.7 \V  9.3 & 1.957 &  8.2 \V  7.5 &\p\;0.4\V  14.0 & 72.3 \V 70.9 & -59.8 \V -52.2 \\
         Anti-Str &2c & 66.4    &  1007.72    &  9.0 \V  9.7 &   {\it 1} &     - \V     - & 2.009 &  6.3 \V  6.5 &      - \V      - & 87.3 \V 83.9 &      - \V      - \\
         Anti-Str &2i & 66.4    &   790.13    &  8.4 \V  6.7 & {\it 0.5} &     - \V     - & 3.733 &\p9.9 \V 15.1 &      - \V      - & 95.0 \V 83.6 &   \hp - \V      - \vspace{0.1cm}\\
         Anti-Rnd &3p     & 66.4    &   995.30    &  7.0 \V  7.2 &     1.002 &  6.7 \V  5.8 & 1.997 &  6.5 \V  6.2 & -21.1 \V -17.5 & 78.7 \V 81.2 & -68.4 \V -60.6 \\
         Anti-Rnd &2c     & 66.4    &   995.18    &  6.9 \V  7.2 &   {\it 1} &     - \V     - & 1.991 &  4.7 \V  5.1 &      - \V      - & 90.9 \V 92.1 &      - \V      - \\
         Anti-Rnd &2i     & 66.4    &   905.53    &  6.2 \V  6.6 & {\it 0.5} &     - \V     - & 2.711 &  7.8 \V  8.1 &      - \V      - & 92.7 \V 90.0 &      - \V      - \\
    \end{tabular}
    \caption{\label{tab:specialsamples}
      Results from the simulation experiments presented in
      Fig.~\ref{fig:specialsamples} for 100 realizations of irregular sampling
      patterns. For each sampling scenario, the experiments are ordered by the 
      inversion for the three Mat\'ern parameters (3p), and the two parameters 
      for correctly (2c) and incorrectly (2i) fixed $\nu$. 
      The columns list the sample means for each of the estimated
      Mat\'ern parameters ($\overline{\theta}_i, i=1,2,3$, that is,
      $\overline{\st}$, $\overline{\nu}$, and $\overline{\rho}$), the parameter
      standard deviations (analytically calculated from eq.~\ref{eq:magic} at
      the mean estimate, followed by the empirical values for the ensemble),
      both listed in per cent of the parameter itself, and the
      parameter correlation in per cent calculated from the normalized
      (analytical $|$ empirical) parameter covariance. All experiment details 
      are as in Fig.~\ref{fig:specialsamples}.}
\end{table}

\begin{table}
    \centering
    \begin{tabular}{ccccccccccccc}
         Exp & Obs    & $w\ofx$& $\overline{\st}\,\mathrm{units}^2$ & $\pm$\%      &$\overline{\nu}$& $\pm$\%        &$\overline{\rho}\,\km$& $\pm$\% & $\{\st,\nu\}$ & $\{\st,\rho\}$ & $\{\nu,\rho\}$ \\\hline
         Ia & full    & unit   & 1.14             &  8.1 \V  7.3 & 0.65           & 2.7  \V 2.6    & 1.788 & 8.0  \V  7.7 & -28.0 \V -32.7 & 90.5 \V 89.5 &\p-63.5\V -68.2 \vspace{0.1cm}\\
         Ib & mix     & unit   & 1.47             & 19.0 \V 10.8 & 0.65           & 2.6  \V 2.8    & 2.436 & 18.9 \V9.6\p & -41.0 \V -16.2 & 95.9 \V 76.3 &\p-63.7\V -43.0 \vspace{0.1cm}\\
         Ic & mask    & unit   & 0.42             &  2.4 \V  2.8 & 0.62           & 2.7  \V 3.7    & 1.601 &  6.4 \V  7.4 & -24.7 \V -24.7 & 87.6 \V 81.6 & -65.0 \V -67.4 \\ 
         Id & anti    & unit   & 0.73             &  4.9 \V  6.8 & 0.64           & 2.6  \V 3.2    & 1.747 &\p7.7 \V 10.9 & -26.6 \V -32.9 & 89.7 \V 91.1 & -63.6 \V -65.0 \\
         Ie & sm.mask & unit   & 0.35             &  2.1 \V  2.5 & 0.66           & 2.8  \V 4.2    & 1.609 &  6.3 \V  7.7 & -25.9 \V -22.9 & 88.0 \V 83.8 & -65.4 \V -67.0 \\
         If & sm.anti & unit   & 0.68             &  4.8 \V  7.2 & 0.66           & 2.7  \V 3.4    & 1.785 &\p8.0 \V 12.1 & -28.5 \V -33.4 & 90.6 \V 92.1 &\p-63.7\V -64.1 \vspace{0.1cm}\\
         Ig & mask    & mask   & 1.13             & 10.2 \V9.2\p & 0.65           & 4.3  \V 4.2    & 1.773 & 11.2 \V 10.4 & -22.0 \V -24.8 & 87.6 \V 87.0 & -59.8 \V -61.7 \\
         Ih & anti    & anti   & 1.14             & 10.2 \V9.2\p & 0.65           & 3.7  \V 3.4    & 1.791 & 11.1 \V 10.0 & -23.9 \V -15.0 & 89.5 \V 87.9 &\p-60.0\V -55.1 \vspace{0.1cm}\\
         Ii & sm.mask & sm.mask& 1.13             & 10.5 \V 10.0 & 0.65           & 4.2  \V 4.6    & 1.771 & 11.2 \V 11.3 & -27.0 \V -34.1 & 88.7 \V 88.9 & -65.2 \V -69.0 \\
         Ij & sm.anti & sm.anti& 1.14             & 10.9 \V 10.0 & 0.65           & 3.8  \V 3.5    & 1.792 & 11.6 \V 10.7 & -32.4 \V -25.0 & 90.5 \V 88.6 & -66.9 \V -63.5 \\
         \hline
         IIa & full   & unit   & 2.84             & 26.9 \V 26.3 & 1.76           & 3.6  \V 3.5    & 2.942 & 11.3 \V 11.1 & -40.5 \V -36.9 & 89.5 \V 84.4 &\p-72.0\V -74.6 \vspace{0.1cm}\\
         IIb & mix    & unit   & 2.60             & 25.1 \V 23.9 & 0.70           & 1.2  \V 1.5    & 3.890 & 31.1 \V 29.0 & -46.2 \V -9.3\p& 97.0 \V 91.6 &\p-65.8\V -36.1 \vspace{0.1cm}\\
         IIc & mask   & unit   & 0.51             &  3.4 \V  2.5 & 0.74           & 1.1  \V 2.8    & 2.665 & 14.2 \V 5.8\p& -50.1 \V -57.3 & 96.9 \V 64.2 & -69.0 \V -45.9 \\
         IId & anti   & unit   & 2.38             & 27.4 \V 25.1 & 0.87           & 1.9  \V 3.7    & 4.676 & 37.2 \V 30.8 & -36.4 \V 15.2\;& 94.2 \V 80.1 & -65.2 \V -24.4 \\
         IIe & sm.mask& unit   & 0.32             &  2.2 \V  1.5 & 1.06           & 1.6  \V 4.9    & 2.314 &  9.6 \V 4.9  & -58.4 \V -63.3 & 96.5 \V 66.5 & -76.8 \V -62.4 \\
         IIf & sm.anti& unit   & 2.55             & 28.4 \V 27.4 & 1.20           & 2.6  \V 5.2    & 4.073 & 24.5 \V 22.5 & -35.5 \V9.2\p\;& 90.9 \V 76.0 &\p-70.4\V -38.3 \vspace{0.1cm}\\
         IIg & mask   & mask   & 2.78             & 22.9 \V 20.8 & 1.81           &18.5  \V17.4    & 2.933 & 13.4 \V 12.4 & -16.0 \V -4.8\p& 74.4 \V 67.9 & -72.1 \V -67.0 \\
         IIh & anti   & anti   & 2.76             & 18.4 \V 19.2 & 1.80           &23.3  \V22.7    & 2.977 & 16.4 \V 15.2 & -18.0 \V 10.8\,\,& 53.3 \V 41.9 &\p-84.4\V -69.5 \vspace{0.1cm}\\
         IIi & sm.mask& sm.mask& 2.82             & 30.2 \V 31.7 & 1.80           &13.9  \V12.1    & 2.918 & 14.5 \V 14.5 &  -4.1 \V  1.8\; & 85.1 \V 85.8 & -51.4 \V -42.7 \\
         IIj & sm.anti& sm.anti& 2.75             & 20.2 \V 21.7 & 1.83           &21.8  \V20.6    & 2.927 & 15.8 \V 14.7 & \p0.9 \V  16.7 & 51.1 \V 50.2 & -76.3 \V -61.8 \\
    \end{tabular}
    \caption{\label{tab:merged}
    Results from the simulation experiments presented in Fig.~\ref{fig:merged}
    for 10 scenarios over 200 realizations of (spatially merged) Mat\'ern
    processes ($I$, $I\!I$). For each process the experiments are ordered as
    labeled (a--j) in Fig.~\ref{fig:merged}. As in
    Table~\ref{tab:specialsamples}, the columns list the sample means for each
    Mat\'ern parameter~$\overline{\theta_i}$, the relative standard deviation
    (analytically calculated from eq.~\ref{eq:magic} at the mean estimate, and
    empirically obtained from the ensemble), and the parameter correlations
    from the (analytical $|$ empirical) parameter covariance. All experiment
    details are as in Fig.~\ref{fig:merged}.\vspace{-1em} }
\end{table}

\end{document}